\documentclass[twocolumn,showpacs,preprintnumbers,amsmath,amssymb,
prd,a4paper,floatfix]{revtex4}

\usepackage{graphics}
\usepackage{graphicx}
\usepackage{epsfig}
\usepackage{color}
\usepackage{longtable}
\usepackage{hyperref}
\usepackage{latexsym}
\usepackage{amsmath}
\usepackage{amsthm}
\usepackage{amsfonts}
\usepackage{amssymb}
\usepackage{dsfont}
\usepackage{bm}
\usepackage{graphics,psfrag}
\usepackage{graphicx,psfrag}
\usepackage{subfigure}
\newcommand{\ov}[1]{\overline{#1}}
\newcommand{\be}{\begin{equation}}
\newcommand{\ee}{\end{equation}}
\newcommand{\bey}{\begin{eqnarray}}
\newcommand{\eey}{\end{eqnarray}}

\newcommand{\DCI}{\ensuremath{D_\srm{CI}} }

\newcommand{\srm}[1]{\textrm{\scriptsize{#1}}}

\newcommand{\E}{\mathrm{e}}
\newcommand{\FC}{\;,}

\renewcommand{\imath}{\mathrm{i}}

\begin{document}
%
\title{Meson and baryon spectrum for QCD with two light dynamical quarks}
\author{Georg P. Engel$^1$, 
C. B.~Lang$^1$, 
Markus Limmer$^1$, 
Daniel Mohler$^{1,2}$, and
Andreas Sch\"afer$^3$\\
\vspace*{2mm}
(BGR [Bern-Graz-Regensburg] Collaboration)
\vspace*{2mm}\\}

\affiliation{
$^1$Institut f\"ur Physik, FB Theoretische Physik, Universit\"at
Graz, A--8010 Graz, Austria\\
$^2$TRIUMF, 4004 Wesbrook Mall Vancouver, BC V6T 2A3, Canada \\
$^3$Institut f\"ur Theoretische Physik, Universit\"at
Regensburg, D--93040 Regensburg, Germany
}

\date{\today}

\begin{abstract}
We present results of meson and baryon spectroscopy using the Chirally Improved
Dirac operator on lattices of size $16^3 \times 32$ with two mass-degenerate
light sea quarks. Three ensembles with pion masses of $322(5)$, $470(4)$ and
$525(7)$ MeV and lattice spacings close to $0.15\,$ fm are investigated. Results
for ground and excited states for several channels are given, including spin two
mesons and hadrons with strange valence quarks. The analysis of the states is
done with the variational method, including two kinds of Gaussian sources and
derivative sources. We obtain several ground states fairly precisely and find
radial excitations in various channels. Excited baryon results seem to suffer
from finite size effects, in particular at small pion masses. We discuss the
possible appearance of scattering states, considering masses
and eigenvectors.  Partially quenched results in the scalar channel suggest the
presence of a 2-particle state, however, in most channels we cannot identify
them.
Where available, we compare our results to results of quenched simulations using the same action.
\end{abstract}

\pacs{11.15.Ha, 12.38.Gc}
\keywords{Hadron spectroscopy, dynamical fermions}

\maketitle

\section{Introduction}

The overwhelming majority of hadronic states in the Particle Data Group's
collection are hadron excitations \cite{PDG08}. QCD as the theory of  strong
interactions should give the whole spectrum  of hadrons, based on the few quark 
mass parameters and a scale.  So far, the lattice regularized form of QCD
provides the only known way to perform ab-initio calculations of these
observables.  Only recently there have been lattice calculations with dynamical
quark masses close to their physical values; most calculations still rely on
extrapolations from unphysically heavy quarks. A reliable determination of the
excited states still remains a hard challenge. The present calculation is
another step in this enterprise.

For decades a lattice realization of chiral symmetry has been an obstacle; quite
early it was observed \cite{Ginsparg:1981bj} that a solution might be Dirac
operators obeying an non-linear relation, the so-called Ginsparg-Wilson (GW)
condition, overlooked for many years. Meanwhile we know several Dirac operators
obeying that condition.  One of them, the overlap operator
\cite{Neuberger:1997fp,Neuberger:1998wv}, has an explicit construction based on
a domain-wall approach  \cite{Kaplan:1992bt,Furman:1994ky} in the limit of
infinite extent of a fifth dimension. Another formulation (perfect fermions) is formally exact 
\cite{Hasenfratz:1993sp}, but, like other approaches, can be constructed only in
an approximate version \cite{Hasenfratz:2005tt}. The lattice version of chiral
symmetry underlying the GW relation has been given in  \cite{Luscher:1998pqa}.
GW-Dirac operators are numerically expensive to construct but have nice
properties like protection from additive mass renormalization or automatic
$\mathcal{O}(a)$ improvement.

Due to the construction method, which, e.g., in the case of overlap fermion, 
involves computation of an operator sign function, simulations with dynamical
GW-fermions are very expensive, typically two orders of magnitude more expensive
than simulations with the simple, improved Wilson Dirac operator.  A problem
especially apparent for GW-fermions concerns tunneling between different
topological sectors. Due to the Atiyah-Singer theorem we know that topological
sectors of the gauge configurations are related to the net number of exact zero
modes of the Dirac operator (counted according to their chirality signature).

The so-called Chirally Improved (CI) Dirac operator is an approximate solution
of the Ginsparg-Wilson equation \cite{Gattringer:2000js,Gattringer:2000qu}. Its
construction is based on a formal parameterization of the Dirac operator
inserted in the GW-equation and solved in truncated from.  This fermion action
has already been used extensively in simulations by the BGR-collaboration in 
the quenched approximation. 
It was found that at least in quenched simulations the
$\mathcal{O}(a^2)$ corrections for baryon masses are small \cite{Gattringer:2003qx} and that
renormalization constants behave similar to an exact chirally symmetric action
\cite{Gattringer:2005ij}.

In this paper we present results of dynamical simulations with two
mass-degenerate light quarks using this action. The parametrization, as well as
details of the simulation and some early results are given in
\cite{Lang:2005jz,Gattringer:2008vj}. The small discretization errors allow us
to use rather coarse lattices in order to save computational costs.

Here we discuss the results concerning several meson and baryon masses derived
on the gauge field configurations of \cite{Gattringer:2008vj}. We present
results of ground states as well as excited states, making use of the
variational method \cite{Michael:1985ne,Luscher:1990ck}. In addition to the
light (dynamical and valence) quarks we also consider another, heavier valence
(strange) quark and include the strange mesons and baryons in our analysis. With
the presently available data we neither perform a continuum limit nor an
infinite volume limit.  This may be justified considering the small
discretization errors and that our values for $m_{\pi}L$ should be large enough
to expect small finite volume effects due to the pion cloud. However, in
addition to pions ``wrapping around the universe'', finite volume effects can
appear when considering large hadrons which may not ``fit in the box''. A
possible interpretation of our results involves such squeezing effects in case
of the excited baryons. Motivated empirically, extrapolation to the physical
pion mass is performed naively, with a fit linear in the pion mass squared.
Preliminary results have been presented in \cite{Engel:2009cq}.

Recent results on light and strange hadron spectroscopy with focus on 
excited states following different approaches can be found in 
\cite{Morningstar:2008mc,Alexandrou:2008bp,Burch:2009wu,
Takahashi:2009ik,Takahashi:2009bu,Fleming:2009wb,
Cohen:2009zk,Mahbub:2009nr,Mahbub:2009cf,Mahbub:2010jz,
Dudek:2009qf,Dudek:2010wm,Bulava:2009jb,Bulava:2010yg,Morningstar:2010ae}.

This paper focuses on results for hadron masses from dynamical CI-simulation
and is organized as follows: In Sect. \ref{SimDetSec} we review the setup and
parameters of the simulation, followed by the discussion of the methods to
extract hadron masses in Sect. \ref{Methods}.   Extrapolation to physical quark
masses and sources for systematic errors are  discussed in Sect.
\ref{SysEffects}.  In Sections \ref{ResultsLight} and \ref{ResultsStrange} we
then present our results on hadron masses.  We conclude in Sect.
\ref{Conclusion}.

\section{Simulation details}
\label{SimDetSec}

\subsection{Fermion action and gauge action}

All details of the simulation method and our checks for equilibration and
autocorrelation are given in \cite{Gattringer:2008vj}. For completeness we
summarize the essential features here.

We use the Chirally Improved Dirac ($\DCI$) operator
\cite{Gattringer:2000js,Gattringer:2000qu}; this is an approximate solution to
the Ginsparg-Wilson equation. It is obtained by insertion of the most general
ansatz for a Dirac operator into the GW equation and comparison of the
coefficients. This leads to an algebraic set of coupled equations, which can be
solved numerically. The accuracy of the approximation depends on the number 
of terms included after the truncation, in our case considering paths to neighbors up to a maximum
path length of 4 lattice units. The paths and coefficients used are found in the
appendix of \cite{Gattringer:2008vj}.  In principle, one could optimize the
coefficients in the parametrization for each gauge coupling and quark mass value
with respect to chiral symmetry. However, defining the setup this way, the
predictive power of the simulation is weakened, and, furthermore, comparison of
different sets of gauge ensembles is more complicated. We therefore decided to
use the same paths and coefficients in all our dynamical runs and thus the bare
Dirac operator is the same in all discussed ensembles. This leads to additive
mass renormalization which is corrected for by determining the PCAC (partially
conserved axial current) mass, also called AWI (axial Ward identity) mass  for
each ensemble.

We include one level of stout smearing \cite{Morningstar:2003gk} as part of the
definition of  $\DCI$ in order to improve the fermion action further.  The
parameters of stout smearing are adjusted such that the value of the plaquette
is maximized ($\rho= 0.165$ in the notation of Ref.  \cite{Morningstar:2003gk}).

It was found that the combination of the $\DCI$ with the tadpole-improved
L\"uscher-Weisz gauge action shows nice properties \cite{Gattringer:2000qu}. We
use this gauge action as discussed in \cite{Gattringer:2008vj}.

\subsection{Algorithm}

We generate the dynamical
configurations with a Hybrid Monte-Carlo (HMC) algorithm \cite{Duane:1987de}, 
with the implementation for  $\DCI$
described in \cite{Lang:2005jz}.  Performance improvement is obtained by
Hasenbusch mass preconditioning \cite{Hasenbusch:2001ne} with two
pseudofermions. Further improvement is achieved by the use of a chronological
inverter by minimal residue extrapolation \cite{Brower:1995vx} and a mixed
precision inverter \cite{Durr:2008rw}. A discussion of the autocorrelation time,
the eigenvalues of the Dirac operator and the topological sector of the
generated configurations is found in \cite{Gattringer:2008vj}. We choose to
analyze every fifth configuration and neglect any remaining weak correlations.
The distribution of the eigenvalues of $\DCI$ indicates that we may simulate
small pion masses on relatively coarse lattices. The algorithm was found to show
frequent tunneling through topological sectors.

The gauge configurations are determined on an SGI ALTIX-4700  (a machine with a
peak rate of 62.3 TFlop/s for 9728 processors) using  for each configuration 
(i.e., one unit of HMC-time) a total amount (summed over the processors used in
parallel) of $\mathcal {O}(60)$ CPUh for set A and up to $\mathcal {O}(120)$
CPUh for set C. The quark propagators for the analysis are computed with a
multi-mass solver and thus the CPU time depends only on the smallest (the
dynamical) quark mass. Computing the quark propagators for one configuration
and 60 sources (five source types, cf., Sect. \ref{QuarkSourceSmearing} and 12
Dirac-color sources each) takes approximately the same amount of CPU time as
for generating one gauge configuration. A recent comparison for different
actions can be found  in Ref.~\cite{Jung:2010jt}.

\subsection{Simulation parameters}\label{SimPars}

We use lattices of size $16^3 \times 32$ at three different values of the gauge
coupling $\beta_{LW}$ and bare quark mass parameter $m_0$, see Table
\ref{SimDetTab}. The lattice spacing is determined via the static quark
potential, using a Sommer parameter of $r_{0,exp} = 0.48\,$ fm. The bare
simulation parameters are chosen such that the lattice spacing is of
approximately the same magnitude in all three ensembles. Hence their physical
volume is of the same size as well ($\approx 2.4\,$ fm).

\begin{table*}[tbp]
\begin{center}
\begin{tabular}{ccccccccc}
\hline
\hline
set&	$\beta_{LW}$&$m_0$& configs 	& $a$[fm]	& $m_{\pi}$[MeV]& $m_{AWI}$[MeV] 	&$m_{\pi}L$ \\
\hline
A&	4.70& 		 -0.050	&100 	&0.151(2)	&525(7)         & 43.0(4)		& 6.4\\
B&	4.65& 		 -0.060	&200 	&0.150(1)	&470(4)         & 35.1(2)		& 5.7\\
C&	4.58& 		 -0.077	&200 	&0.144(1)	&322(5)         & 15.0(4)		& 3.7\\
\hline
\hline
\end{tabular}
\end{center}
\caption{Bare parameters of the simulation: Three ensembles (A,B,C), at different gauge coupling $\beta_{LW}$ and quark mass parameter $m_0$.
The number of configurations, lattice spacing from the static potential assuming a Sommer parameter of 0.48 fm, the pion mass, the (non-renormalized) AWI-mass and the dimensionless product of the pion mass with the physical lattice size are given. 
For more details see \cite{Gattringer:2008vj}.}
\label{SimDetTab}
\end{table*}

Considering the chiral extrapolation, many calculations make use of the
so-called mass independent scheme (cf., the discussion in \cite{Aoki:2000kp}).
In this scheme the lattice spacing is determined (for fixed bare gauge coupling)
in the chiral limit and assigned to all ensembles with that gauge coupling.
Since so far we only have one mass value at each gauge coupling, we make use of
a mass dependent scheme, differing by $\mathcal{O}(a)$ corrections.

\section{Methods in hadron spectroscopy}
\label{Methods}

\subsection{Variational method}

Excited state contributions are suppressed by $\mathcal{O}(\E^{-\Delta E})$ in
the hadron correlators. However, the so-called variational method 
\cite{Michael:1985ne,Luscher:1990ck} allows to extract excited states in
principle.  This method has been used extensively by the BGR-collaboration
\cite{Burch:2004he,Burch:2005wd,Burch:2006dg,Burch:2006cc,Gattringer:2008be,Gattringer:2008vj}
and has gained popularity in recent years.

A simple hadron interpolating field operator with the right quantum numbers will
have a correlation function that asymptotically decays with $e^{-Et}$ where $E$
denotes the ground state energy. However, at finite time distance there will be
contributions from excited states embedded in the continuum of scattering
states. On a lattice with finite spatial extent the corresponding energy
spectrum is discrete. In case there are no dynamical quarks, the eigenstates may
be interpreted as  bound states of the valence quarks in the interpolator. In
the fully dynamical situation such a simple interpretation is not possible, 
since mixing with all many particle states with the same quantum numbers can
occur. The eigenenergy levels are related to the scattering phase shifts of the
coupling channels \cite{Lu86,Lu86a,Lu91,Lu91a} and the space of lattice hadron
interpolators has to be large enough, in order to represent the possible intermediate
states with sufficient quality.

Given a set of $N$ interpolators (with given quantum numbers) in the variational
method one constructs  the corresponding correlation matrix
\begin{eqnarray}
C_{ij}(t) 		&=& 	\langle 0 \vert O_i(t)  O_j^{\dag} \vert 0\rangle \FC\nonumber \\
			&=& 	\sum_{k=1}^N \langle 0 \vert O_i \vert k\rangle \langle k \vert O_j^{\dag} \vert 0\rangle\,\E^{-t E_k}.
\label{VarMethEq}
\end{eqnarray}
The idea is  to offer a basis of convenient interpolators, wherefrom the system
chooses the linear combinations closest to the physical eigenstates.
Diagonalization of the correlation matrix of the interpolators thus allows to
disentangle the physical states, if the state of interpolators is sufficiently
complete. The generalized eigenvalue equation
\begin{eqnarray}
C(t)\, \vec{v}_k 	&=& 		\lambda_k(t,t_0)\, C(t_0)\, \vec{v}_k \FC\nonumber   \\
\lambda_k(t,t_0) 	&\propto& 	\E^{-(t-t_0)\,E_k} \left( 1+\mathcal{O}(\E^{-(t-t_0)\,\Delta E_k}) \right)\ ,
\end{eqnarray}
gives the energies of the eigenstates, where $\Delta E_k$ is the distance of
$E_k$ to the closest state. In the interval $t_0\leq t \leq 2 t_0$ it would be
determined by the distance to the first neglected state  $E_{N+1}$
\cite{Blossier:2009kd}. However, we use small values for $t_0$ (1 or 2), since
otherwise the quality of the diagonalization decreases. We therefore determine
the eigenvalues in a larger window of $t$-values.

The corresponding eigenvectors represent the linear combinations of the given
interpolators which are closest to the considered physical states at each time
slice. Hence they may be used to derive some information on the composition of
the physical modes \cite{Glozman:2009rn,Glozman:2009cp}.

Obviously, the number of interpolators should be large enough, they should be
independent and have overlap primarily with the low modes of the theory, in
order to reduce contamination from highly excited states. In actual
calculations, including more interpolators unfortunately increases
the statistical noise in the diagonalization. Thus, the optimal choice is
usually to use only those interpolators, which show good overlap with the low
physical modes.

\subsection{Quark source smearing}\label{QuarkSourceSmearing}

Hadron correlation functions are built from quark propagators $D^{-1}$, which
are computed by inverting the Dirac operator on a given quark source. Extended
sources improve the signal and also allow for a larger operator basis in the
variational method. We use three different kinds of sources:  narrow
($0.27\,$fm),  wide ($0.55\,$fm) and a ($P$ wave like) derivative source.

The sources are computed using Jacobi smearing
\cite{Gusken:1989ad,Best:1997qp}: A point-like source is smeared out by
applying a polynomial of the hopping term,
\begin{eqnarray}
S_{\kappa,K} 		&=& \sum_{n=0}^K\kappa^nH^n S_0\,,  \\
H(\vec{n},\vec{m})	&=& \sum_{j = 1}^3 \Big(U_j\left(\vec{n},0\right) 
\delta\left(\vec{n} + \hat{j}, \vec{m}\right) \vspace{-6pt}\\
&&  \phantom{\sum_{j = 1}^3}+ U_j\left(\vec{n}-\hat{j\,},0\right)^\dagger \delta\left(\vec{n} - \hat{j},
 \vec{m}\right) \Big)\,,\nonumber
\end{eqnarray}
where $S_0$ denotes the point source.  The resulting source shape is
approximately Gaussian. The parameters $\kappa$ and $K$ are tuned (for each
ensemble of configurations) such as to ensure approximately the same source
width in all ensembles. Narrow (wide) sources will be denoted by quark
subscripts $n$ ($w$) in the remainder of this paper.

The derivative sources, $S_{\partial_i}$, are obtained by applying the covariant
difference operators on the wide source, $S_w$ \cite{Gattringer:2008be},
\begin{eqnarray}
P_i(\vec{x},\vec{y})	&=&	
U_i(\vec{x},0)\delta(\vec{x}+\hat{i},\vec{y}\,) 
- U_i(\vec{x}-\hat{i},0)^\dagger\delta(\vec{x}-\hat{i},\vec{y}\,)\FC\nonumber\\
S_{\partial_i}		&=&	P_iS_w\FC
\label{genEigValProb}
\end{eqnarray}
where $\hat{i}$ is one of the spatial directions. The derivative sources were
found to be crucial for some states \cite{Gattringer:2008td}, as will be
confirmed in this paper as well. In the following, derivative sources are
indicated by the  subscript $\partial_i$ of the quark field.

\subsection{Constructing the interpolators}

As already mentioned, we construct several interpolators in each channel in
order to be able to extract excited states using the variational method. All
sources are located in a single time slice and built on configurations  which
have been  HYP-smeared in the spatial directions three times
\cite{Hasenfratz:2001tw}.  The main motivation for link smearing is suppression
of UV-fluctuations which  manifest themselves, e.g., in the distribution of the
plaquette. Thus, the parameters of the spatial HYP-smearing  have been
optimized by a trade-off between a maximum  average plaquette and a maximum of
the minimum plaquette, partly following the arguments of
\cite{Hasenfratz:2001hp,Hasenfratz:2007rf}. We obtained the parameters
$\alpha_1=0.8$ and $\alpha_2=0.4$, where $\alpha_1$ is the parameter in the last
step of the smearing algorithm, where the center link is smeared.

The center positions of the quark  sources are  shifted for subsequent
configurations in order to decrease statistical correlation of the data.  The
interpolators at the sink are projected to zero momentum, thus for sufficiently
many configurations the sum projects to propagators of zero momentum hadrons due
to translation invariance. Tables of the interpolators are found in Appendix
\ref{Interpolators}. 

There exists another approach for interpolator construction developed recently,
called ``distillation'' \cite{Peardon:2009gh}, which we do not follow here.
Recent results on hadron spectroscopy following this approach are found in
\cite{Dudek:2010wm,Bulava:2010yg}.

\subsubsection{Meson interpolators}

We consider isovector-mesons, thus there are no disconnected diagrams. Using spatially 
isotropic sources ($n$ and $w$), the quantum numbers of an interpolator are
determined by the combinations of the spinor components (Dirac content). This
restricts the meson to just a few (non-exotic) channels of spin $\leq 1$. A way
to enlarge the basis of interpolators and access higher spin states is given by
considering the direct group product of spinor and spatial structure. The
decomposition to the irreducible representations then leads to interpolators
with definite quantum numbers 
\cite{Lacock:1996vy,Liao:2002rj,Basak:2005aq,Dudek:2007wv,Petry:2008rt}. We realize a
non-trivial spatial structure by using the derivative sources, which transform
according to the lattice spin irreducible representation $T_1$.

Depending on the quark content and the implementation of the derivative sources,
symmetrization of the interpolators is needed in order to have definite
$C$-parity. Hence, light meson interpolators are symmetrized properly, while
symmetrization is omitted in the strange meson sector (see Appendix
\ref{Interpolators}).  Our strange meson correlator calculation omits cross
correlation matrix elements corresponding to interpolators with different
$C$-parity quantum numbers in the limit of degenerate quark masses. Therefore,
when analyzing strange mesons, we have to restrict ourselves to subsets of
interpolators sharing the same $J^{PC}$ quantum numbers in the limit of
degenerate quark masses.

\subsubsection{Baryon interpolators}

For the construction of baryon interpolators we use only Gaussian smeared quark
sources ($n$, $w$). In case of the nucleon, $\Sigma$ and $\Xi$ we use three
different Dirac structures, in case of the $\Delta$ and $\Omega$ only one. Since
a baryon is built from three valence quarks, there are $2^3=8$ possible smearing
combinations. If there is isospin symmetry, some of the resulting 8 interpolators
are very similar to others, which we thus prune from the considered set of
interpolators. We end up with 18 interpolators in the nucleon channel, 6 in the
$\Delta$ and $\Omega$ channels and 24 in the $\Sigma$ and $\Xi$ channels (see
Appendix \ref{Interpolators}). We project to definite parity in each channel.

\subsubsection{Energy levels}\label{EnergyLevels}

In full QCD calculations the single hadron states couple to channels with  two
or more hadrons, like the even number of pions in the $\rho$ sector.  Although
the original hadron is projected to its rest frame, the scattering states have
internal relative momenta. For finite spatial extension the admissible values of
the momenta depend on the spatial size and the (Euclidean) discrete energy
levels are related to the phase shift of the scattering states.  In the elastic
region this relationship has been discussed in  \cite{Lu86,Lu86a,Lu91,Lu91a}. 

Neglecting further interactions of the hadronic bound states, the energy level
$E(A,B)$ for two free hadrons reads
\begin{eqnarray}
 E \left( A(\vec p),B(-\vec p) \right) &=&\\
 &&\hspace*{-20mm}
  \left( \sqrt{m_A^2+|\vec p|^2} + \sqrt{m_B^2+|\vec p|^2} \right) \left( 1 + \mathcal{O}(ap) \right) 
   \, \, .\nonumber
\end{eqnarray}
The hadrons $A$ and $B$ have back-to-back momenta since the whole state is
projected to zero momentum. In the infinite volume limit, there is a continuum
of scattering states. In a finite box, the momentum $\vec p$ can take only
discrete values, determined by the boundary conditions, $a \vec p =
2\pi(n_x,n_y,n_z)/L$, where $L=16$ in the present work. In the $S$ wave, the
lowest 2-particle state level thus shows vanishing relative momentum, while in
the $P$ wave, the lowest 2-particle state level has a momentum of $a |\vec
p| = 2\pi/L$.

The lowest energy levels of the fictitious two-free-hadron-state for each
ensemble are indicated in the figures using symbols $\times$ and +, provided the energy
levels are in the range of our investigation and can be estimated from our
results. The corresponding non-correlated statistical uncertainty, neglecting the
hadronic interaction and finite volume effects, is of the magnitude of 5 to 60
MeV. For clarity, these error bars are suppressed in the figures if the error is
smaller than 30 MeV, and furthermore, we always omit the error bar in case of
continuous curves of many-particle states in the figures.

In many of our hadron correlators, due to the parameters of the
simulation and the resulting pion masses, two-particle intermediate
states will have an energy larger than the ground state energy. As an example, 
the $\rho$ could
formally couple to two pions with relative momentum (to build the $P$ wave)
but this is forbidden for kinematical reason in our case, even for the lightest
pion mass of ensemble C.  There may be a slight shift of the lowest energy level
due to avoided level crossings, though.  However, the higher levels could well
be due to two-particle states.  In the $\rho$ sector one would expect such an
energy level between the ground state and a possible $\rho'$.   In most other
cases the expected two-particle state  levels are at least close to other
observed possible resonance state levels.  Except for partially quenched results
in the light scalar channel, we do not observe such scattering state levels, or,
if they are there, we cannot  distinguish them from single particle states.

A possibility to shed some light on the nature of the state is to monitor the
eigenvectors $\vec v_k$ of Eq.~\ref{genEigValProb} of the state when varying parameters of the simulation. Ideally,
one compares the eigenvectors for several dynamical simulations, but also
partially quenched data can yield some information. Since effects from partial
quenching can shift the energy level, corresponding results may also allow to
extract further information about the state. 

For example, in the light and strange scalar channels we find that our {\em
partially} quenched results are well described by partially quenched formulae
of 2-particle states \cite{Prelovsek:2004jp}.  However, at the dynamical point
our data do not allow for a unique interpretation. Both, the resonance and the
scattering state may be present and contribute to the measured energy level. 

In the case of the negative parity baryon channels, positive parity baryons and
pseudoscalars might form scattering states whose energy levels are consistent
with our results. While the extracted masses slightly favor the scattering
states, the eigenvectors do not allow for an interpretation in terms of a level
crossing and thus do not confirm the picture of a scattering state at small
pion masses, either.

An explanation for missing scattering states would be weak coupling to the
interpolators considered. In case of the $S$ wave, there is a noteworthy
amplitude already at small momenta, while in case of the $P$ wave higher
momenta are needed.  This may explain why we see possibly a 2-particle signal
in the $S$ wave of  $a_0$ ($J^{PC}=0^{++})$, but not, e.g., an additional
energy level between the ground state and $\rho'$ in the $P$ wave of $\rho$
($1^{--})$. Consideration of explicit two-hadron interpolators may help.

It is known that channels with two or more particles are suppressed by factors
$\mathcal{O}(1/L^3)$ \cite{Mathur:2006bs,Prelovsek:2008rf}. This suppression
comes on top of the generic suppression of the excited states.  Even in
kinematical situations where already the ground state couples to scattering
states, it turned out, that one has to include both, one- and two-particle
states,  in the set of interpolators in order to see clear signals
\cite{GaLa93}.  For such attempts, see 
\cite{Moore:2006ng,Bulava:2009ws,Bali:2009er,Prelovsek:2009bk,Prelovsek:2010kg}.
Unfortunately, the computation of cross-correlations of one- and two-particle
states intrinsically includes disconnected diagrams, which are technically
demanding and thus not considered here.

\section{Comparing to Experiment}\label{SysEffects}

\subsection{Chiral Extrapolation}
\label{ChPT}

We simulate at pion masses larger than the physical one. Therefore we have to
rely on some extrapolation towards small quark masses, if we want to make
predictions at the physical point. The analytic form of the chiral extrapolation
depends  on the path taken in the parameter space of, e.g., the lattice
spacing,  the quark mass, the gauge coupling and the volume. In chiral
perturbation theory (ChPT) \cite{We79,Gasser:1984gg} the quark
mass is the only varying parameter.  The lattice spacing and the volume are
assumed to remain constant. In the mass-independent scheme (cf., the discussions
in \cite{Aoki:2000kp,De:2008xt})  the scale (the physical value of the lattice
constant) is set by extrapolating the lattice spacing towards the chiral limit
along some path in parameter space (e.g., constant bare gauge coupling). The
extrapolated value is then used for all ensembles along that path.

Since we have only one ensemble at each value of the coupling, we cannot use 
this mass-independent scheme.  Hence, our path of extrapolation in parameter
space implies a mass-dependent scheme and formulae of ChPT would have to be
adjusted.  Nevertheless, we assume our path to be close to the one in the
mass-independent scheme, and expect that the analytic form of the chiral
extrapolation should be similar, although with different expansion 
coefficients. Therefore, we perform chiral fits linear in the pion mass squared
for all results and discuss possible other fit forms in certain cases. Note that
the fits include only the three dynamical points, the partially quenched points
are left out. In the figures, the solid black curve shows the chiral
extrapolation, the dashed lines delimit the region of one standard deviation. The results of
the chiral fits are summarized in Fig.~\ref{collection_masses}.  

\subsection{Systematic Effects}

We set the scale  via the static potential and the Sommer parameter (Sect.
\ref{SimPars}) with $r_0=0.48$ fm.  In the literature also other values have
been used. We have only one dynamical quark mass for each value of the gauge
coupling and thus  cannot extrapolate to the chiral limit in order to define
the scale in the mass independent scheme. More dynamical points would be
desirable for a more reliable chiral extrapolation and will be included in
future calculations. 

The strange quark is considered as a valence quark only. In view of results
including full strange quark dynamics (e.g., \cite{Durr:2008zz}) we find,  at
least for the ground states, no noticeable difference in the mass range
considered here. 

Discretization effects have been discussed for baryons in the quenched
simulation, where, due to the improvement of the used action, only small
$\mathcal{O}(a^2)$ corrections have been identified \cite{Gattringer:2003qx}.
In oder to confirm this for the dynamical simulation we would have to perform
our study at several lattice spacings and lattice volumes. This is a future
task.  In the extrapolation to the physical quark masses, using the
mass-dependent scheme, we disregard discretization errors and discuss finite
volume effects only qualitatively.

The physical size of our spatial lattice volume is $\approx2.4\,$fm. Finite
volume effects due to ``pions wrapping around the world'' are expected to be 
small in ensemble A and B ($m_{\pi}L > 4$), however, they could be significant 
in ensemble C ($m_{\pi}L \approx 3.7$). Squeezing of large hadrons, which do
not ``fit in the box'', may be an obstacle in the case of  excited baryons.
Indeed, we seem to observe such effects. Studies with a larger volume are
planned.

A subtle issue is the choice of interpolators for the various hadron states.
For sufficiently large and complete sets and exhaustive statistics the
variational method would produce the eigenmodes sufficiently well. In our case
we are restricted to few interpolators and modest statistics. We attempt to
optimize this situation by choosing suitable subsets of interpolators, as
discussed, motivated by plateaus in the eigenvector components and exponential
decay of the eigenvalues in some fit range. This brings in certain systematic
effects, which will be reduced only by enhancing both, statistics and the set
of interpolators. We discuss our choice for all hadrons considered
subsequently. Indeed, in some cases we find sizeable dependence on the chosen
sets.

Another possible systematic influence comes from choosing $t_0$ in the
variational method and the fit range for the generalized eigenvalues. In
principle, that impact can be estimated by choosing several values of $t_0$ and
varying the fit range. For the final fit one should then choose a window where
this impact is negligible. However, in practice the corresponding choices are
restricted by the given signal-to-noise ratio for coarse lattices and weak
signals. We use $t_0=1$ in most (and $t_0=2$ in a few) cases and perform a fit
in the maximum range possible, which is, however, often limited to 3 or 4
points for noisy observables.

\section{Results: Light Hadrons}
\label{ResultsLight}

Following the idea of the variational method, a large basis of interpolators
should be optimal. In practice, one finds that increasing the number of
interpolators also implies an increased statistical noise. One thus has to
find an optimal balance when choosing the basis for the variational method. 

In each channel we take the subset of interpolators, which yields the optimum
plateaus of effective masses. In order to find this subset, we first look at the
diagonal elements of the correlation matrix, which represent the autocorrelation
of the interpolators used. From that data one has a first hint which
interpolators are candidates for such an optimal subset.  To be more concrete,
we extract information about the interpolators such as to which physical state
they couple dominantly,  how strongly they are affected by contamination of
excited states and until which time separation the signal of the
corresponding physical state is reliable. 

Usually, the so obtained set of interpolators includes a number of interpolators
which are known to have rather large overlap with each-another, due to, e.g.,
their common Dirac structure. This in turn means that they are far from being
orthogonal and thus not well-suited to build a basis of the variational method.
Thus, we proceed by singling out subsets, applying the criterion of
maximum orthogonality, i.e., least overlap. Doing so, we obtain a number of
candidate subsets, all of which are analyzed with the variational method. The
fit range of the plateau is chosen from the first point without contaminations
from higher states until the plateau breaks down or the noise gets too large. 
This qualitative criterion is made more precise by choosing an optimal $\chi^2$
as quantitative criterion for the fit.

The actual values of the energy levels are then determined by an overall
exponential, correlated fit to the eigenvalues at all time slices in the fit
range.

This range is validated by comparison with the corresponding plateau of the
eigenvectors. The variational method simultaneously yields results for several
states using one set of interpolators. However, we find that in many cases the
signal can be improved by considering different sets for different states. The
extracted masses of such different sets agree well within error bars, but the
noise is reduced. Theoretically, one would expect to find improvement by joining
both sets of interpolators, but in practice this means an increase of the noise,
which in most cases is the stronger effect. We show a plot of the effective mass
of the $\rho$ meson as one example for consistent results for two sets (see
Fig.~\ref{rho_consistency}).

We present results on three ensembles of gauge configurations (cf.,  Table
\ref{SimDetTab}). In all plots, filled symbols denote dynamical results and open
symbols denote partially quenched results, where the valence quark mass is
always larger than the sea quark mass. The symbols $\times$ and + represent
energy levels of free scattering states, neglecting hadronic interaction and finite
volume effects. The scale of the vertical axis is set by the lattice spacing, as
discussed, unless stated otherwise. The scale of the abscissa is set using the
results for the pion mass squared. The tables of the corresponding interpolators
are found in the appendix. The chosen subset of interpolators is stated in the
caption of the figures.

\subsection{Mesons with light quarks}

We simulate two mass-degenerate light quarks and thus use the symmetrized meson
interpolators found in  Appendix \ref{Interpolators}.

\subsubsection{The 0$^{-+}$ channel: $\pi$}
\begin{figure}[tbp]
\includegraphics[width=85mm,clip]{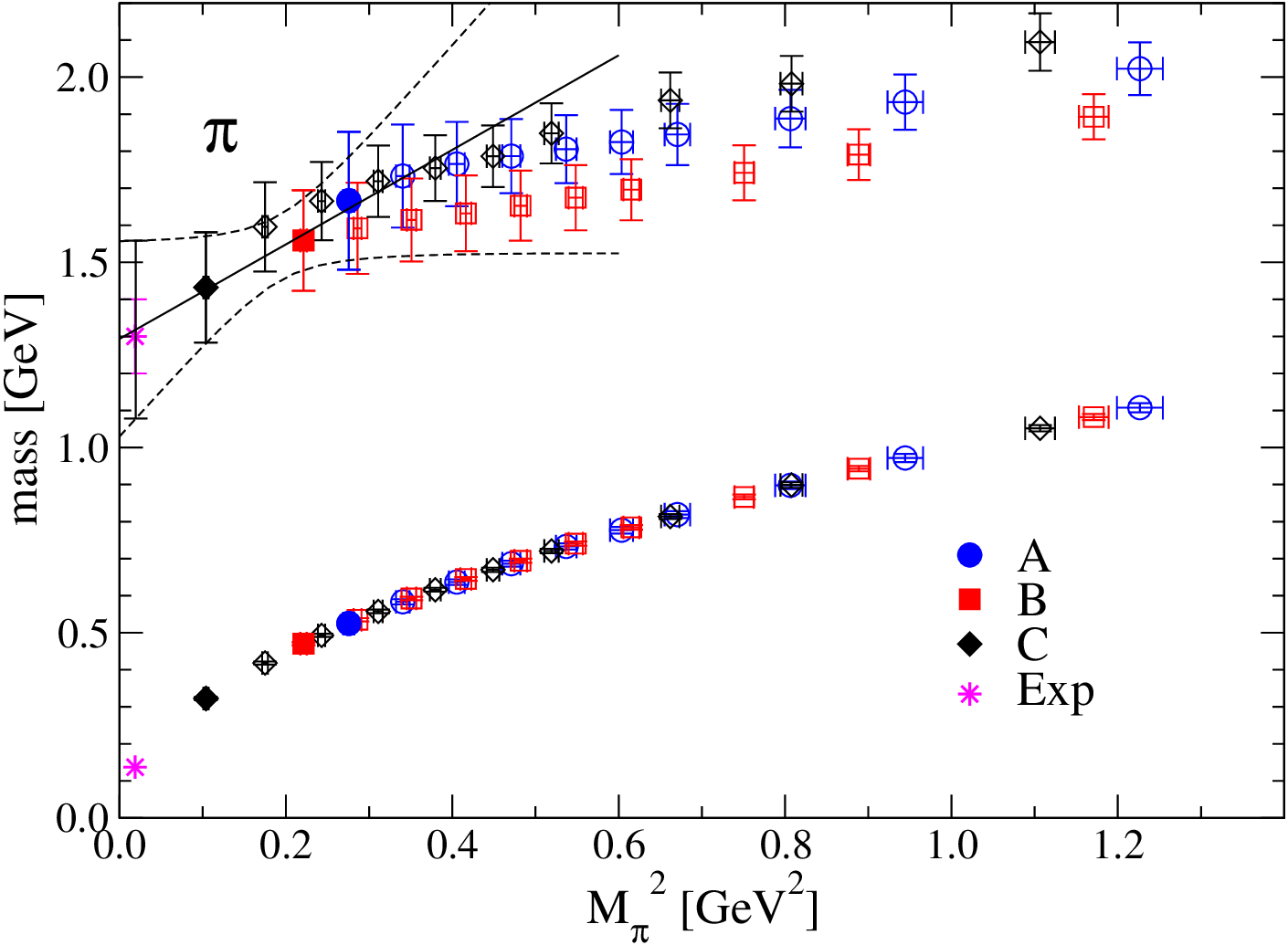}
\caption{Mass plot for the 0$^{-+}$ channel (pion), ground state and first
excitation.  Filled symbols denote dynamical results, open symbols denote 
partially quenched results. Parameters of the
three ensembles A, B and C are found in Table \ref{SimDetTab}. The list of the
interpolators is found in the appendix. Results for the pion ground state are
used to set the scale on the abscissa, here and in other figures as well. The
ground state is measured using interpolator (1), the excitation using
interpolators (3,5,9,10) in ensemble A, (1,6,9,10) in B and (3,7,8,11) in C.} 
\label{pi}
\end{figure}
\begin{figure*}[tbp]
\centering
\subfigure{
\includegraphics[width=85mm,clip]{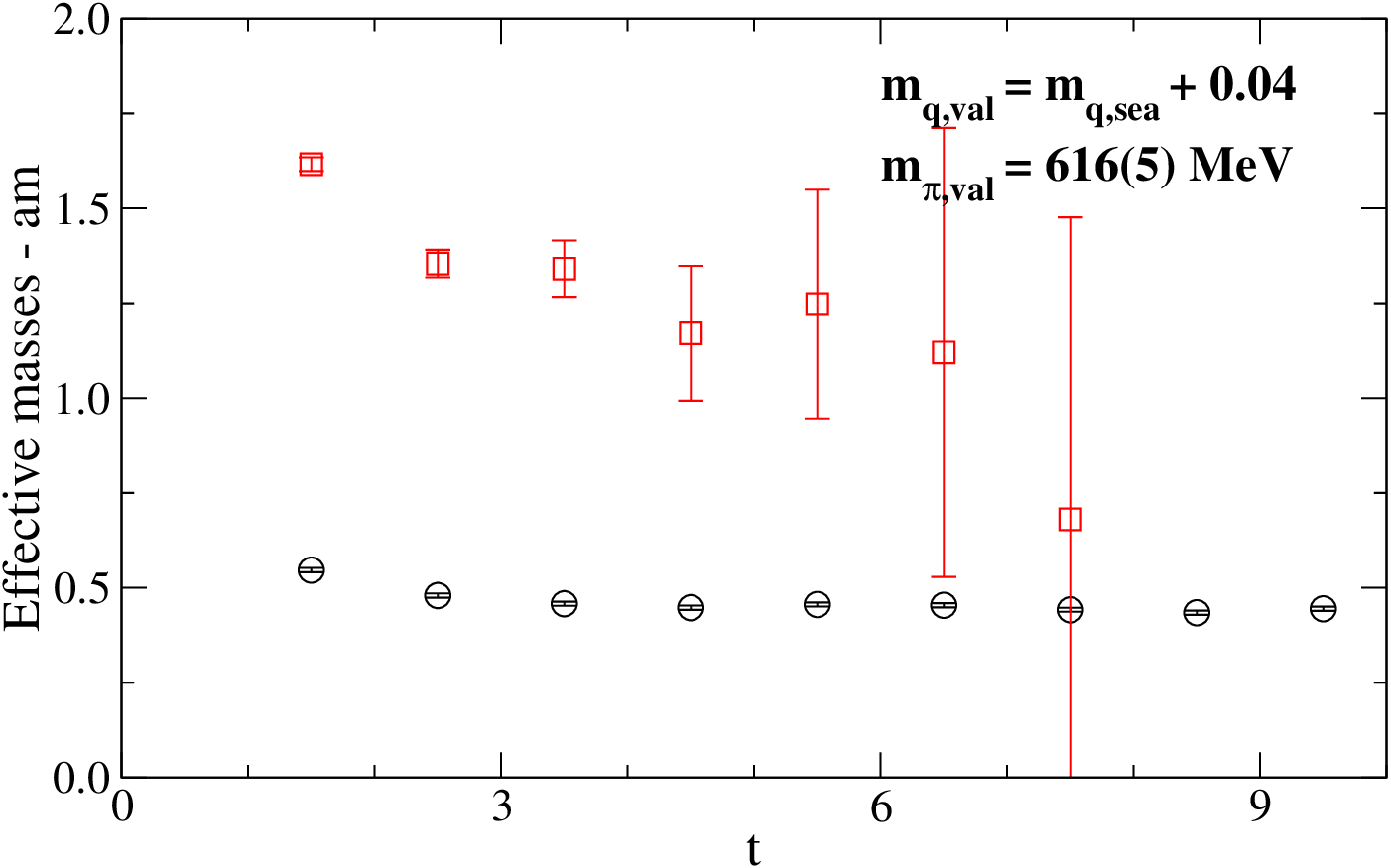}
\includegraphics[width=85mm,clip]{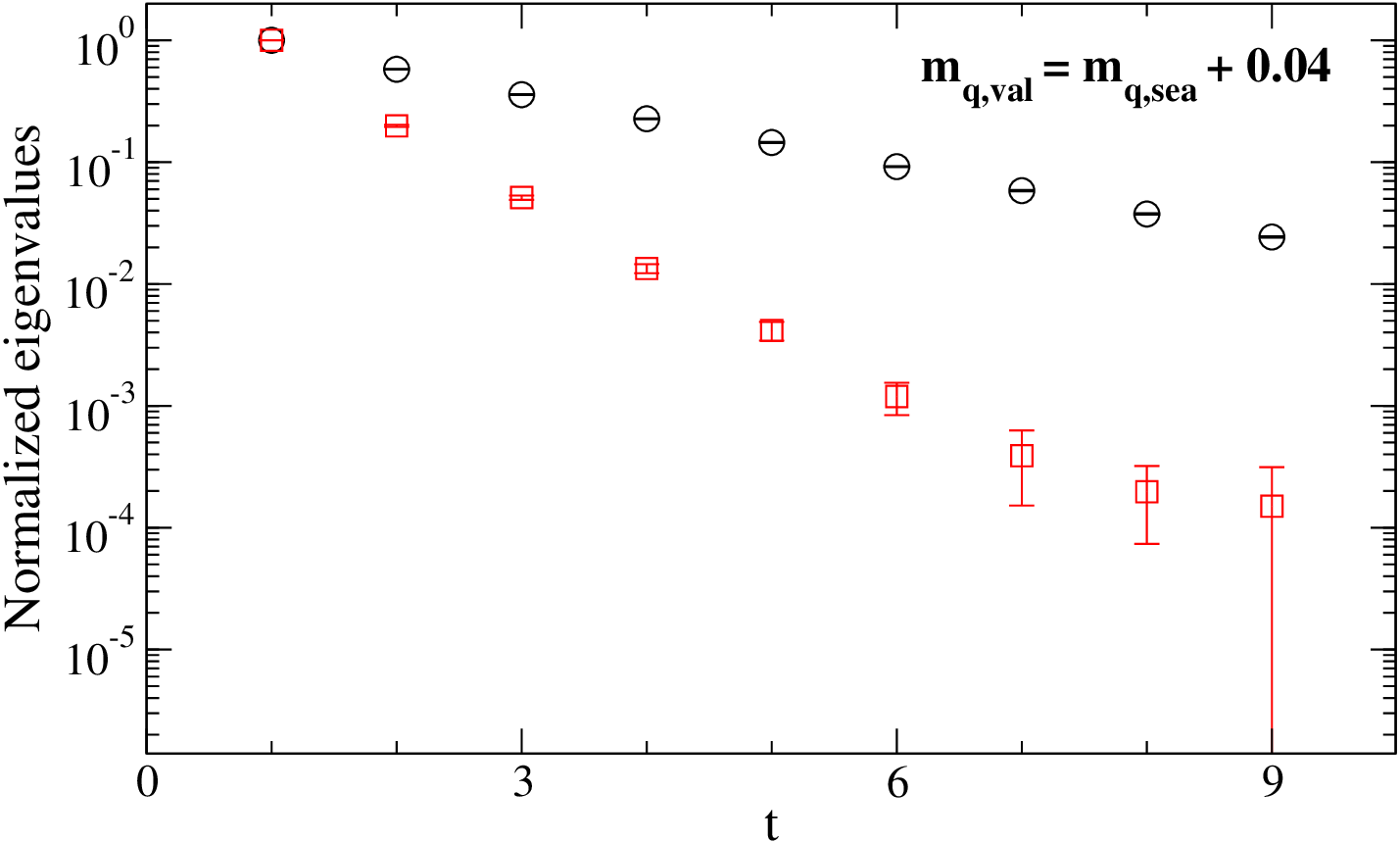}}
\subfigure{
\includegraphics[width=85mm,clip]{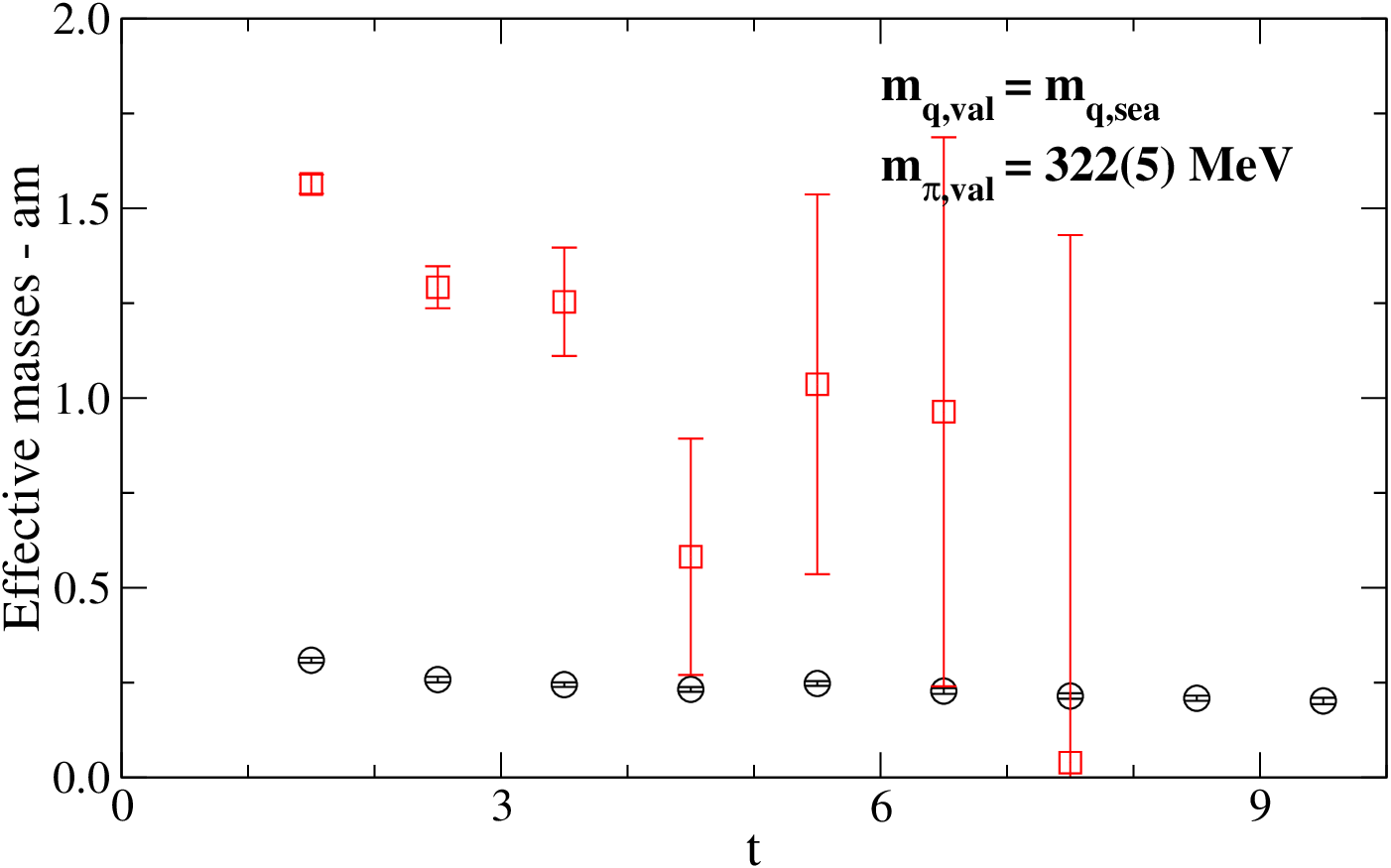}
\includegraphics[width=85mm,clip]{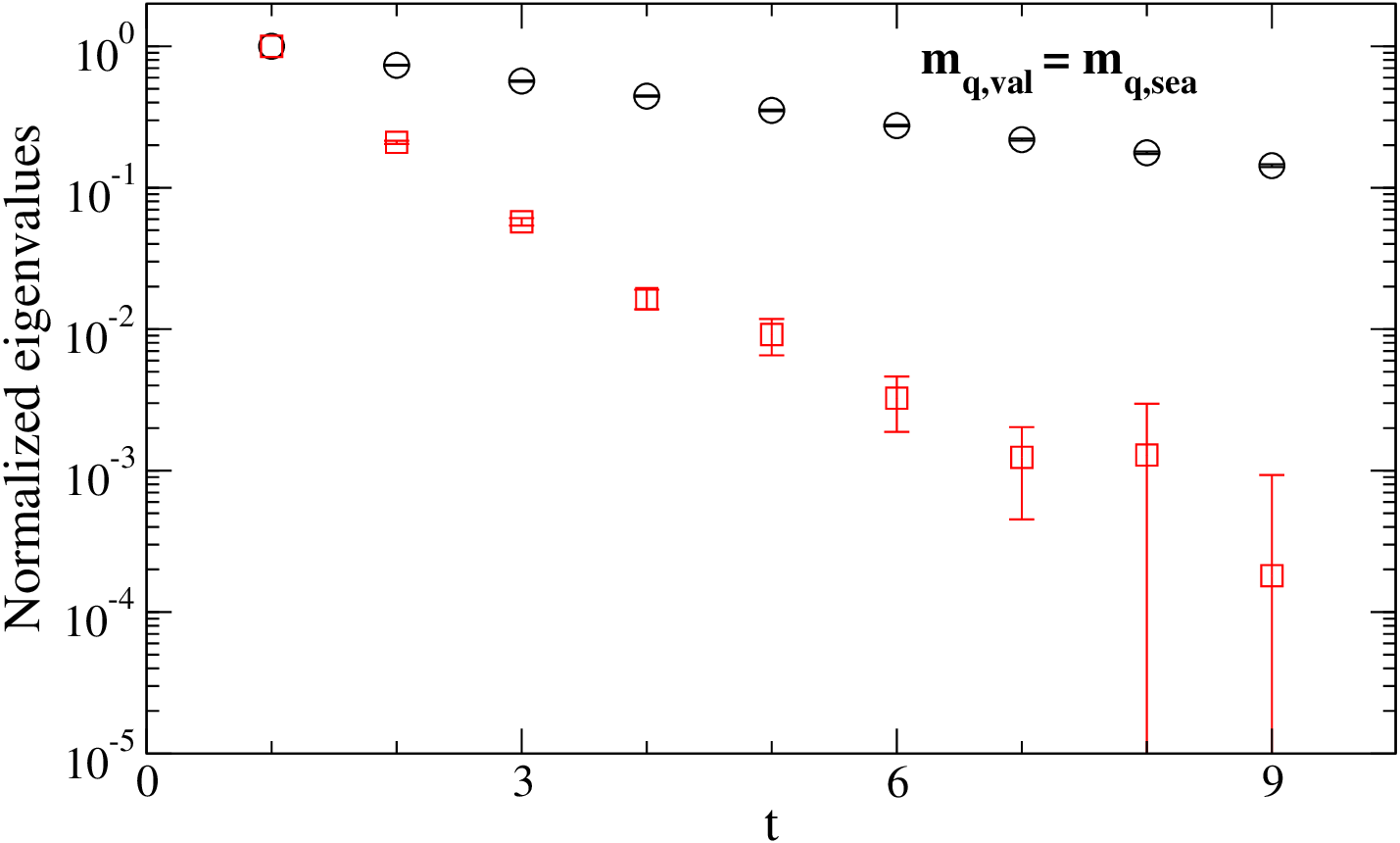}}
\caption{Effective mass plots (lhs) and normalized eigenvalues (rhs, logarithmic
scale) for the 0$^{-+}$ channel (pion) at the dynamical point (bottom) and at a
partially quenched point (top) of ensemble C, using interpolators (3,7,8,11).
Ground state and the first excitation data are shown. The corresponding valence
quark mass parameter and the fictitious pion mass are indicated in the figure.
The eigenvalues suggest that the first excitation can be fitted from $t=3$ to
$t=7$, at $t=8$ we find a kink at partially quenched data and even a loss of the
signal at the dynamical point.  The effective masses are proportional to the
derivative of the logarithm of the eigenvalues, thus showing huge error bars and
very bad plateaus in case the eigenvalues are tumbling. The fit to the
excitation at the dynamical point is motivated by the plateau at partially
quenched points and the almost stable behavior of the eigenvalues. We remark
that the final fit is an overall exponential fit of the eigenvalues for all
points in the chosen range $3\leq t \leq 7$.}
\label{pi_masses_eigvals}
\end{figure*}
In Fig.~\ref{pi} the mass of the pion ground state is shown as function of 
$M^2_\pi$ to indicate the
statistical errors. In the variational method, the observation range of the
excited states in this channel is limited by the backwards-running pion (see
also the discussion in \cite{Gattringer:2008vj}). This leads to short plateaus
for small pion masses and weakens the signal in our simulations. Nevertheless,
we find a clear signal of the first excited state, compatible with experimental
data.  The signal weakens towards smaller quark masses, thus the fit at the
dynamical point is motivated by the plateau at partially quenched points and the
behavior of the eigenvalues (see Fig.~\ref{pi_masses_eigvals}). We find no
indication of an intermediate three pion state signal; a $\pi\rho$ signal
would be expected at even higher energy values. The excited state signal
is expected to improve when using a lattice with a larger time extent, in
particular in the case of ensemble C. We do not see a reliable and stable signal
of the second excitation.
The quenched results using the same action \cite{Gattringer:2008be} showed,
within the errors, similar behavior as set A.

\subsubsection{The 1$^{--}$ channel: $\rho$}
\begin{figure}[tbp]
\includegraphics[width=85mm,clip]{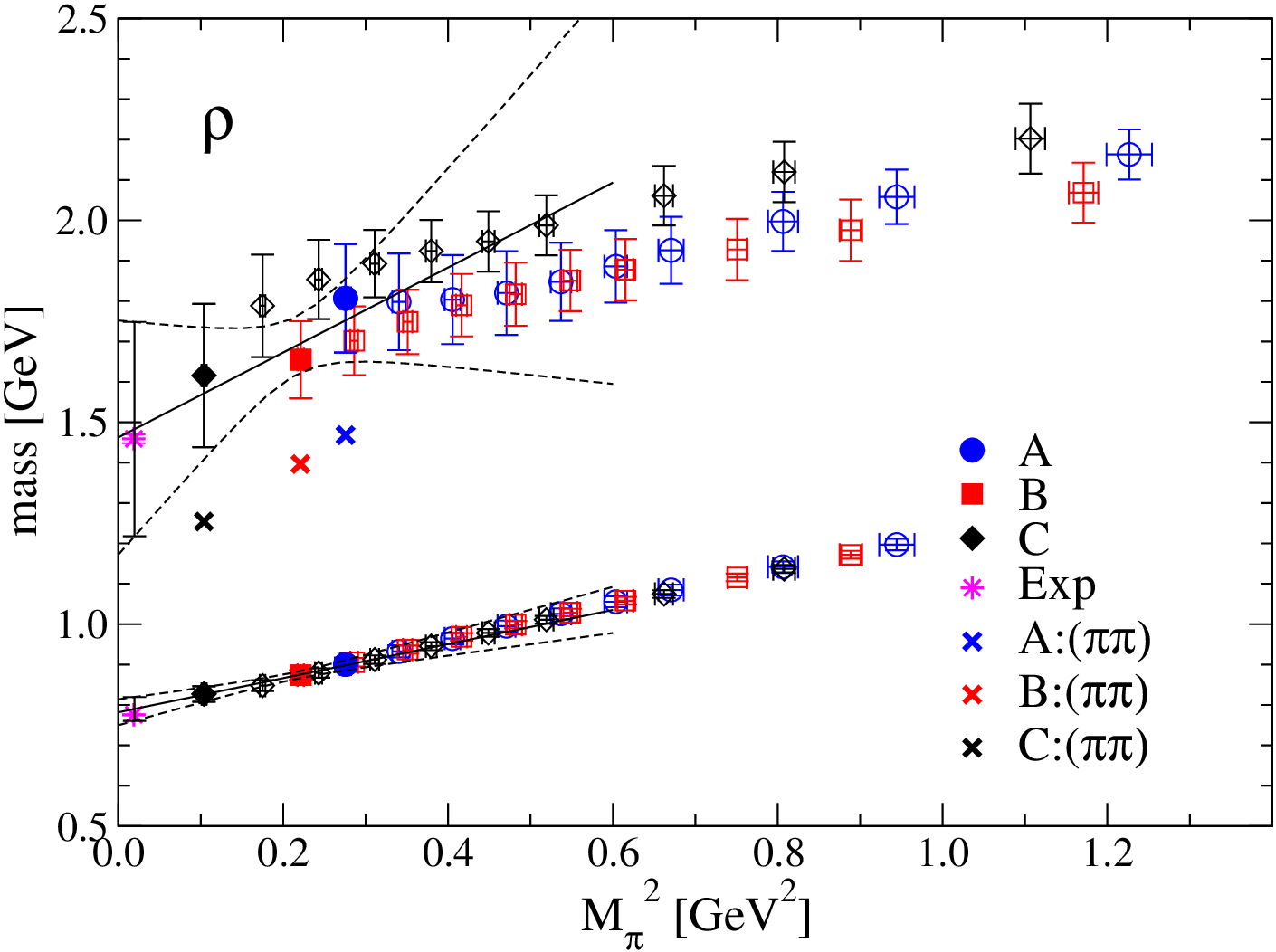}
\caption{Mass plot for the 1$^{--}$ channel ($\rho$), ground state and first
excitation. The ground state is measured using interpolators (1,4), the
excitation using interpolators (7,8,13,14) in ensemble A, (6,11,14) in B and C.
The estimated energy level of the $P$ wave scattering state $\pi\pi$ lies
below the first excitation in all three ensembles. The scattering state is not
observed, the reason may be too weak coupling to the used interpolators. The
coupling may be especially weak in case of $P$ wave scattering states. The
statistical error of the 2-particle state (based on the errors of the
particle masses involved) is of the magnitude of 5-10 MeV and
therefore not visible in the figure.} 
\label{rho}
\end{figure}

\begin{figure}[tbp]
\includegraphics[width=85mm,clip]{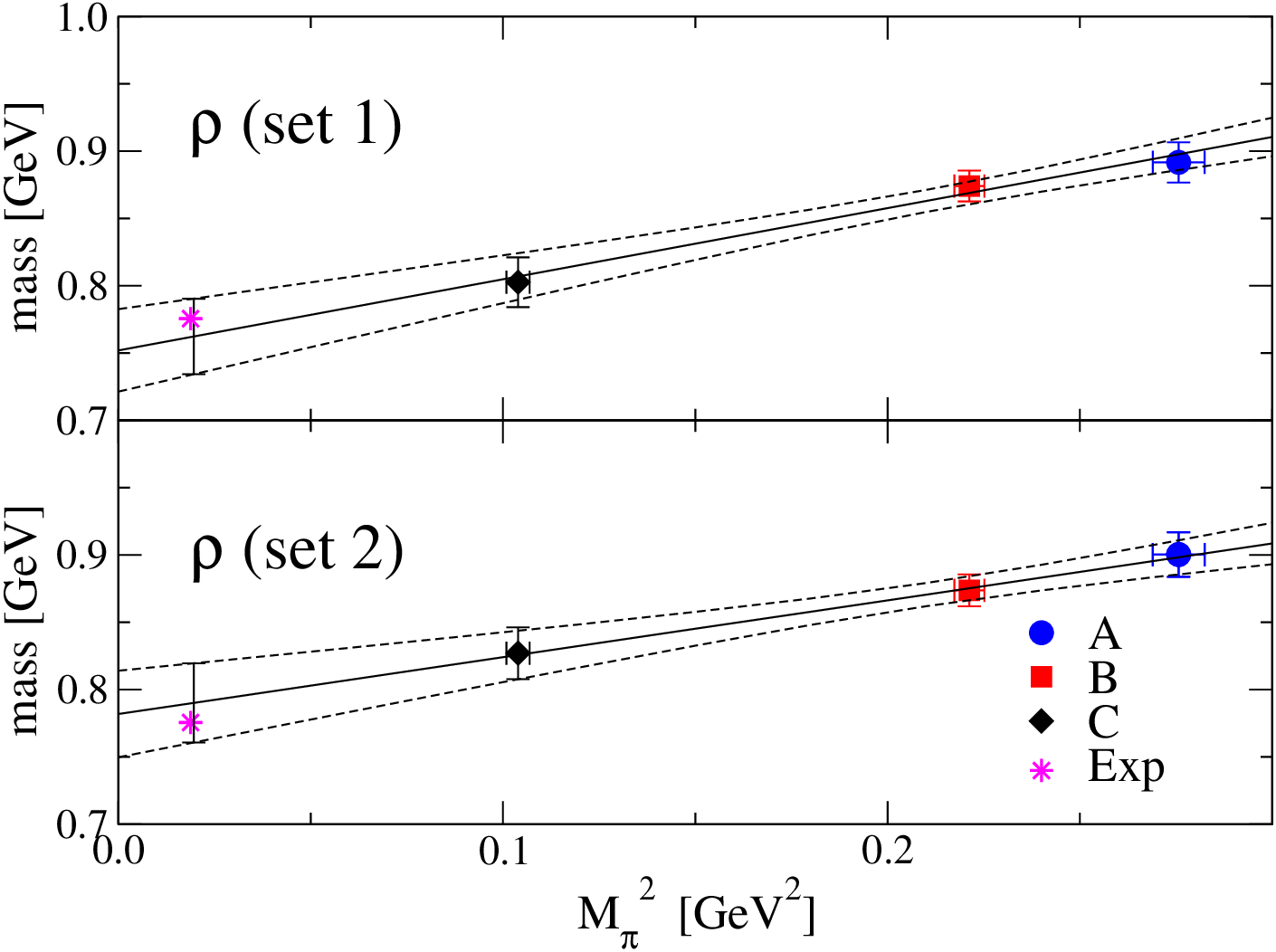}
\caption{Zoom in the mass plot for the 1$^{--}$ channel ($\rho$), ground state. 
The different sets of used interpolators are checked for consistency in the ground state.
Top: Interpolators (7,8,13,14) in ensemble A, (6,11,14) in B and C.
Bottom: Interpolators (1,4).}
\label{rho_consistency}
\end{figure}

We find an excellent plateau for the ground state and an excited state signal
compatible with experimental data (see Fig.~\ref{rho}).  Here, the excitation
signal improved using $t_0=2$, which may indicate more contamination  from
excited states in this particular channel. We decided to extract the ground
state result from the set of interpolators (1,4), which displays a better
plateau than the combination used for the excited state. The consistency of the
ground state from (1,4) and the sets chosen for the excitation is shown
in Fig.~\ref{rho_consistency}.  In ensemble B we furthermore see a signal of the
second excitation compatible with the $\rho$(1700). The physical $\rho$ meson is
a resonance which can decay into two pions with relative momentum. The energy of
the corresponding lowest scattering state is determined by the mass of the pion
and the minimum non-vanishing momentum, defined by the finite spatial extent of
the lattice. On the lattice, for our ensemble parameters the energy of this
scattering state would be above the mass of the $\rho$ meson. Hence the ground
state of the channel is dominated by the $\rho$ meson, which does not decay and
is therefore called ``stabilized''. One might expect, however, a two-pion
intermediate state between the $\rho$ ground state energy and the signal
associated to $\rho'$, but no such state is observed here. It is suggested that
the coupling of the used interpolators to two-particle states is strongly
suppressed in the $P$ wave. 

In principle, T$_1$ interpolators may couple to spin 3 states with energy levels
close to $\rho'$. However, from the naive continuum limit of our interpolators
we expect such coupling to be small.

Using only ``gaussian-type'' interpolators (i.e., without derivative sources), 
the energy levels of the first excitation are found to deviate slightly from
the results presented here. A possible reason could be mixing with the nearby
higher excited states, or an early breakdown of the plateau which can
complicate the identification of a correct fit range. Moving the used fit range
3--6 to 2--5 the excitation level increases.

Comparing with quenched results using the same action \cite{Gattringer:2008be},
we find that the dynamical $\rho$ ground state comes out significantly lighter
than its quenched counterpart, which, however, is partially due to the
different Sommer parameter value used in the quenched analysis ($r_0=0.5$ fm).
Again, the first excitation of the quenched simulation is compatible with set A
of the dynamical case. The dynamical points B and C indicate a steeper slope
pointing towards the experimental results.

The isospin singlet vector meson $\phi$ has mainly $s\overline{s}$ content and
disconnected parts are suppressed (``Zweig-forbidden'', cf., the decay channels
in the experiments). Thus we extract the $\phi$ meson mass considering
partially quenched results of the $\rho$ correlator. This is discussed in Sect.
\ref{strangemass}.

\subsubsection{The 0$^{++}$ channel: $a_0$}
\begin{figure}[tbp]
\subfigure{\includegraphics[width=85mm,clip]{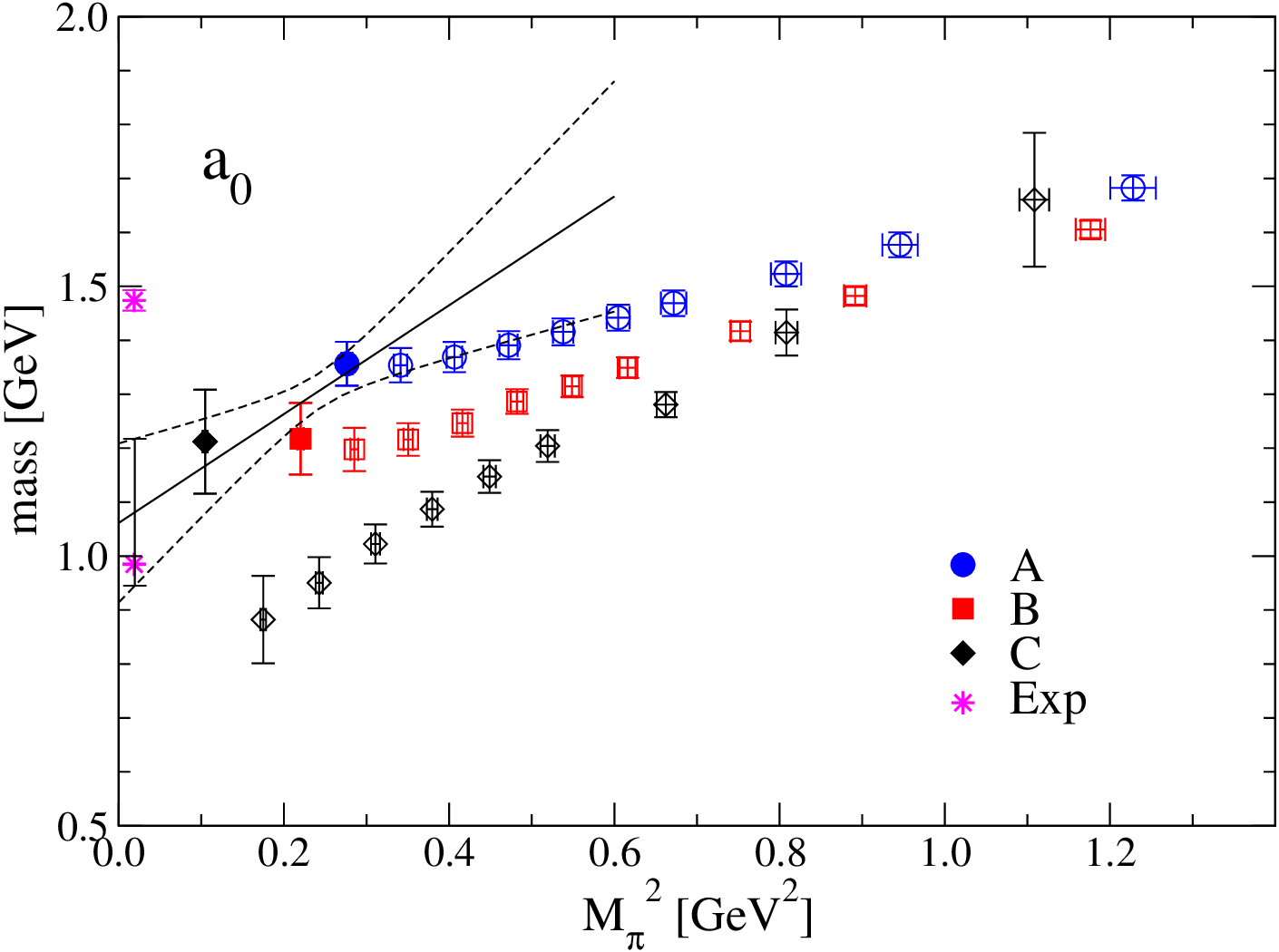}}
\subfigure{\includegraphics[width=85mm,clip]{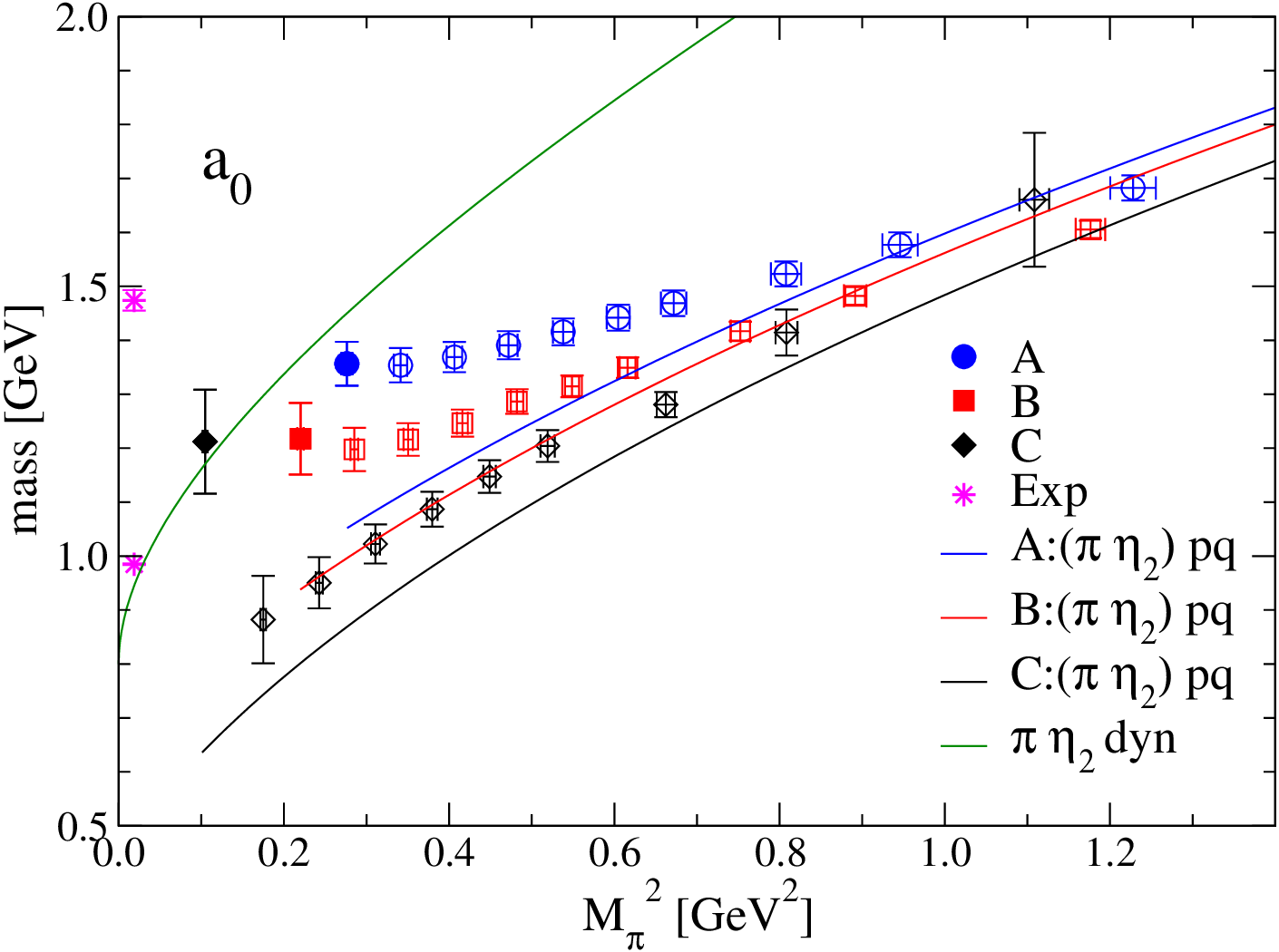}}
\caption{Mass plot for the 0$^{++}$ channel. Interpolator (8) is used throughout.
Top: Ground state of $a_0$ and chiral extrapolation relying on a 1-particle
interpretation are shown. Bottom: Estimate of the two-particle state
$\pi\eta_2$ for various parameters shown. The blue, red and black curve (online
version) show a prediction of the partially quenched (``pq'') scattering state
$\pi\eta_2$ for $m_{val}\gg m_{sea}$ in ensemble A, B and C, respectively. The
green curve (online version) shows an estimate of the dynamical (``dyn'')
scattering state $\pi\eta_2$ ($m_{val}=m_{sea})$.  Interaction of the bound
states and finite volume effects are neglected in the energy level of the
two-particle state. For clarity, the corresponding statistical error is omitted
in the figure. The partially quenched data suggests an interpretation
in terms of the two-particle state $\pi\eta_2$, while no clear statement about
the particle content can be made at the dynamical point.} 
\label{a0}
\end{figure}

The scalar channel is difficult as usual. The mass plateau is short and varies
depending on the chosen set of interpolators. The ground state mass of $a_0$ is
very close to the lowest energy level of the (dynamical) two-particle state
$\pi\eta_2$ (see Fig.~\ref{a0}), as discussed in
\cite{Jansen:2008wv,Jansen:2009hr}. 

We find large effects due to partial quenching close to the dynamical point  at
small pion masses  which are very obvious in ensemble C (see Fig.~\ref{a0}).  
The partially quenched data do not smoothly extrapolate to the dynamical point.
An explanation has been offered in \cite{Prelovsek:2004jp}:  The partially
quenched states may couple to pairs of pseudoscalars (composed of valence and
sea quarks), leading to unphysical contributions that cancel only in the fully
dynamical case.  We find that our partially quenched data are well described by
the partially quenched formulae of the scattering state, and thus interpret the
partially quenched data as the 2-particle state $\pi\eta_2$. We stress that our
results do not allow for a clear interpretation as 1- or 2-particle state at the
physical point. The question of the coupling of our interpolators to the
scattering states has already been discussed in Sect. \ref{EnergyLevels}. 
Note that the scattering state  $K \bar K$  cannot appear in our simulations
since it involves strange quark loops.

In quenched simulations scattering states cannot contribute to the signal when
using 1-particle interpolators. Also, in particular in the light scalar
channel, ghosts may appear and complicate the spectroscopy. A strategy to
disentangle the contributions has been discussed in \cite{Burch:2005wd}. The
ground state energy level in quenched simulations with the same action
\cite{Burch:2006dg,Gattringer:2008be} was extracted only at larger pion
masses.   There it was essentially compatible with our dynamical data of set A,
extrapolating to the $a_0(1450)$ rather than to $a_0(980)$. The spectroscopy of
the light scalar channel appears to benefit from sea quarks.

\subsubsection{The 1$^{++}$ channel: $a_1$}

\begin{figure}[tbp]
\includegraphics[width=85mm,clip]{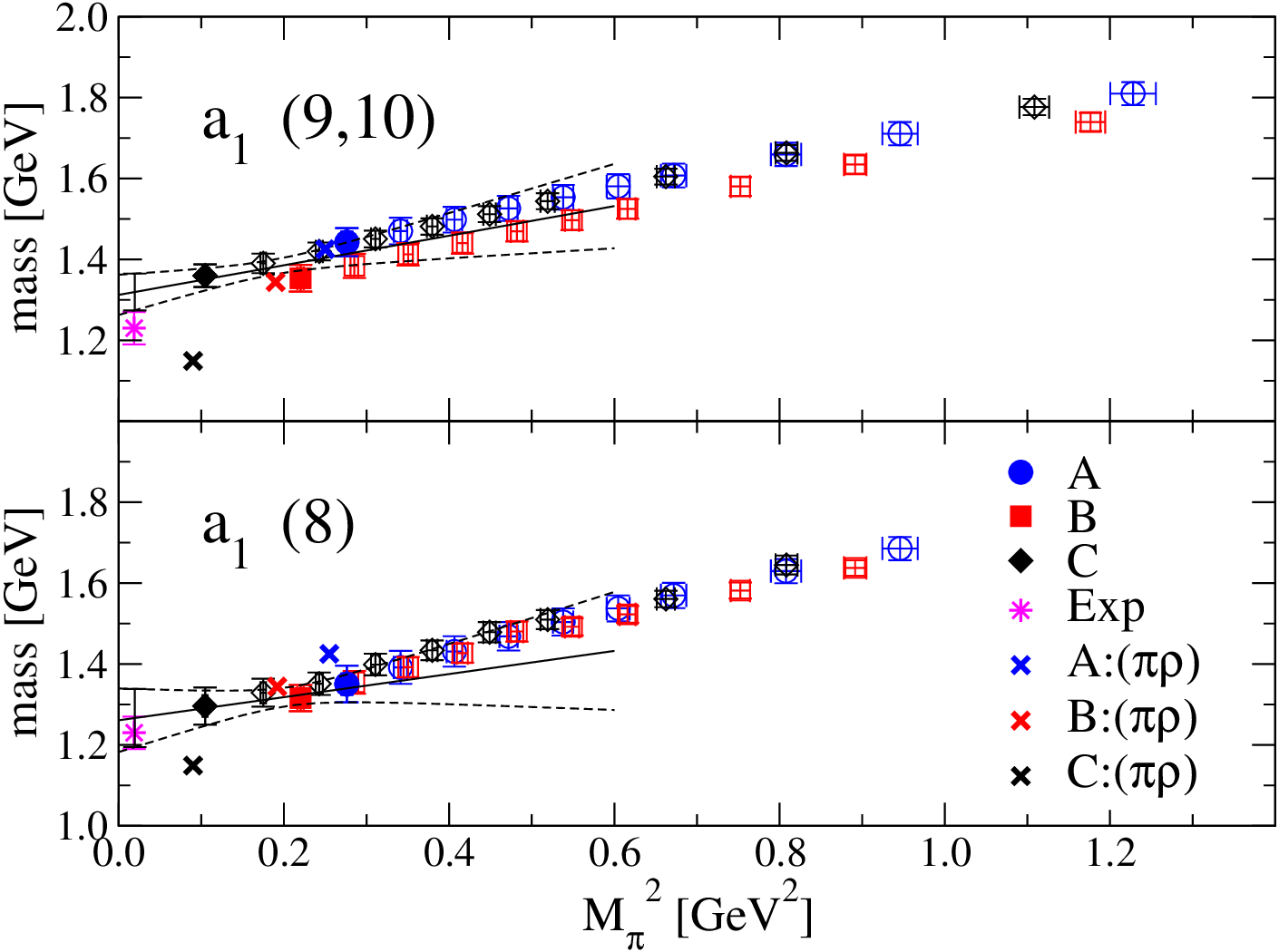}
\caption{Mass plot for the 1$^{++}$ channel ($a_1$), ground state. Results using
interpolators (9,10) are shown in the top plot, results using interpolator (8) at the bottom.
The uncertainty stemming from the choice of interpolators is discussed in the
text. Note that the $a_1$(1260) is a very broad resonance. The estimated energy
level of the $S$ wave scattering state $\pi\rho$ is below the ground state in
ensemble C and comparable to the ground state mass in ensemble A and B. For
better identification, we display the scattering states slightly shifted to the
left. We may interpret that we do not observe the scattering state, at least in
ensemble C.}
\label{a1}
\end{figure}

Considering the pseudo vectors, we encounter some practical difficulties.
Gaussian shaped interpolators do not yield reliable signals and derivative
interpolators are needed. We obtain short plateaus even for the ground state
and there appears an uncertainty associated with different choices of
interpolators. We show results from two sets of interpolators to illustrate this
issue (see Fig.~\ref{a1}).  In principle, such a situation is possible if the
interpolators couple to different physical states or show different
discretization effects. In the absence of such effects, the results from
different sets should agree within error bars and, indeed, the error bars
overlap. Note that the $a_1$(1260) is a very broad resonance, hence scattering
states may complicate the spectroscopy. Based on experiments one could expect an
$S$ wave $\pi\rho$ energy level below the $a_1$ level at least for ensemble
C, which we do not observe. 

The energy levels of quenched simulations using the same action \cite{Gattringer:2008be} appear a bit
high.  However, they are compatible with the results presented in this work if one considers the
uncertainty associated with different choices of interpolators.

\subsubsection{The 1$^{+-}$ channel: $b_1$}

\begin{figure}[tbp]
\includegraphics[width=85mm,clip]{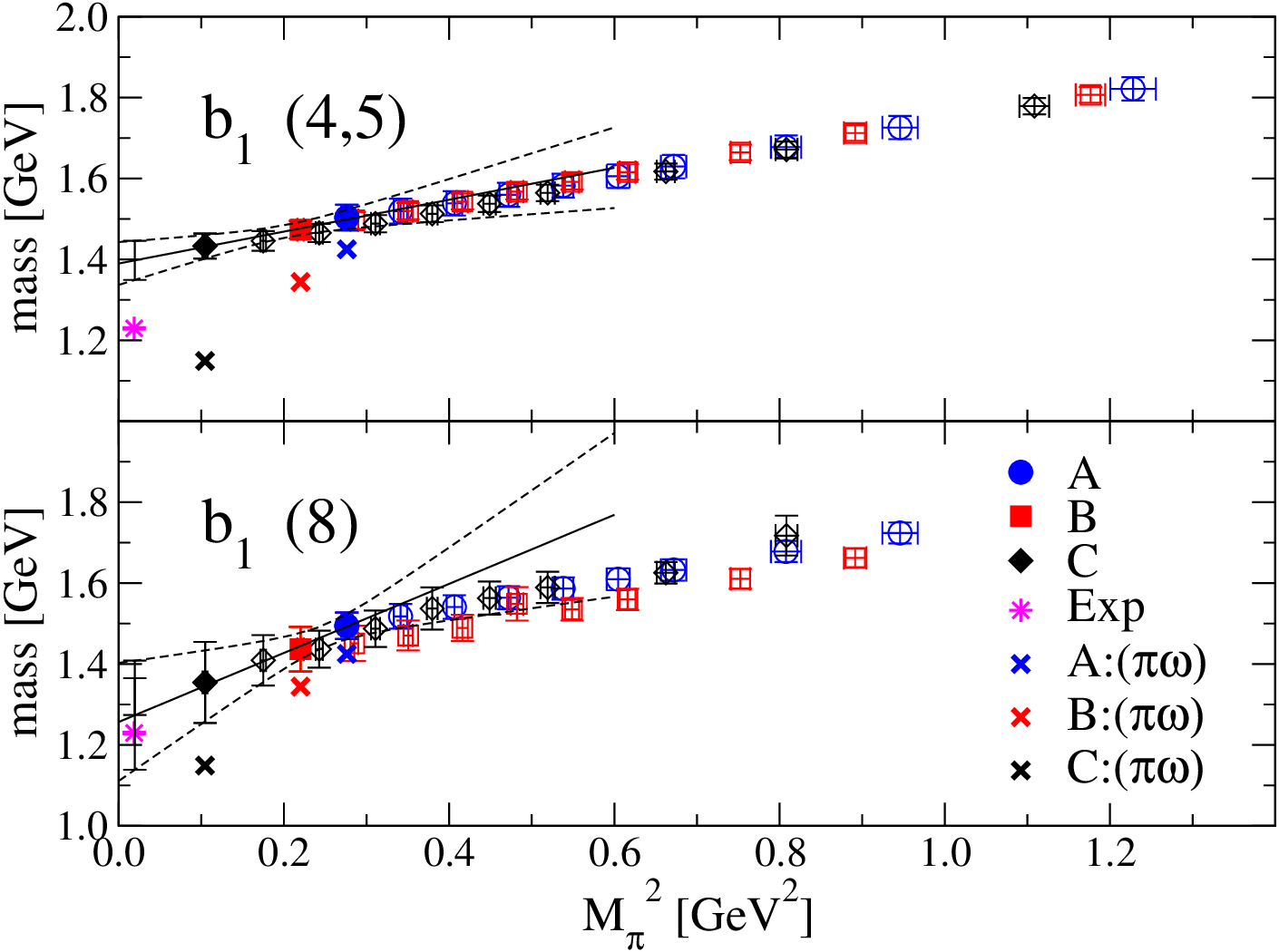}
\caption{Mass plot for the 1$^{+-}$ channel ($b_1$), ground state. Results using
interpolators (4,5) are shown in the top plot, results using interpolator (8) at
the bottom. The uncertainty stemming from the choice of interpolators is discussed
in the text. The energy level of the $S$ wave scattering state $\pi\omega$ is
estimated using the approximation $m_\omega=m_\rho$. For better identification, we
display the scattering states slightly shifted to the left. The energy level is
below the ground state, in particular in ensemble C.} 
\label{b1}
\end{figure}

Here the situation is similar to the 1$^{++}$ channel.  The usage of derivative
interpolators is mandatory, nevertheless weak plateaus are obtained and one
finds an uncertainty which is even larger than in the $a_1$ channel. Again, we
show results from two combinations of interpolators (see Fig.~\ref{b1}). Using
interpolators (4,5) one gets a result rather far from the experimental
$b_1$(1235), while the result of interpolator (8) is compatible with experiment
within a large error bar. Both results are consistent within error bars. The
energy level of the $S$ wave scattering state $\pi\omega$ is estimated using
the approximation $m_\omega=m_\rho$. However, this rough estimate does not
allow for any precise statement about the particle content of the measured
state.

\subsubsection{The 2$^{++}$ channel: $a_2$}

\begin{figure*}[tbp]
\subfigure{\includegraphics[width=85mm,clip]{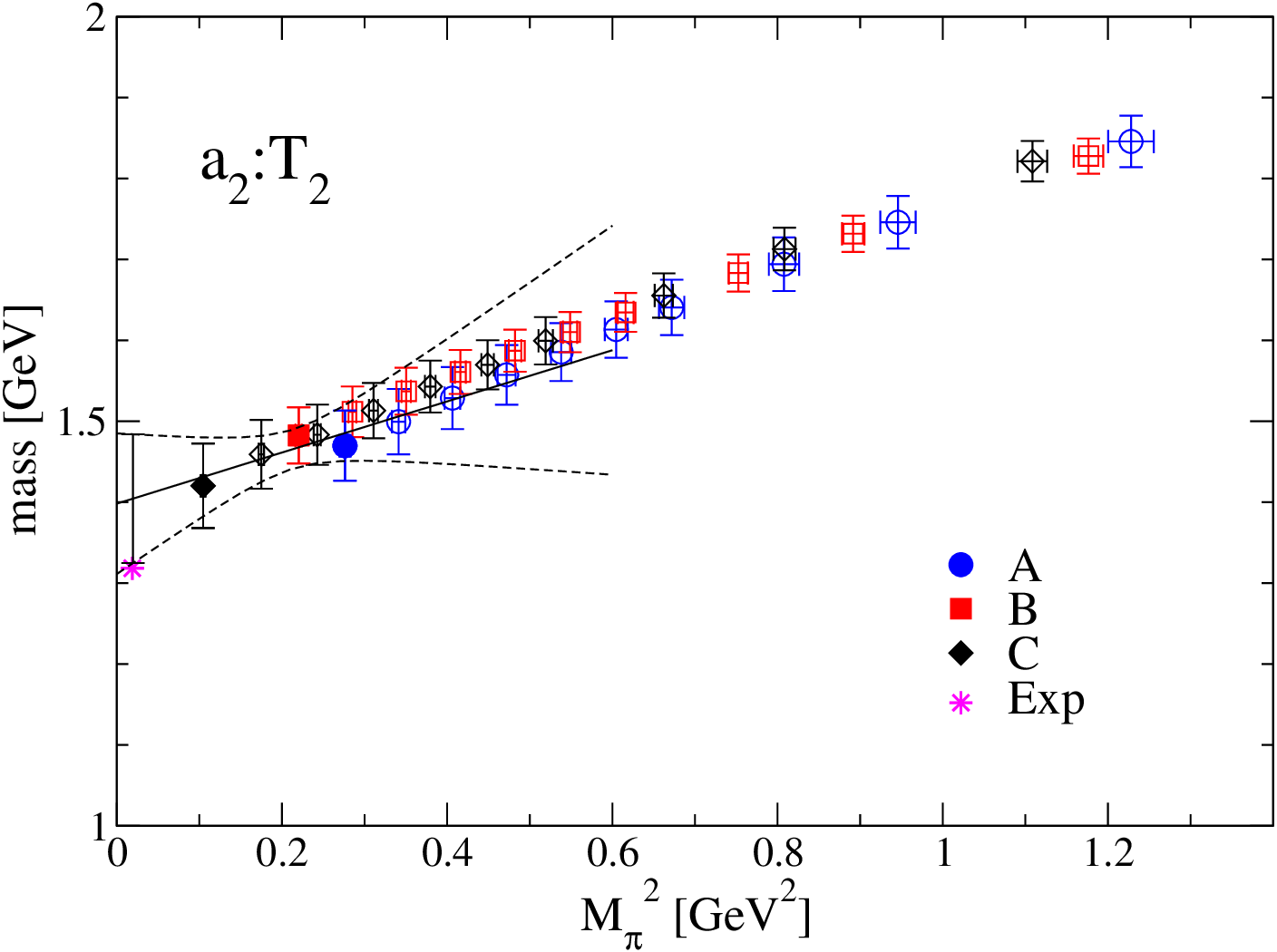}}
\subfigure{\includegraphics[width=85mm,clip]{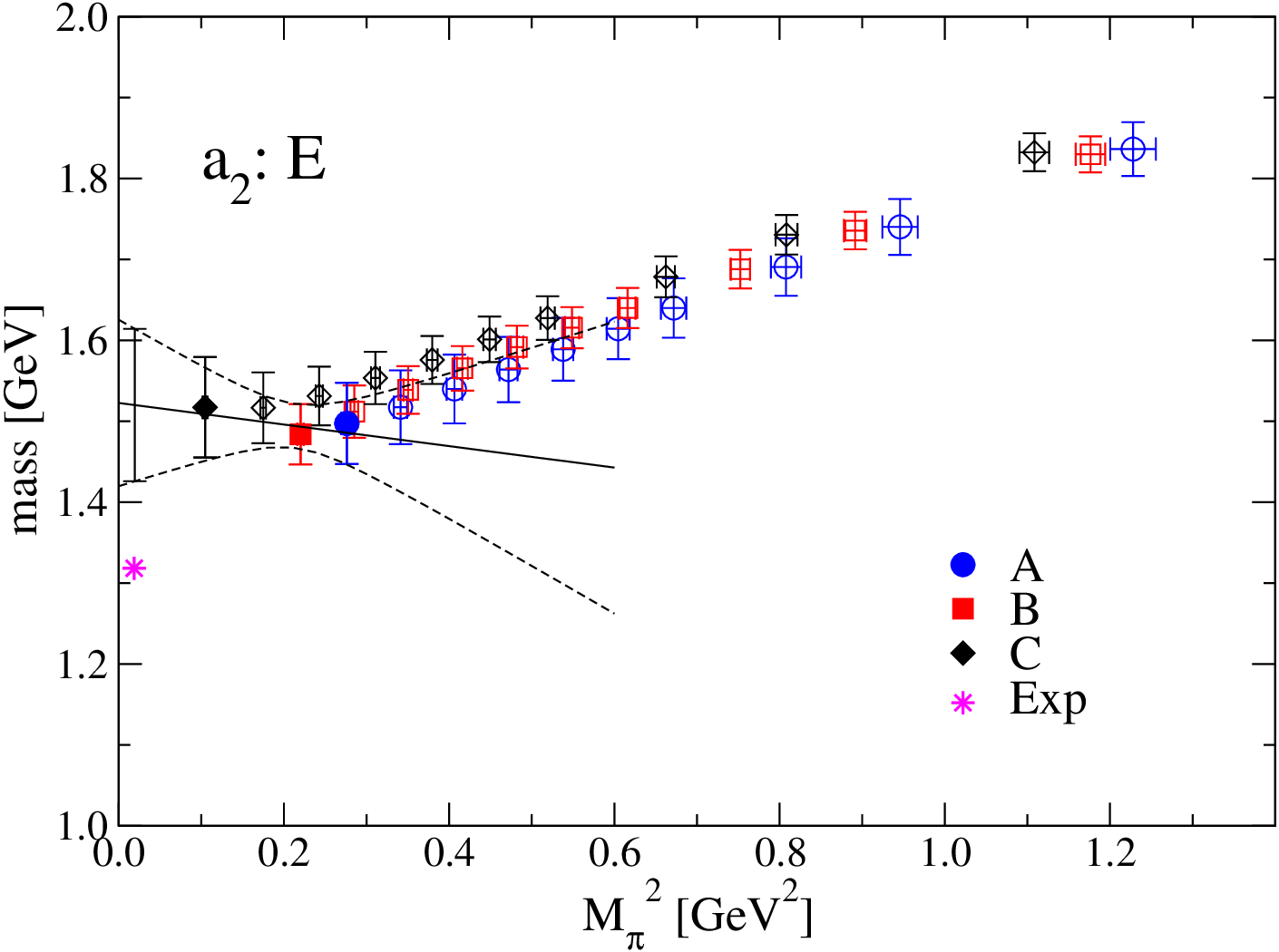}}
\caption{Mass plot for the 2$^{++}$ channel ($a_2$), ground state.
Lhs: Irrep T$_2$, using interpolators (1,2,3,4) in ensemble A and B, (2,3) in C.
Rhs: Irrep E, interpolators (1,2,5) are used throughout.
The result of ensemble C likely suffers from small statistics, leading to a negative slope of the chiral fit.} 
\label{a2}
\end{figure*}

In the spin $2$ channels we encounter for the first time the situation to have
orthogonal irreducible representations (irreps) on the lattice, which couple to the same
spin state in the continuum limit. Hence we are able to compare
results from these different irreps. Fig.~\ref{a2} shows the mass of 2$^{++}$ in irrep T$_2$
and E. In both representations using only one interpolator does not yield a
reliable signal and employing the variational method is necessary. The resulting
mass of T$_2$ agrees with the experimental $a_2$(1320) within one $\sigma$. The
resulting mass of E, however, comes out too high, where the reason seems to be a
large mass of ensemble C, leading to a negative slope of a linear fit.

\subsubsection{The 2$^{--}$ channel: $\rho_2$}

\begin{figure*}[tbp]
\subfigure{\includegraphics[width=85mm,clip]{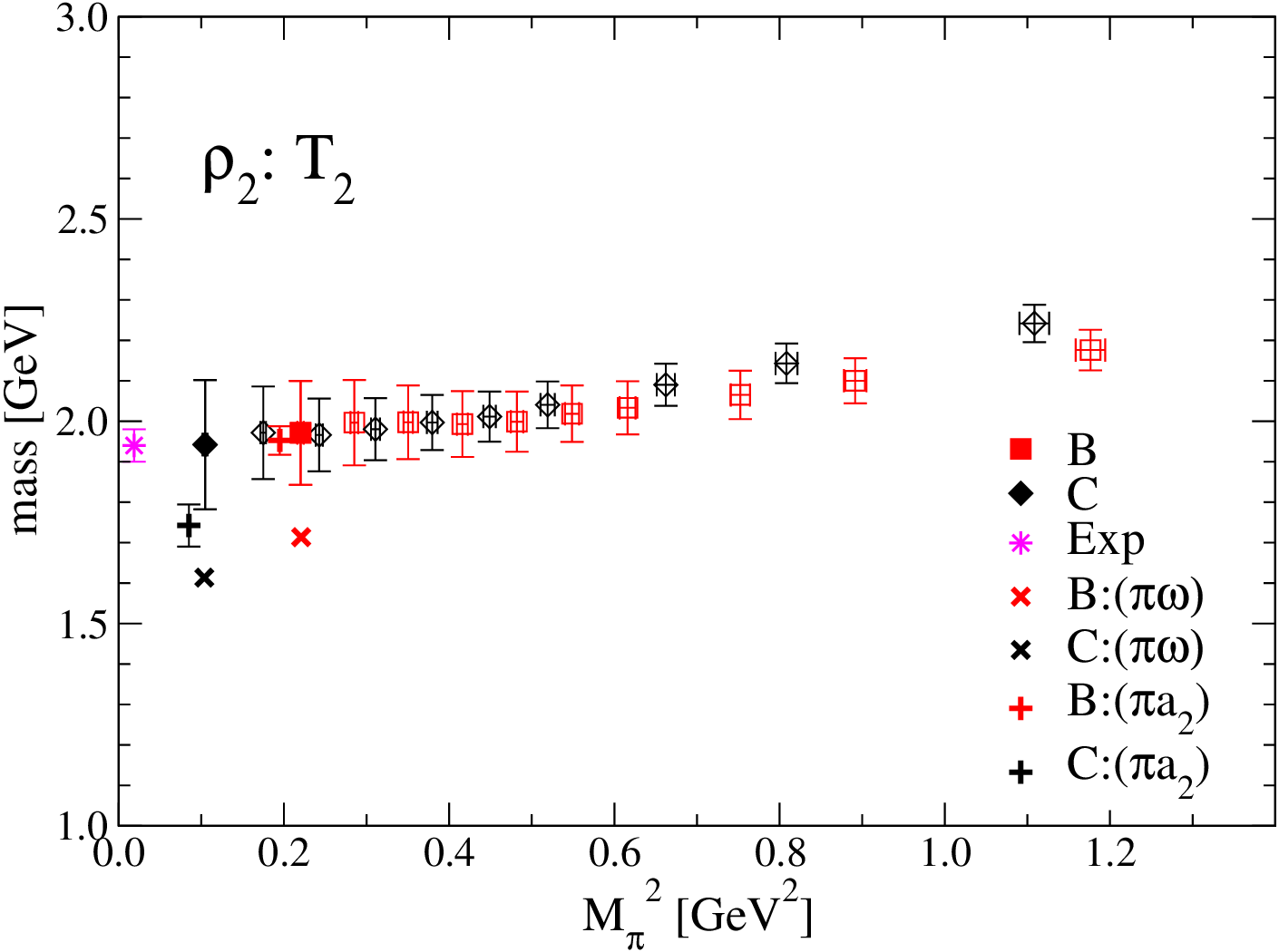}}
\subfigure{\includegraphics[width=85mm,clip]{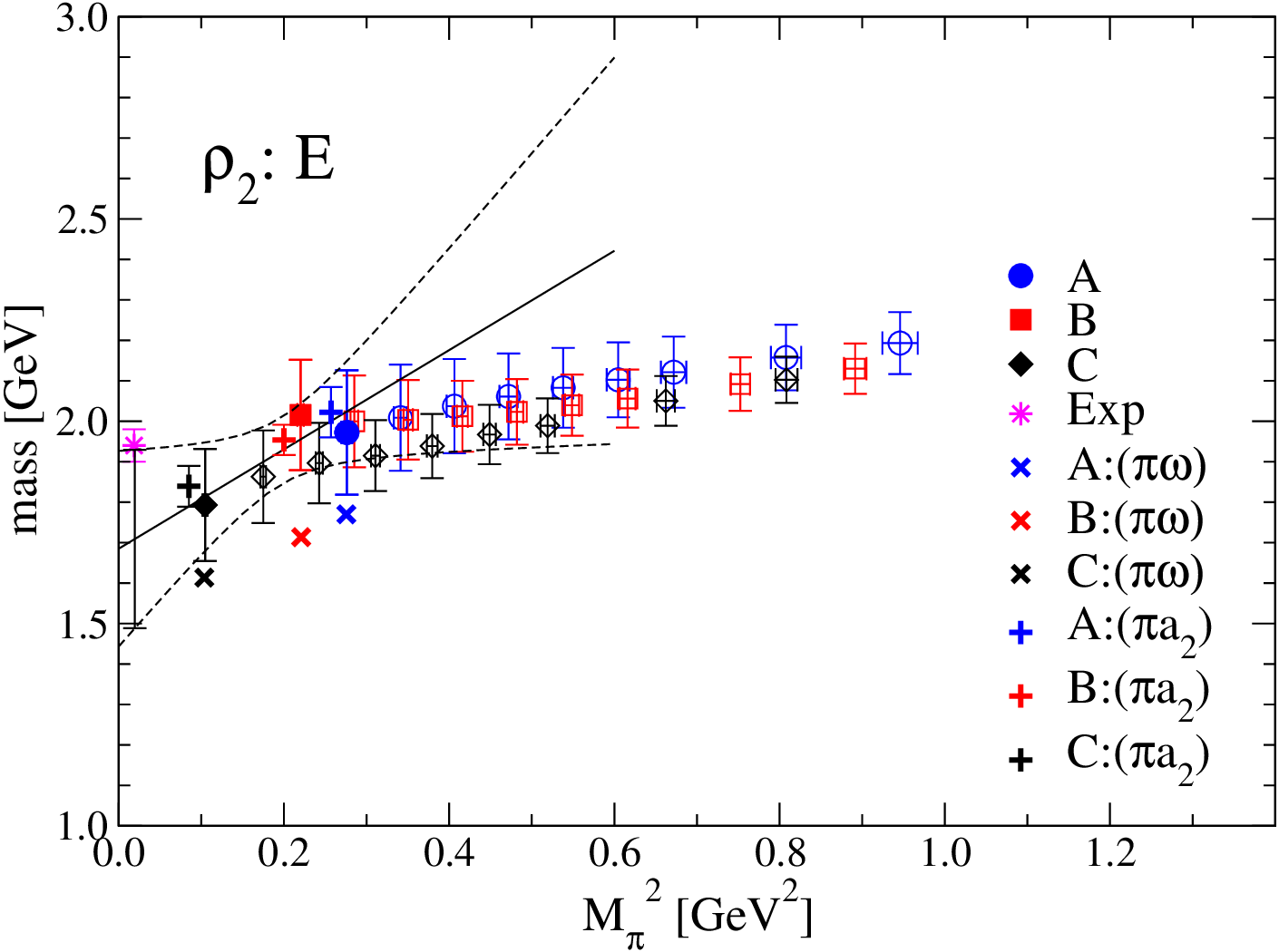}}
\caption{Mass plot for the 2$^{--}$ channel ($\rho_2$), ground state.
Lhs: irrep T$_2$, interpolator (1) is used in ensemble B and C. Rhs: irrep E,
interpolator (1) is used throughout.  The energy level of the $P$ wave
scattering state $\pi\omega$ is estimated using the approximation
$m_\omega=m_\rho$. It is found to be below the ground state in all three
ensembles. The higher energy level of the $S$ wave scattering state $\pi a_2$
is in better agreement with the observed states. However, the estimate does
not allow to identify the particle content of the measured state.} 
\label{rho2}
\end{figure*}

Unfortunately, in the irrep T$_2$ we can extract a mass only in two of the 3
ensembles (see Fig.~\ref{rho2}). Enlarged statistics will be necessary in
ensemble A in order to observe a reliable mass plateau. Also in the
other ensembles and in the irrep E the mass plateaus are short and the fit
ranges are partly motivated by the clearer plateaus of partially quenched mass
results. Nevertheless, the mass obtained is consistent with the experimental
$\rho_2$(1940)  (see Fig.~\ref{rho2}).  An estimate of the scattering states
appearing in this channel is shown in the figure. It seems possible that the
measured state involves contributions from the $S$ wave scattering state
$\pi a_2$, however, the rough estimate does not allow for any clear statement.

Note that in the continuum limit the irrep T$_2$ couples also to spin $3$ and
irrep E to spin $4$ states. Hence a signal of $\rho_3$(1690) could be seen in
T$_2$, but not in E. However, the interpolators used have naive continuum limit
of spin $2$, which we consider to be the reason for the missing of a signal of
$\rho_3$(1690).  We stress that also in this channel higher statistics is
necessary for a reliable extrapolation to the physical point.

\subsubsection{The 2$^{-+}$ channel: $\pi_2$}

\begin{figure*}[tbp]
\subfigure{\includegraphics[width=85mm,clip]{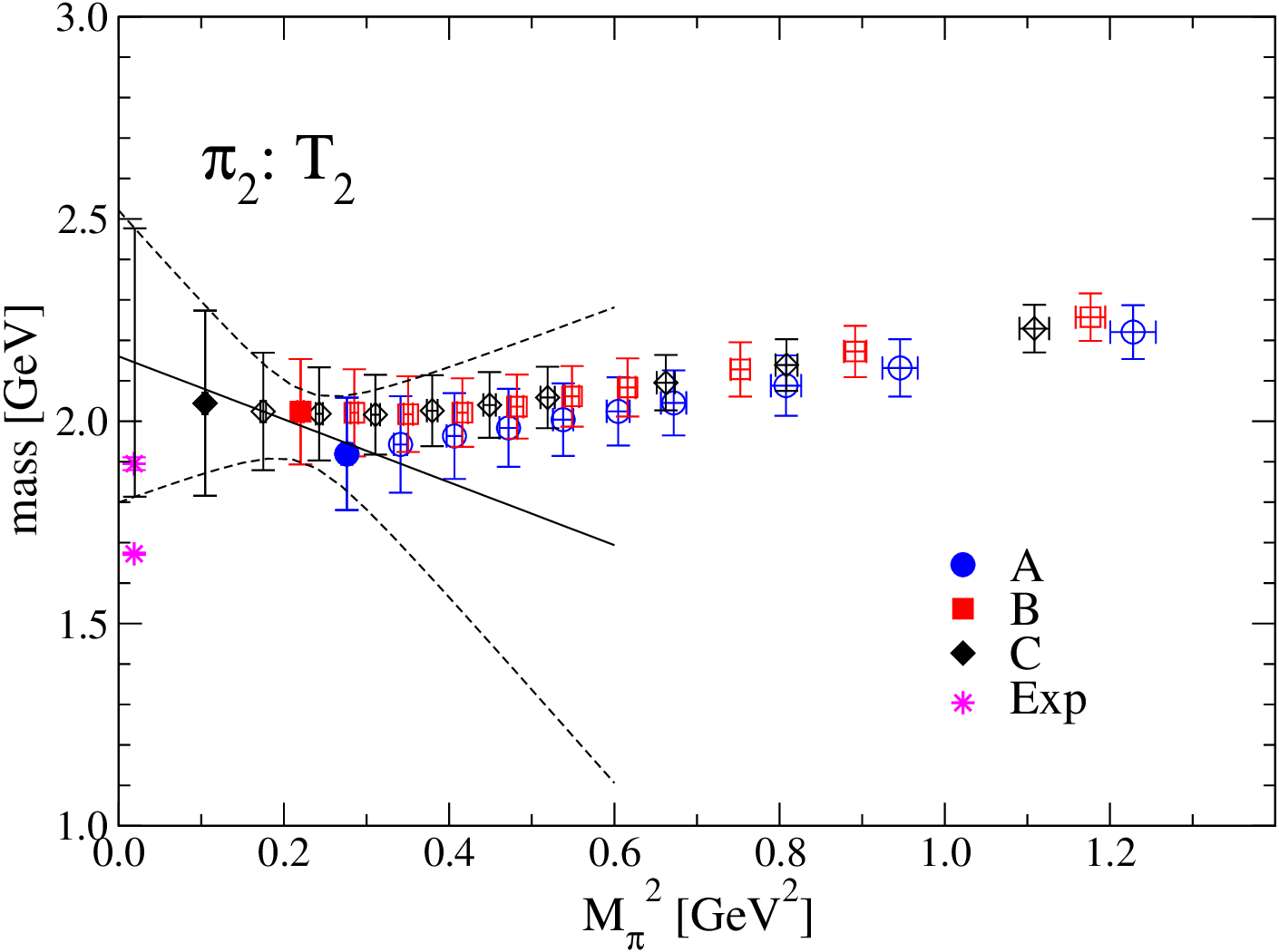}}
\subfigure{\includegraphics[width=85mm,clip]{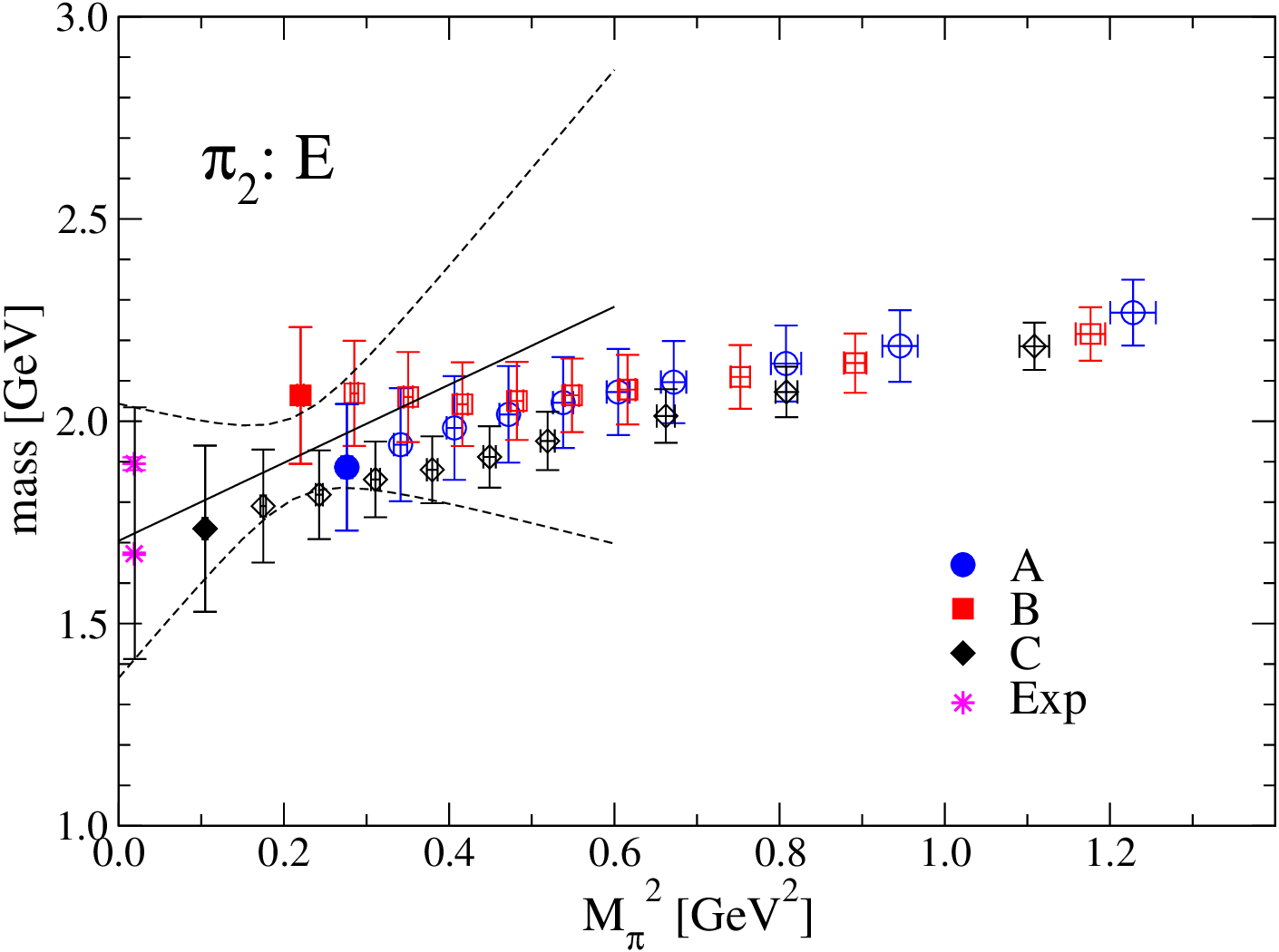}}
\caption{Mass plot for the 2$^{-+}$ channel ($\pi_2$), ground state.
Lhs: Irrep T$_2$, interpolator (1) is used throughout.
Rhs: Irrep E, interpolators (1,2) are used throughout.} 
\label{pi2}
\end{figure*}

Here we encounter rather large statistical errors and additional uncertainty
associated with different sets of interpolators  (see Fig.~\ref{pi2}.)
Nevertheless, all results are consistent within $2\sigma$. Due to the large
error bar, we find the linear chiral extrapolation compatible with both, the
$\pi_2$(1670) and the $\pi_2$(1880). As in the other spin $2$ channels, the
results would benefit from higher statistics.

\subsubsection{Exotics}

So-called exotic states have quantum numbers which cannot be constructed using a
naive quark model. Most of the known exotic particles are found above $2\,$ GeV,
but also some lower ones are known, e.g. $\pi_1$(1400) and $\pi_1$(1600) in the
1$^{-+}$ channel.  We implemented interpolators with exotic quantum numbers
using derivative sources. Unfortunately, the obtained data are very noisy and we
found it impossible to perform a fit of the effective masses. Since we observe a
very weak signal in the 1$^{-+}$ channel, we hope that a fit can be done when
larger statistics becomes available.

\subsection{Baryons with light quarks}

Our baryon interpolators are built from Gaussian sources only (no derivative sources) and can be found
in Appendix \ref{Interpolators}. All interpolators are projected to definite
parity in each channel.

\subsubsection{Nucleon positive parity}

\begin{figure}[tbp]
\subfigure{\includegraphics[width=85mm,clip]{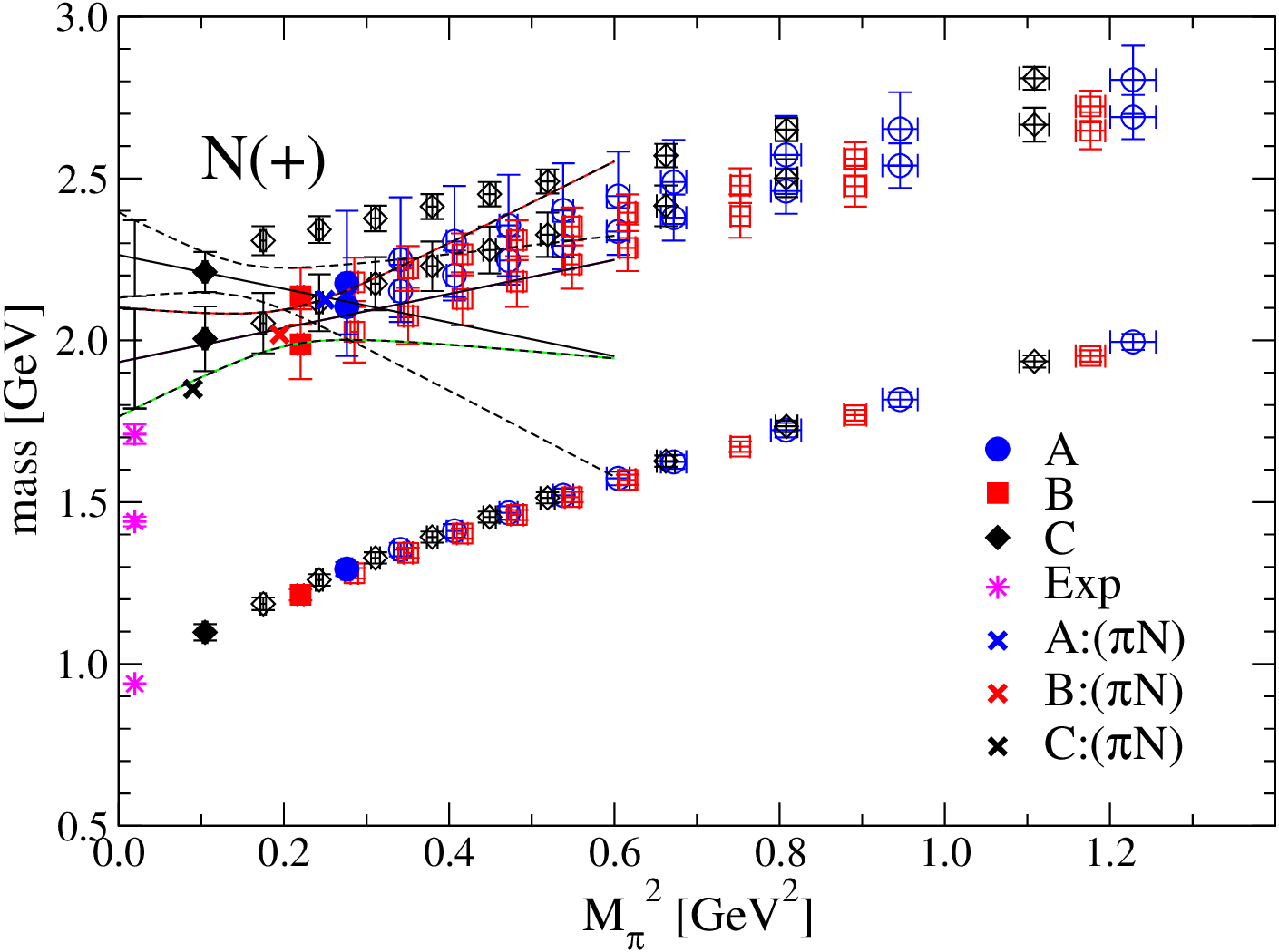}}
\subfigure{\includegraphics[width=85mm,clip]{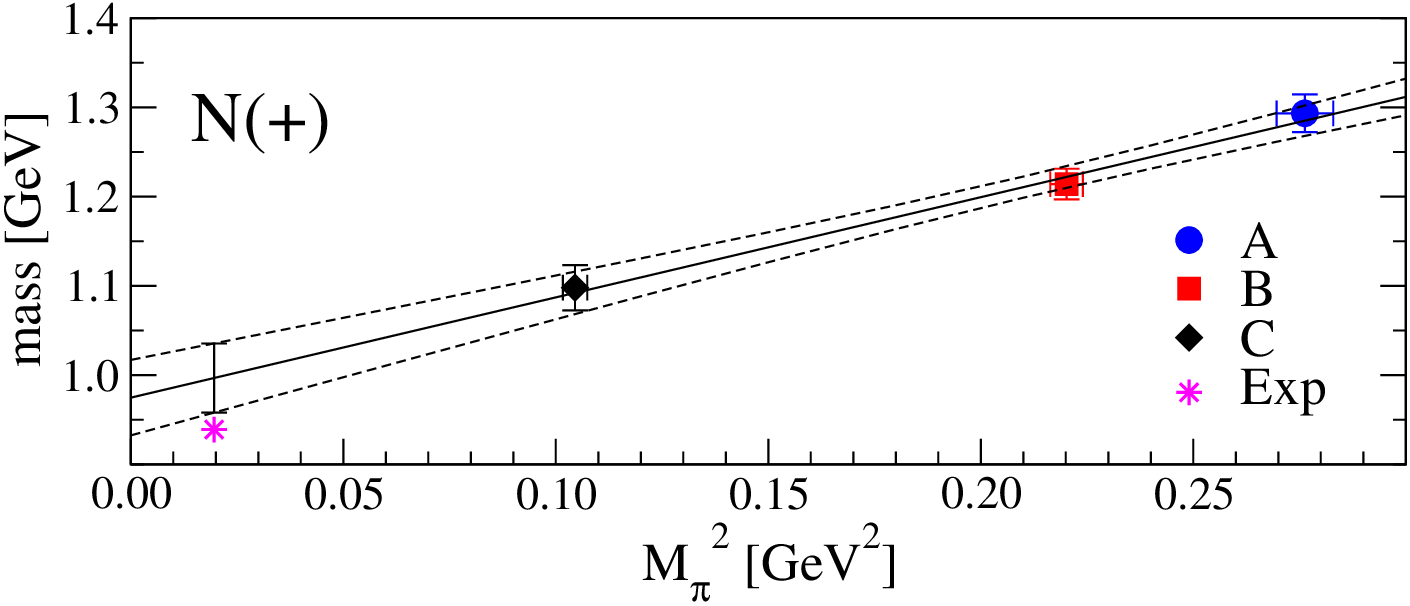}}
\caption{Mass plot for the nucleon positive parity channel. Top: Results for
ground state and two excitations. The energy level of the $P$ wave scattering
state $\pi N$ is close to the first excitation in all three ensembles. For
clarity, we display the scattering states slightly shifted to the left. Bottom:
Fit to the ground state mass of the nucleon positive parity channel.
Interpolators (1,2,13,14) are used in ensemble A, (1,2,4,14,15,18) in B and C.} 
\label{nucleon_pos}
\end{figure}

The nucleon positive parity ground state can be extracted to good precision  as
usual (see Fig.~\ref{nucleon_pos}).  A conventional fit linear in the pion mass
squared yields a nucleon mass a bit higher than the experimental value.  In
\cite{WalkerLoud:2008bp,WalkerLoud:2008pj} it was found that a simple fit linear
in the pion mass of the nucleon positive parity ground state agrees well with
experiment. Indeed such an extrapolation agrees well with experiment also in
our case. However, a reliable clear distinction would only be possible using
more data from dynamical simulations and higher statistics. Therefore, we stick
to the expectation of the analytic behavior being close to the pion mass squared
as suggested from ChPT and thus we quote the corresponding fit linear in the
pion mass squared in the conclusions. 

The first excitation would be compatible with the energy level of the $P$
wave scattering state $\pi N$. However, following the arguments of Sec.
\ref{EnergyLevels}, we believe to see an almost pure 1-particle state.

The first nucleon excitation, the so-called Roper, comes out several hundred
MeV too high  in dynamical simulations
\cite{Bulava:2009jb,Engel:2009cq,Bulava:2010yg}. There are several possible
explanations for this. On one hand, whereas the negative parity ground state
baryons are orbital excitations (according to the quark model), the Roper is a
radial excitation. Thus its size may be substantially larger  than that of
the nucleon and affected by squeezing due to the spatial volume size.
Another reason may be a strong influence from the $\pi N$ channel, which may
not be properly represented by our set of interpolators (cf., the discussion
in Sect. \ref{EnergyLevels}). It may be necessary to explicitly include such
meson-nucleon interpolators, which then, however, poses the technical
challenge of backtracking quark loops.

Comparing to the corresponding quenched simulations \cite{Burch:2006cc}, we do
not observe a significant qualitative difference of the results in  the nucleon
positive parity channel. Our data are also in agreement with quenched and
dynamical results  from other groups (e.g.,
\cite{Cohen:2009zk,Mahbub:2009cf,Bulava:2010yg}).

\subsubsection{Nucleon negative parity}

\begin{figure}[tbp]
\includegraphics[width=85mm,clip]{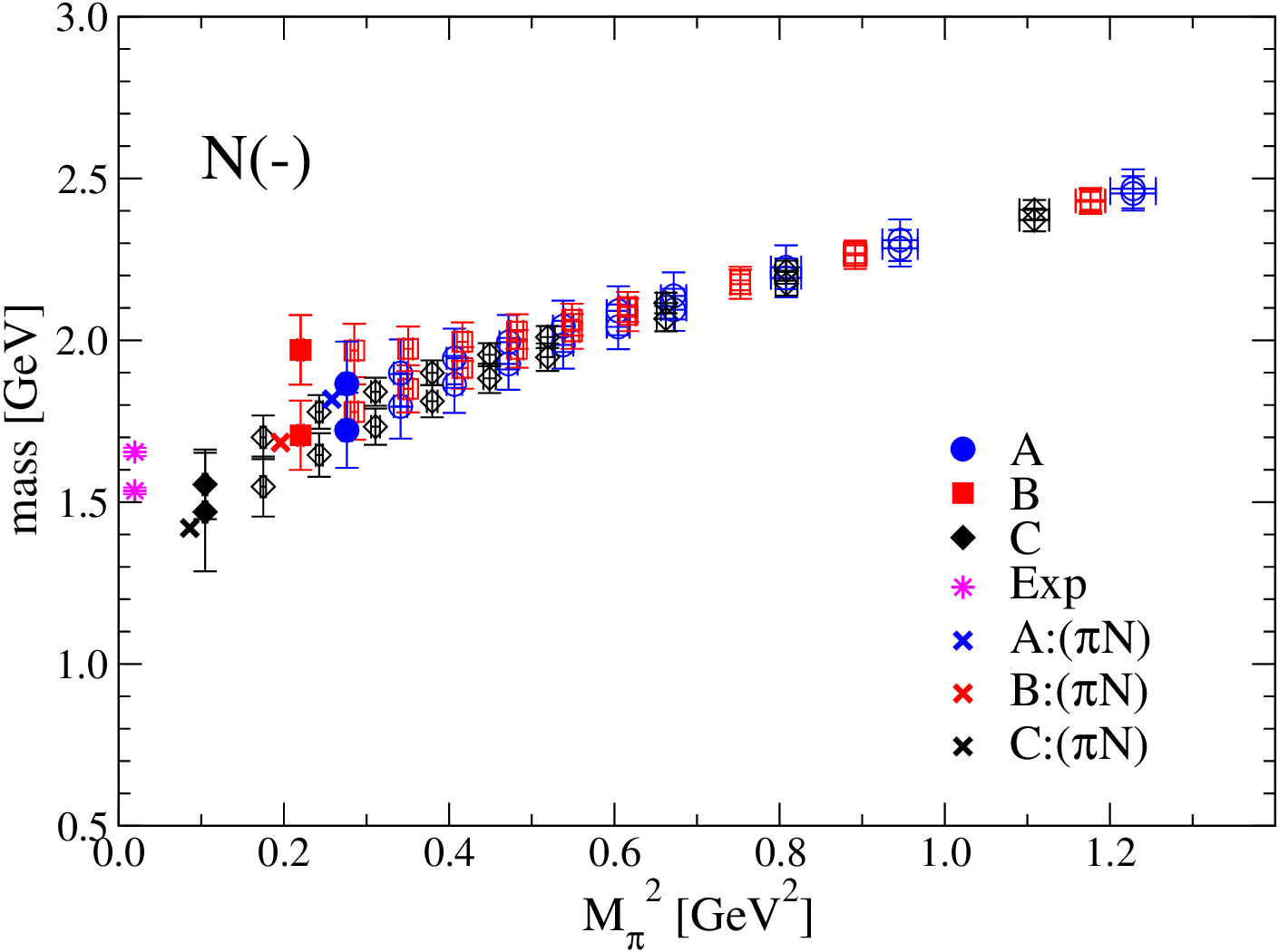}
\caption{Mass plot for the nucleon negative parity channel, ground state and
first excitation. The energy level of the $S$ wave scattering state $\pi N$ is
close to the first excitation in A and close to the ground state in B and C.
Naively, this could be interpreted as a hint for a level crossing of the
1-particle ground state and the scattering state. Also the low results in case
of ensemble C suggest such an interpretation. However, the eigenvectors
contradict this picture (see Fig.~\ref{nucleon_neg_vectors}). Interpolators
(2,7,9) are used in ensemble A, (1,3,7,8,9) in B and (1,7,8,9) in C. For
clarity, we display the scattering states slightly shifted to the left and omit
the chiral extrapolation in the Figure.} 
\label{nucleon_neg}
\end{figure}
\begin{figure}[tbp]
\includegraphics[width=85mm,clip]{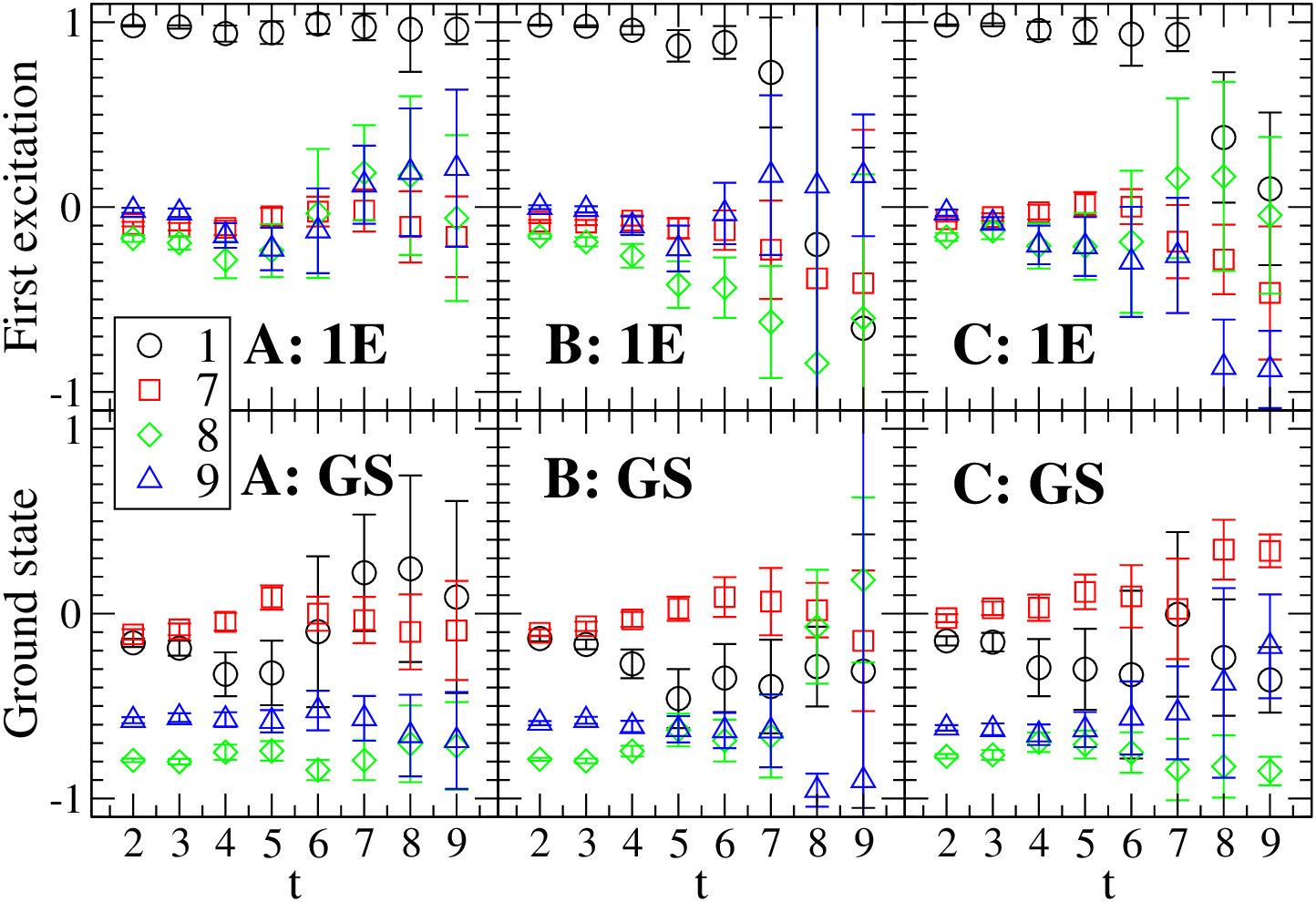}
\caption{Normalized eigenvectors of the nucleon negative parity channel, ground
state and first excitation at the dynamical point.  To allow for a direct
comparison, the same set of interpolators (1,7,8,9) is used in all three
ensembles A, B and C.  The corresponding fits of the eigenvalues are performed
in the time range $4\leq t \leq 7$. In all three ensembles, the ground state is
dominated by the second Dirac structure ($\chi_2$), while the first excitation
is an almost pure $\chi_1$ state (see Table \ref{tab:baryoninterpolators1}). One
may conclude that no level crossing of the lowest two states is observed for
pion masses in the range between 320 and 520 MeV. Note that in order to extract
masses, different sets are used in A and B, which yield more stable plateaus.} 
\label{nucleon_neg_vectors}
\end{figure}

In general, the signal is rather noisy in negative parity baryon channels. In
the case of the nucleon, we also find that the backwards-running positive parity
nucleon (and possibly also back-to-back scattering states
\cite{Takahashi:2008fy}) limits the fit window to the interval $3\leq t \leq 7$.
However, we even find a signal of the first excitation, which is close to the
ground state, both being compatible with experimental data (see
Fig.~\ref{nucleon_neg}).  We find that all our negative parity baryon ground
states come out somewhat too low. 

In nature, the $S$ wave state $\pi N$ (assuming free particles) is below the
one-particle  ground state in the nucleon negative parity channel. This may also
be true at slightly larger pion masses, e.g., in ensemble C. That would explain
the low results in C and thus the low chiral extrapolation as well. At large
pion masses, the scattering state becomes heavier than the 1-particle ground
state and one expects a level crossing to take place in between. Indeed, our
results on masses in the three ensembles are compatible with such a picture.

However, in contrast to the scalar channels, we can extract useful information
from the eigenvectors in the nucleon and sigma negative parity channels. The
reason is that several interpolators are needed for a good signal and that two
states are observed.  We find that in all three ensembles the ground state is
dominated by the second Dirac structure ($\chi_2$), while the first excitation
is an almost pure $\chi_1$ state  (see Fig.~\ref{nucleon_neg_vectors} and
Table~\ref{tab:baryoninterpolators1}). This property is seen even more clearly
for partially quenched points, where the plateau is more stable. One may
conclude that no level crossing of the lowest two states is observed for pion
masses in the range of 320 to 520 MeV.

Another hint comes from the comparison with old quenched results using the same
action \cite{Burch:2006cc}. Using only 3-quark nucleon interpolators, no
scattering states can appear in quenched simulations. Hence, the eigenvectors in
the quenched simulations can clearly be identified with 1-particle states. We
stress that in the quenched approximation ghosts appear at low pion masses, thus
a reliable comparison to dynamical simulation can be done only at large pion
masses. We assume the pions of ensemble A to be heavy enough to allow for a
comparison with quenched simulations. By subsequent comparison of dynamical
simulations at decreasing pion masses, in principle conclusions down to physical
pion masses are possible. Unfortunately, direct comparison of the quenched
eigenvectors with our dynamical ones is impossible, since the corresponding
interpolators differ slightly (e.g., in the width of the Gaussian source).
However, qualitatively, we find the same Dirac structure content of the two
lowest states. From that, one may conclude that both our lowest two states are
1-particle states, which is also suggested from the missing level crossing of
the eigenvectors.

So, while the extracted mass values slightly favor the interpretation of a
scattering state in ensemble C,  the eigenvectors tell a different story.  Since
the line of arguments based on the eigenvectors seems to be more reliable, we
believe to see almost pure 1-particle states and quote the corresponding chiral
extrapolation in the summary.

In the quenched simulation with the CI-action \cite{Burch:2006cc}, 
nucleon negative parity masses have been extracted only at larger pion masses.
The bending down at low pion masses observed in the present work can thus not be
compared directly.

Note that in order to extract masses, different sets are used in A and B than
the ones shown in Fig.~\ref{nucleon_neg_vectors}. In ensemble C, the fit range
is motivated by tracing of plateaus of partially quenched results.

\subsubsection{$\Delta$ positive parity}

\begin{figure}[tbp]
\includegraphics[width=85mm,clip]{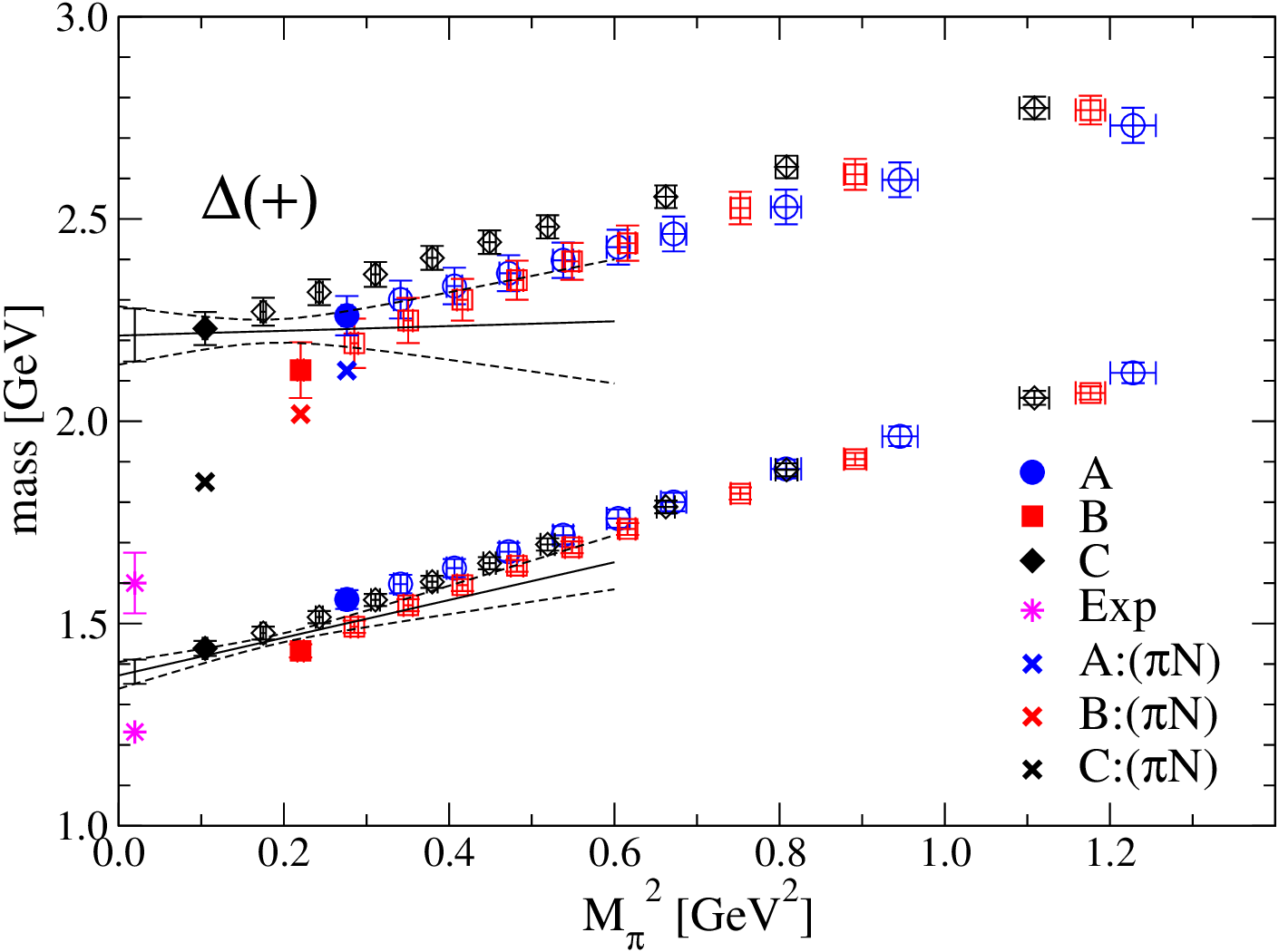}
\caption{Mass plot for the $\Delta$ positive parity channel, ground state and
first excitation. The energy level of the $P$ wave scattering state $\pi N$ is
close to the first excitation in A and B, but clearly separated from any
measured state in C. Interpolators (1,2,3) are used in ensemble A and C, (2,4,6)
in B.} 
\label{delta_pos}
\end{figure}

In the $\Delta$ positive parity channel, we find the masses being too high, in
particular the first excitation (see Fig.~\ref{delta_pos}), although they are
lower than in the quenched analysis \cite{Burch:2006cc}. Since the statistical
error is fairly small, systematic errors and finite volume effects seem to be
responsible. As will be discussed below, this channel -- for large values of the
valence quark mass -- is used  to identify the strange quark mass parameter.
Analogous to other $P$ wave scattering states, $\pi N$ seems to be missing in
our simulation.

Quenched results using the CI-action \cite{Burch:2006cc}  have shown a similar
systematic upwards shift of the masses. Our $\Delta$ positive parity ground
state is compatible with other  groups (e.g., \cite{Bulava:2010yg}). However,
there the first excitation is fairly close to the ground state,  which we do
not observe here. A possible reason could be the larger basis of interpolator
used in that work. Hence, we cannot exclude possible systematic errors
associated with the  choice of interpolators in this channel.

\subsubsection{$\Delta$ negative parity}

\begin{figure}[tbp]
\includegraphics[width=85mm,clip]{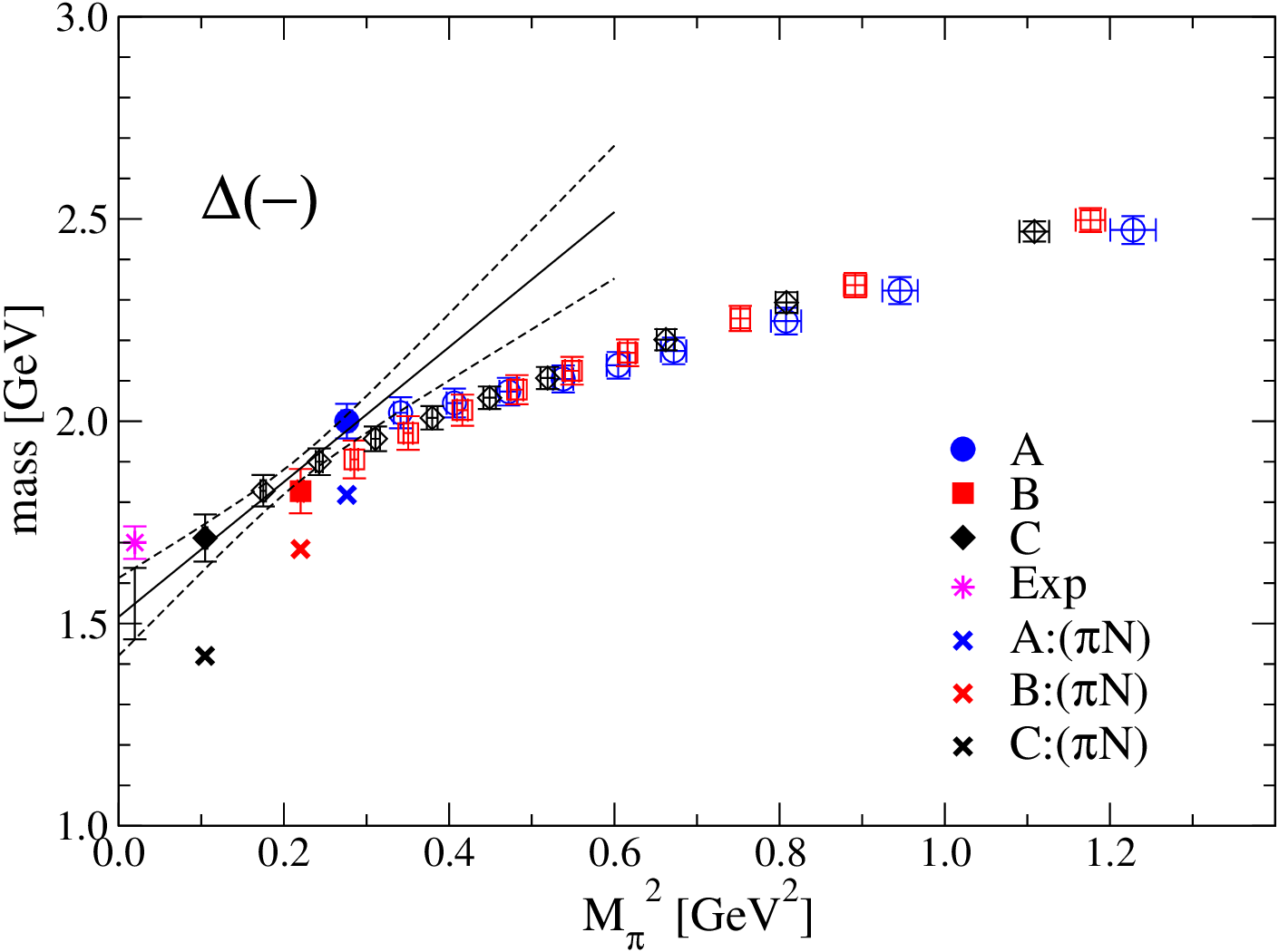}
\caption{Mass plot for the $\Delta$ negative parity channel, ground state. The
energy level of the $S$ wave scattering state $\pi N$ is below the ground
state in all three ensembles. We display the scattering states slightly shifted
to the left. Interpolator (3) is used in ensemble A, (2,3) in B and C.} 
\label{delta_neg}
\end{figure}

There is a clear signal found of the $\Delta$ negative parity channel ground
state (see Fig.~\ref{delta_neg}).  The energy level of the $S$ wave
scattering state $\pi N$ is below the ground state in all three ensembles. The
results seem to be in better agreement with an interpretation in terms of a
1-particle state.

Quenched results of \cite{Burch:2006cc} have been extracted only at rather 
large pion masses, making a comparison to present results uninstructive.

\section{Results: Strange Hadrons}
\label{ResultsStrange}

We extract the strange quark mass parameter by identification of a partially
quenched $\Delta$ (i.e., valence quark of larger mass)  with $\Omega$(1670), as
described in Sect. \ref{strangemass}. The interpolators used for strange meson
spectroscopy are listed in Appendix \ref{Interpolators}. Since $C$-parity is an
exact symmetry for mesons only in the case of mass-degenerate quarks, we
consider interpolators without projection to definite $C$-parity for the strange
mesons. This means that we do not perform a symmetrization of the interpolators
in the strange meson sector.  Our strange meson correlator calculation lacks
cross correlations between interpolators of different $C$-parity quantum numbers
in the limit of degenerate quark masses. Therefore, when analyzing strange
mesons, we have to restrict ourselves to subsets of interpolators sharing the
same $J^{PC}$ quantum numbers in the limit of degenerate quark masses.

In all plots shown for strange hadrons, the full symbol represents a hadron
where the valence strange quark mass is determined from the $\Omega$(1670) and
the light quark has the mass of the dynamical quarks. The open symbols denote
partially quenched data, where only the light valence quark mass varies. We
thus neglect the effect of a dynamical strange quark, motivated by the
dominance of light quark loops over strange quark loops. 

A quantum field theory, where the sea quark masses do not agree with the
valence quark masses,  is ``sick'', as can be seen, e.g., by the appearance of
ghosts which violate the spin-statistic theorem. As
discussed in \cite{Prelovsek:2004jp}, the correlators are {\em strongly}
affected by partial quenching if the valence quarks are lighter than the sea
quarks. In our simulations we consider only valence quarks heavier than or
equal to the sea quarks. 

Clearly, it would be desirable to include strange sea quarks in the simulation.

\subsection{Setting the strange quark mass}
\label{strangemass}

\begin{figure}[tbp]
\includegraphics[width=85mm,clip]{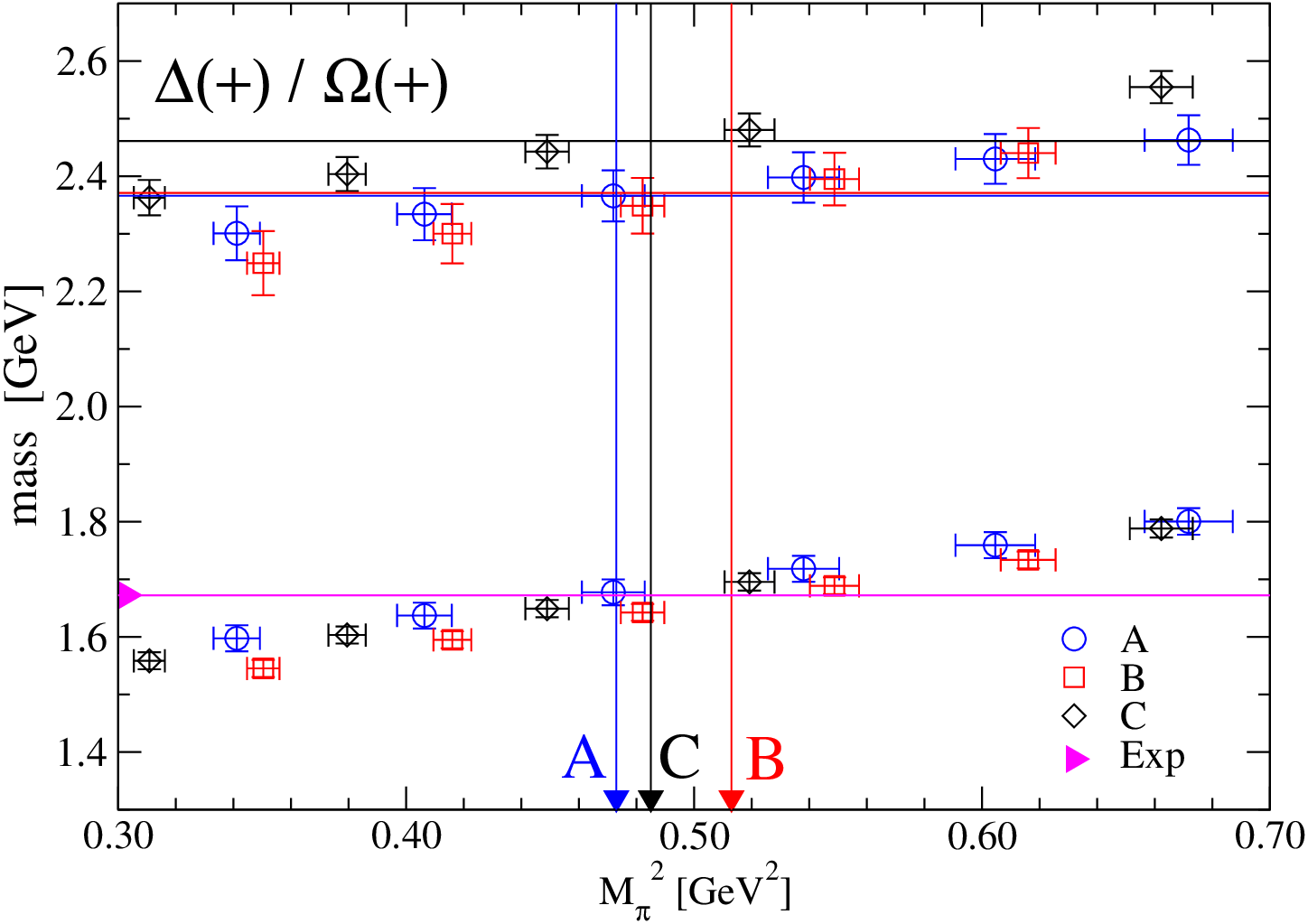}
\caption{Extracting the strange quark mass parameter by identifying a partially
quenched $\Delta$ with $\Omega$(1670). The lowest (magenta) horizontal line
represents the $\Omega$(1670). Crossing this horizontal line with the partially
quenched $\Delta$ mass curves of ensembles A, B and C defines the strange quark
mass parameter of the three ensembles, illustrated by the three vertical lines
in the corresponding colors. The resulting bare quark mass parameters are
-0.020, -0.015 and -0.022, for ensembles A, B and C, respectively. The
calculation for the excited $\Omega$ around $2400$ MeV is expected to be too
high due to finite volume effects.}
\label{omega}
\end{figure}

\begin{figure}[tbp]
\includegraphics[width=85mm,clip]{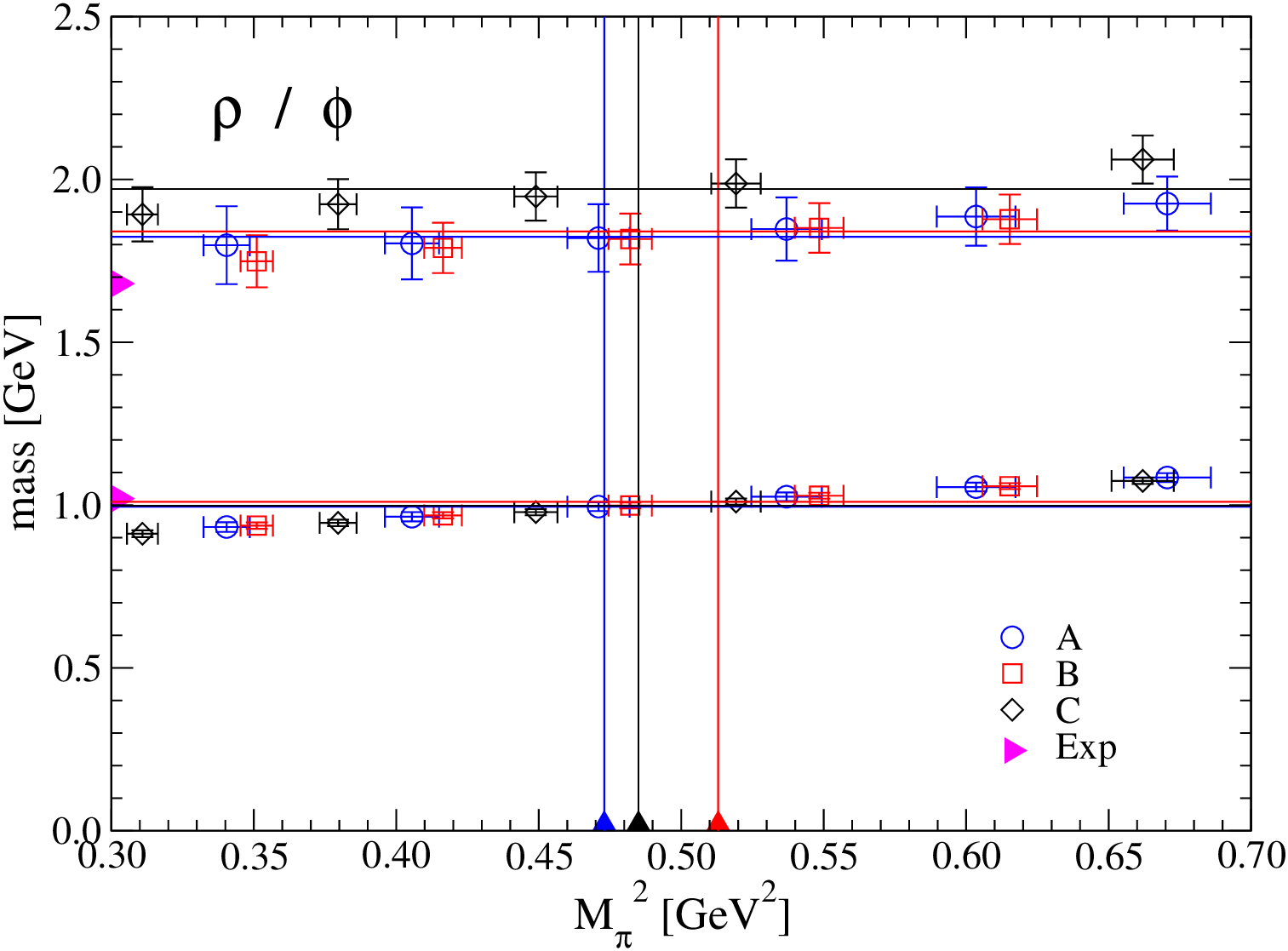}
\caption{Cross-check of the obtained strange quark mass parameter: The partially
quenched $\phi$ from the ground state of the 1$^{--}$ channel fits the
experimental $\phi$(1020) very nicely.  The result for the excited $\phi$ is 
higher than the experimental value, the deviation may be due to the neglected
disconnected diagrams or simply due to the weak signal.}
\label{phi}
\end{figure}

The $\Omega$(1670) consists of three strange valence quarks and  shows weak
dependence on the light quark masses. $\Omega$ and  $\Delta$ share the same
$J^{P}$ quantum numbers, they differ only in their flavor content. Therefore, we
use our partially quenched results in the $\Delta$ positive parity channel to
identify the strange quark mass parameter for our ensembles A, B and C (see
Fig.~\ref{omega}). We decided to choose strange quark mass parameters which fit
the experimental $\Omega$ within our error bars, allowing only for parameter
values in discrete steps of $0.05$ additive to the sea quark mass.  In case of
ensemble A the obtained mass parameter perfectly fits one of the already
available quark propagators. In case of ensembles B and C, we decided to
recompute the quark propagator at the strange quark mass instead of
interpolating between the two adjacent values.

Another possibility to extract the strange quark mass parameter would be to
apply the same recipe to a partially quenched $\phi$ in the 1$^{--}$ channel.
The decay channels of $\phi$(1020) suggest that it is dominated by its
($s\bar{s}$) flavor content.  We use $\phi$ as a cross-check for the strange
quark mass obtained via $\Omega$(1670). The result fits the experimental
$\phi$(1020) very nicely (see Fig.~\ref{phi}), indicating that our approach is
consistent.  The excited state of $\Omega$ is assumed to suffer from finite size
effects. The signal of the excitation of $\phi$ suffers from neglected
disconnected diagrams and poor statistics. Thus these two levels are not
appropriate for further checks of the strange quark mass. However, the ground
state levels of $\Sigma$ and $\Xi$ positive parity may be regarded as additional
affirmative cross-checks (see Figs. \ref{sigma_pos} and \ref{xi_pos}).

Even in the case of the $\Omega$, one could expect finite size effects, since
they show up in the $\Delta$ positive parity channel for both, ground and
excited state. However, using heavier quarks, finite size effects are expected
to become less and less important in the ground state. The cross-check using
$\phi$ verifies this expectation.

\subsection{Mesons with strange quarks}

\subsubsection{The 0$^{-}$ channel: $K$}

\begin{figure}[tbp]
\includegraphics[width=85mm,clip]{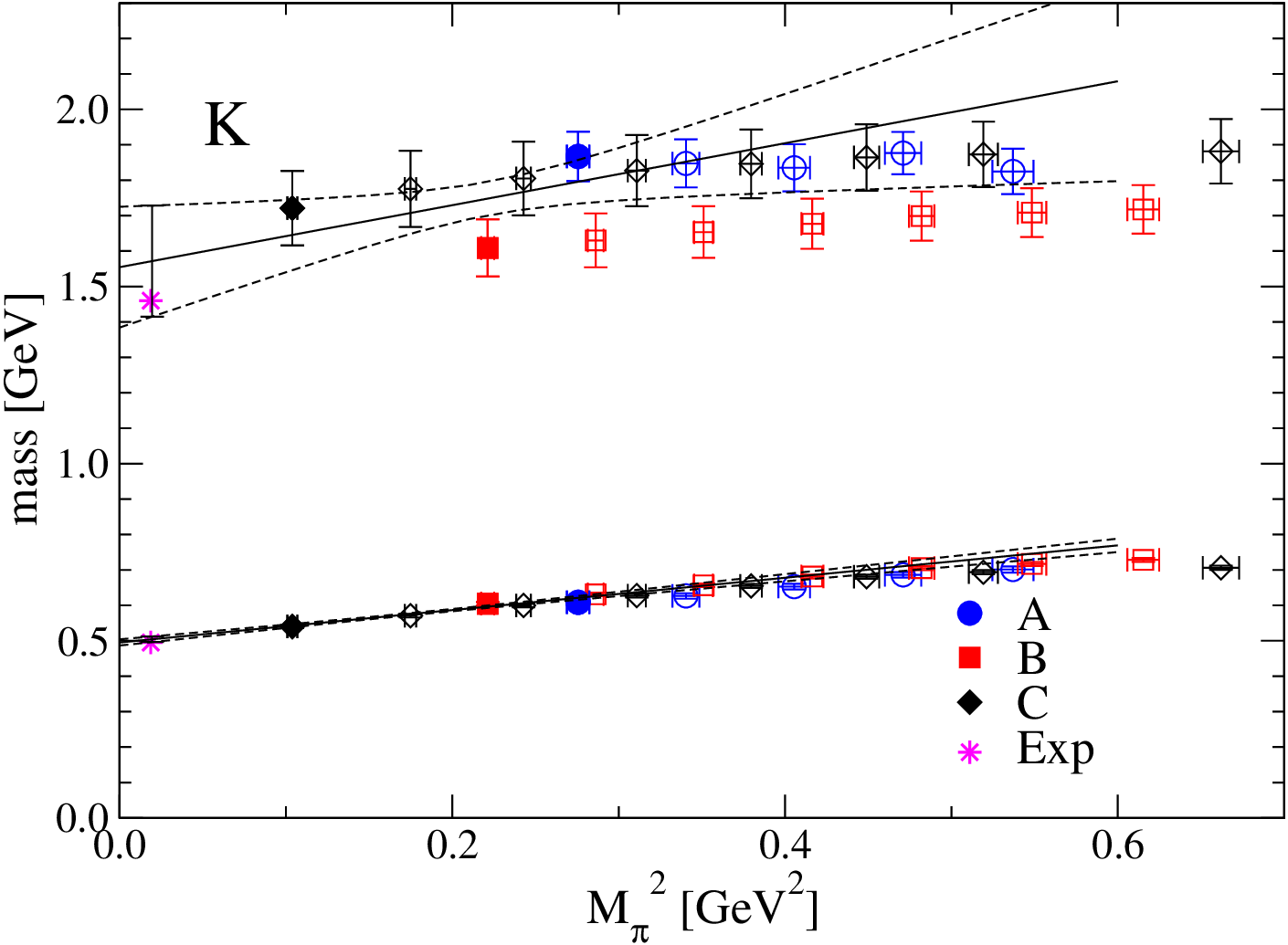}
\caption{Mass plot for the 0$^{-}$ channel ($K$), ground state and first
excitation. Ground state measured using interpolator (4), excitation measured
using (4,8,11,14,17) in A, (4,7,11,14,17) in B and C. Note that the strange
meson interpolators are not symmetrized (see Appendix \ref{Interpolators}). The
$P$ wave scattering state $\pi K^*$ is assumed to be suppressed and therefore
not indicated, since at the used simulation parameters its energy is comparable
to the one of the first excitation.} 
\label{0-}
\end{figure}

In the strange meson channel 0$^{-}$ we find a very accurate determination of
the $K$ ground state, and a fairly reliable result of the first excitation which
is compatible with experiment (see Fig.~\ref{0-}). Comparing to the pion
channel, we find that the signal improves in case of a heavier valence quark, as
expected.

Comparing to the corresponding quenched simulations \cite{Burch:2006dg},
we do not observe a qualitative difference.

\subsubsection{The 0$^{+}$ channel: $K^*_0$}

\begin{figure}[tbp]
\subfigure{\includegraphics[width=85mm,clip]{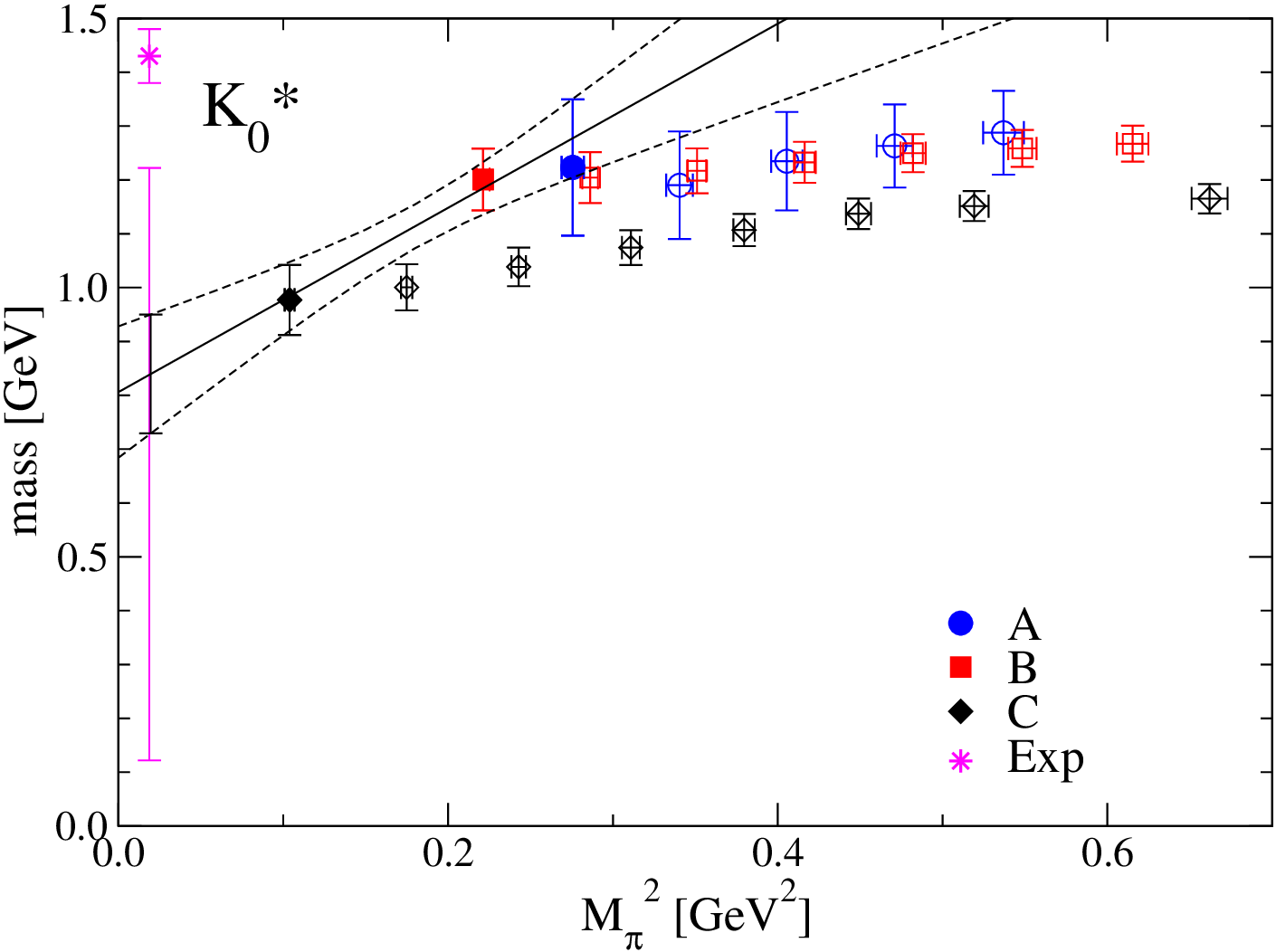}}
\subfigure{\includegraphics[width=85mm,clip]{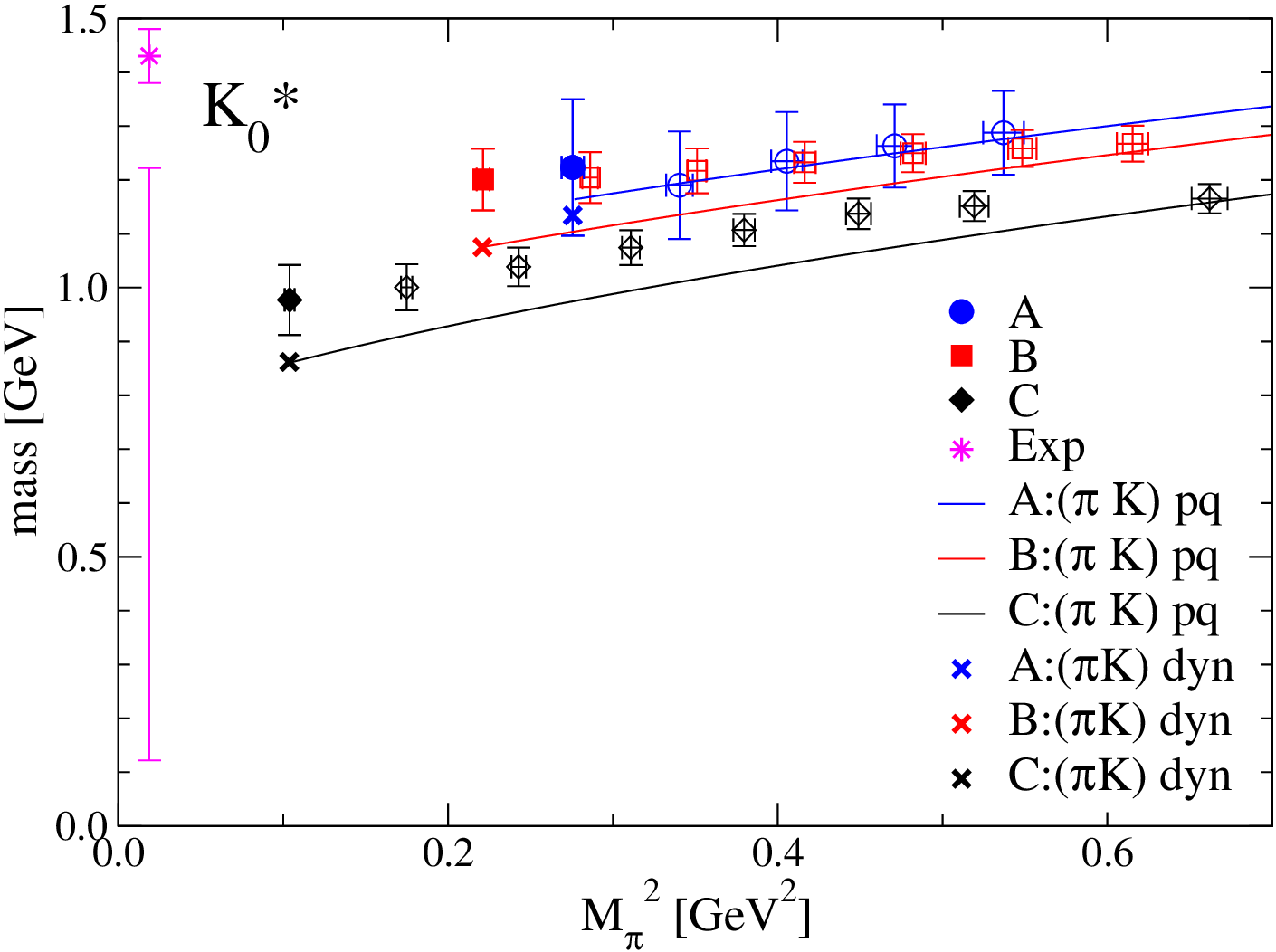}}
\caption{Mass plot for the 0$^{+}$ channel, measured with
interpolator 13.  Since $K_0^*$(800) is a very broad
resonance, thus we show its experimental width only. Top:
Ground state and chiral extrapolation according to a 1-particle interpretation
($K^*_0$) are shown.  Bottom:  The crosses represent an estimate of the
dynamical $S$ wave 2-particle state $\pi K$. The blue, red and black curve
(online version) show a prediction of the corresponding partially quenched
(``pq'') state $\pi K$ for $m_{val}\gg m_{sea}$ in ensemble A, B and C,
respectively. Since there are no strong effects from partial quenching in this
channel, the partially quenched prediction almost hits the dynamical one. Note
that in the partially quenched case only the light valence quark varies and the
strange valence quark mass is held fixed. Thus, considering the scattering
state, the partially quenched pion mass varies, while the $K$ mass is constant
along the partial quenching. The results do not allow for a clear statement
about the particle content of the state.}
\label{0+}
\end{figure}

The strange scalar channel is peculiar similar to the light scalar channel. 
Furthermore, the lowest experimental state in this channel, $K_0^*$(800) or 
$\kappa$, is not established and a very broad resonance with a width of more
than 80\% of its mass. It is thus not clear whether this state is expected to be
observed in our simulation. The lowest established resonance listed by the
Particle Data Group is $K_0^*$(1430). At the parameters we use, an appearance of
a low-lying scattering state $\pi K$ is possible. The lowest momentum
two-particle state (assuming free particles) is indicated in the figure,
together with the result and the experimental $K^*_0$(800) and $K^*_0$(1430)
(see Fig.~\ref{0+}).  The result is compatible with both, the resonance
$K^*_0$(800) and the scattering state.  Hence we cannot definitely exclude
either of the possibilities.  Comparison to the results of the light scalar
channel $0^{++}$ ($a_0$) may suggest that the scattering state contributes at
least at partially quenched points. Higher statistics and another volume would
be desirable in order for a more clear distinction.  If we use, e.g., 
interpolators (12,13), the eigenvectors look very similar for different valence
light quark masses, which means that the state we extract remains roughly the
same over the range of partial quenching we investigate. This can be interpreted
as an indication that the level crossing of a scattering state and the
resonance, if it exists, is not located in the mass range of our investigation. 

\subsubsection{The 1$^{-}$ channel: $K^*$}
\begin{figure}[tbp]
\includegraphics[width=85mm,clip]{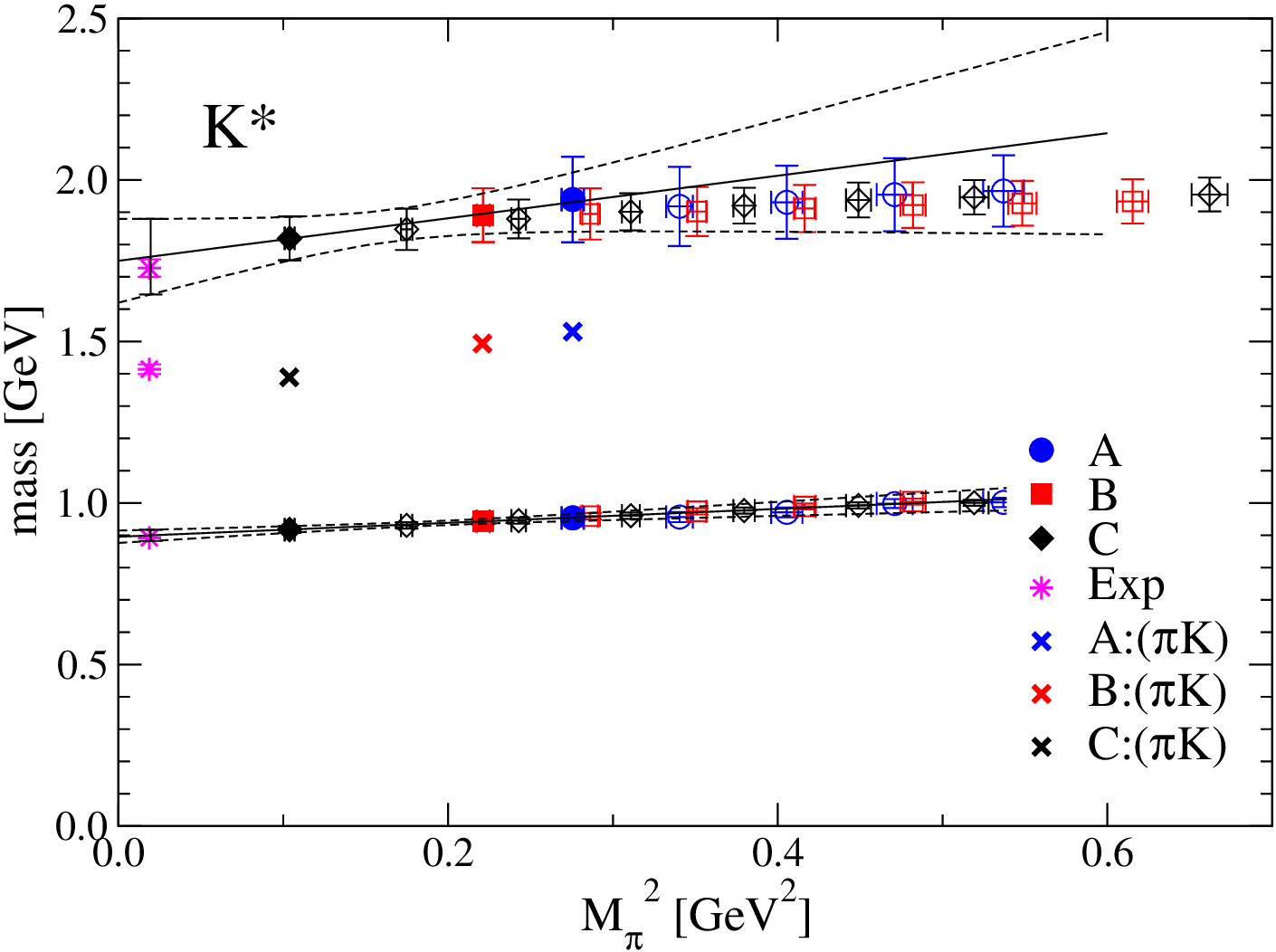}
\caption{Mass plot for the 1$^{-}$ channel ($K^*$), ground state and excitation.
Ground state measured using interpolators (1,5), excitation measured using
(1,5,17,18,19). The energy level of the $P$ wave scattering state $\pi K$ is
well separated from any observed state in all three ensembles.}
\label{1-}
\end{figure}

We obtain an accurate result of the ground state $K^*$(892) and a signal of an
excited state (see Fig.~\ref{1-}). Interestingly, the excitation fits the
$K^*$(1680) way better than the $K^*$(1410).  Compared to the light meson
sector, mixing of $J^{P+}$ and $J^{P-}$ interpolators can appear in the strange meson
sector, which is expected to shift the energy levels. This effect is due to the
missing isospin symmetry when the involved quarks are not mass-degenerate and
thus it grows when increasing the mass difference between the light and the
strange quark. However, $1^{-+}$ interpolators are exotic, hence it is not clear how
much they contribute to the final state. It is possible that
our interpolators do not show overlap with the $K^*$(1410), because we restrict
ourselves to the non-symmetrized interpolators corresponding to the light meson
channel $1^{--}$.  Another possible reason would be a tetraquark dominance of
the $K^*$(1410) (which is generally not expected) or just too small statistics.
However, quenched results using bilinear quark sources have been compatible with
the $K^*$(1410) \cite{Burch:2006dg}. 

We are planning to investigate this channel
more thoroughly using a larger basis in the variational method and a larger set
of ensembles in the future. We do not observe the $P$ wave scattering state
$\pi K$ in our simulation.

\subsubsection{The 1$^{+}$ channel: $K_1$}
\begin{figure}[tbp]
\includegraphics[width=85mm,clip]{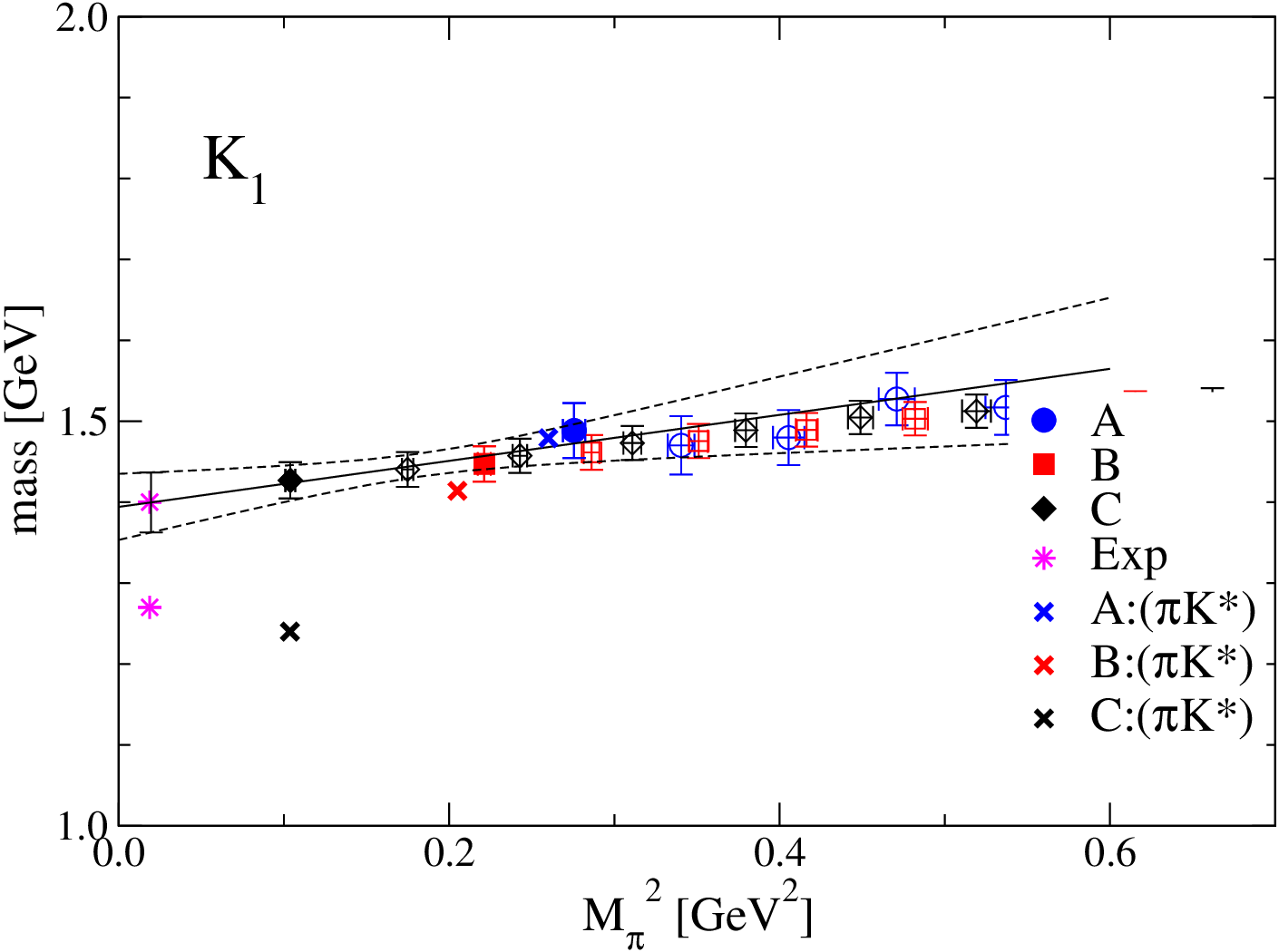}
\caption{Mass plot for the 1$^{+}$ channel ($K_1$). Measured using interpolators
(13,14,15). The energy level of the $S$ wave scattering state $\pi K^*$ is
close to the ground states of A and B, but much lower than the ground state of
C.}
\label{1+}
\end{figure}

Similar to the strange 1$^{-}$ channel, we seem to miss the ground state
$K_1$(1270) in the strange 1$^{+}$ channel (see Fig.~\ref{1+}). Our result is in
good agreement with the experimental first excitation $K_1$(1400).  In this
channel the situation is clearer, due to missing isospin symmetry, one expects a
mixing from the 1$^{++}$ and 1$^{+-}$ interpolators. In contrast to the latter case,
these are both non-exotic. So far, we cannot say how strong the mixing is in our
simulations, however, it was found that the mixing is weak down to pion masses
of 400 MeV \cite{Dudek:2010wm}. Again, a more thorough investigation is hoped to
shed some light on this issue. The results of ensemble C suggest that we do not
observe the $S$ wave scattering state $\pi K^*$.

\subsubsection{The 2$^{-}$ channel: $K_2$}

\begin{figure*}[tbp]
\subfigure{\includegraphics[width=85mm,clip]{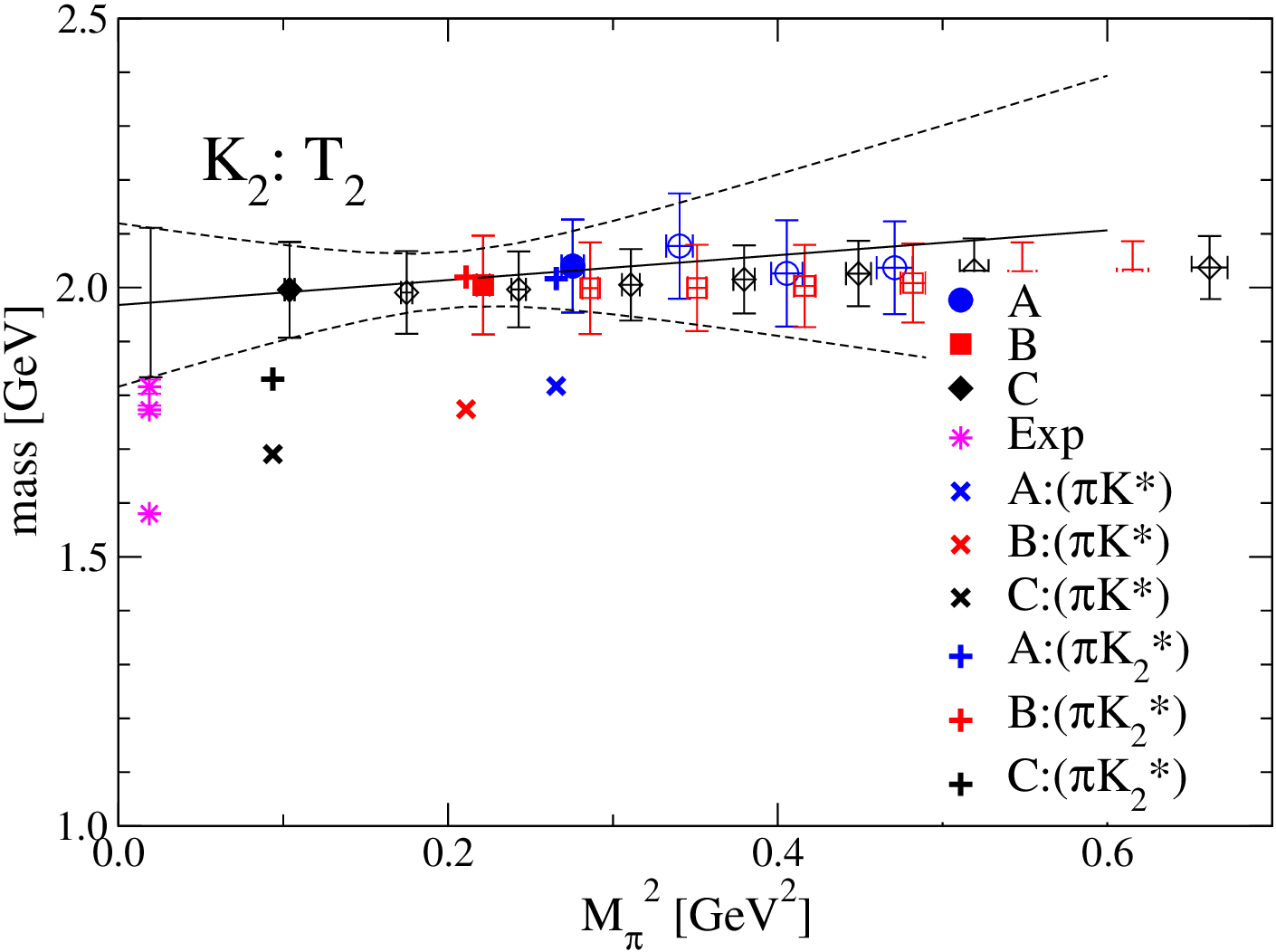}}
\subfigure{\includegraphics[width=85mm,clip]{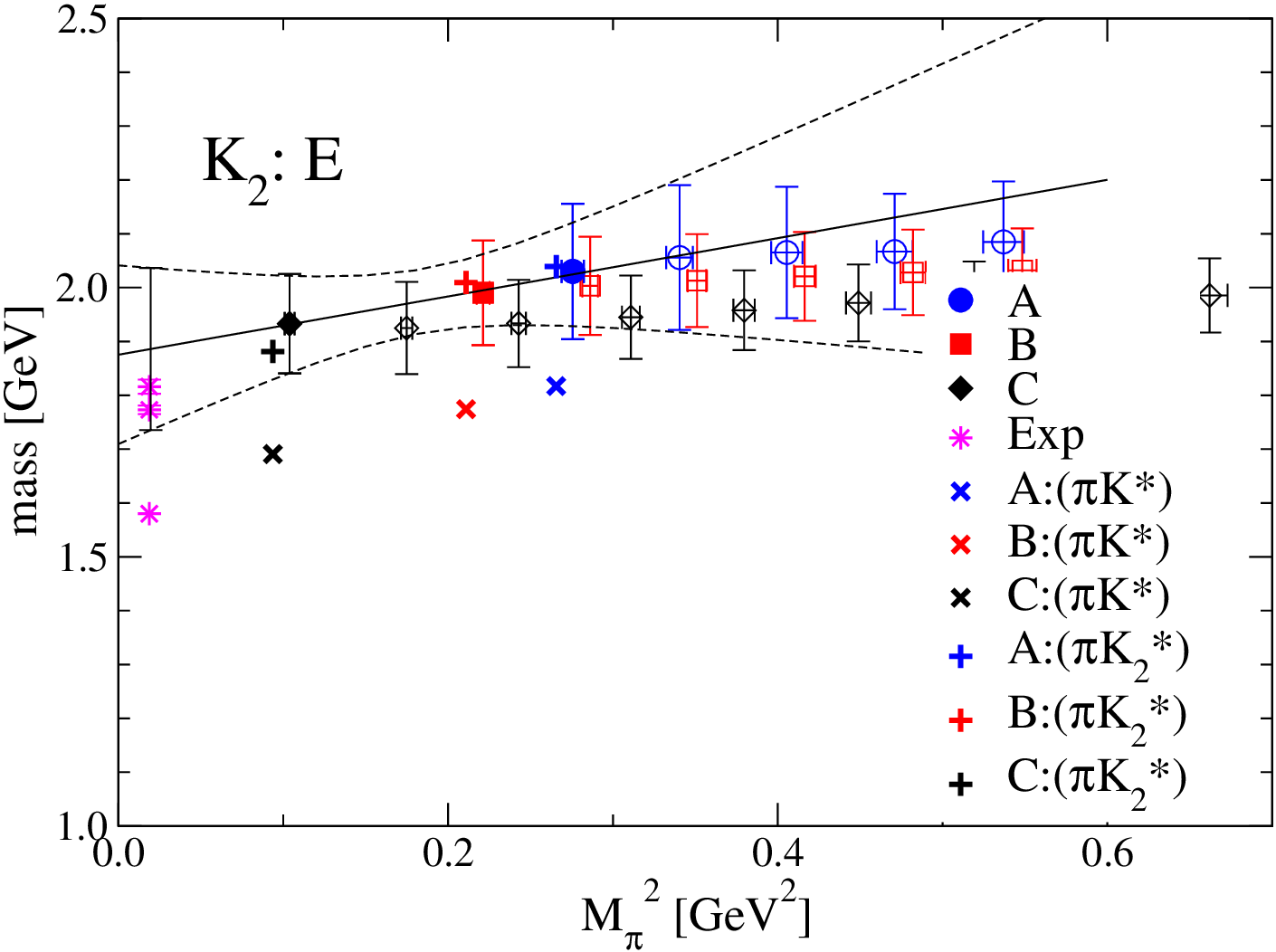}}
\caption{Mass plot for the 2$^{-}$ channel ($K_2$), ground state. Lhs: Irrep
T$_2$, measured using interpolators (1,3). Rhs: Irrep E, measured using
interpolators (1,3). The experimental $K_2$(1580) is not completely confirmed so
far. The energy levels of the scattering states $\pi K_2^*$ ($S$ wave,
dominant in the experiment) and $\pi K^*$ 
($P$ wave) are shown.}
\label{2-}
\end{figure*}

As already discussed, there are two different orthogonal lattice irreducible
representations for spin 2 channels. Again, we analyze the irreps independently
and discuss the results. Similar to the corresponding light meson channels, we
encounter large error bars in the strange 2$^{-}$ channel  (see Fig.~\ref{2-}). 
The results are compatible with both the experimental $K_2$(1770) and the
$K_2$(1820) in both representations, albeit with an error of slightly more than
one $\sigma$ in case of T$_2$. The $K_2$(1580) is not completely confirmed
experimentally so far. The lowest possible scattering state ($\pi K^*$ in $P$ wave) 
is not observed in our simulation. Like in the spin 1 channels, we hope that enlarging
the basis will improve the signal.

\subsubsection{The 2$^{+}$ channel: $K^*_2$}
 \begin{figure}[tbp]
\includegraphics[width=85mm,clip]{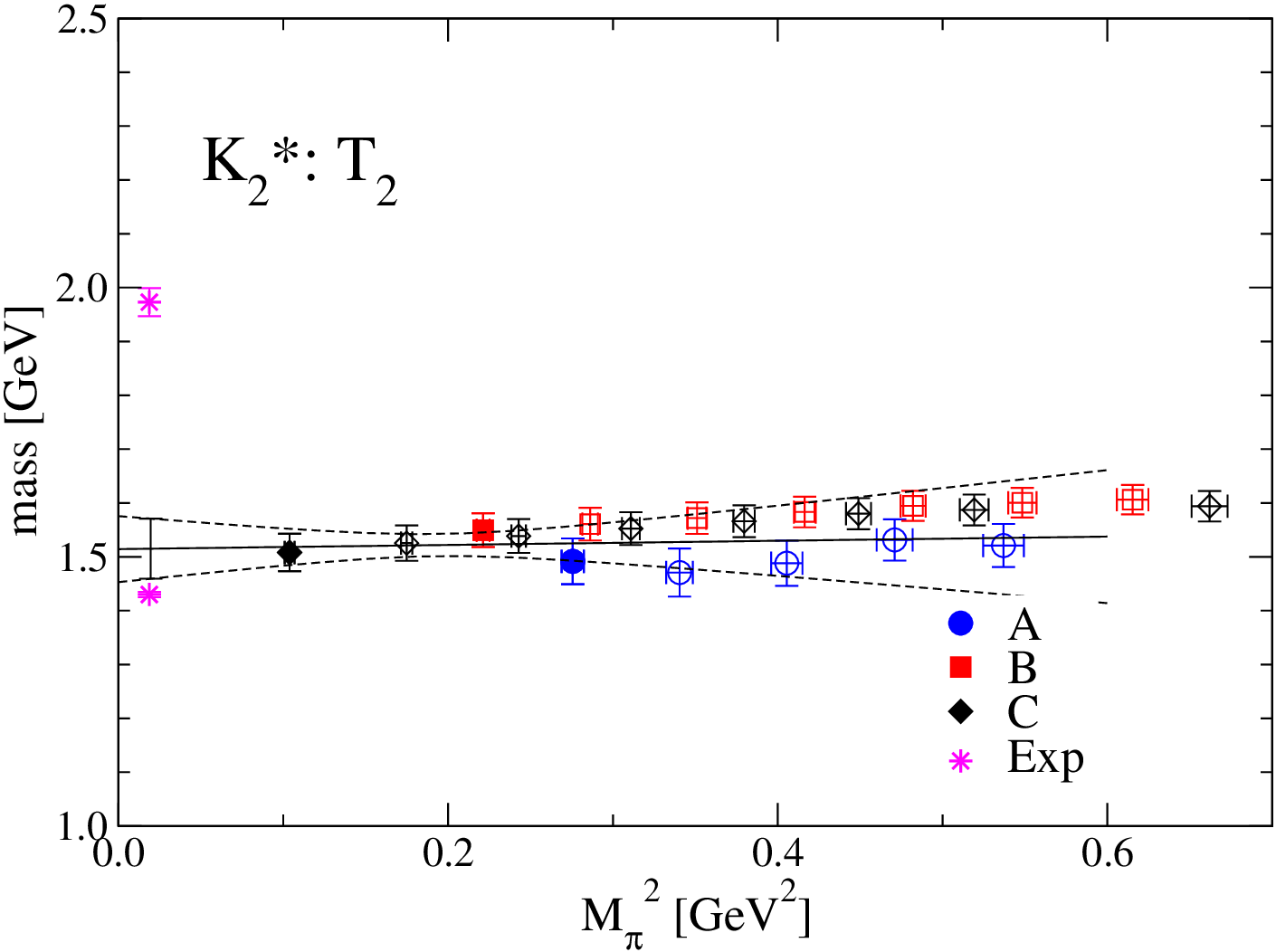}
\caption{Mass plot for the 2$^{+}$ channel ($K^*_2$) in irrep T$_2$, ground
state. Measured using interpolators (1,7) in A, (3,5) in B, (2,4,5,6) in C.}
\label{2+}
\end{figure}

\begin{figure}[tbp]
\includegraphics[width=85mm,clip]{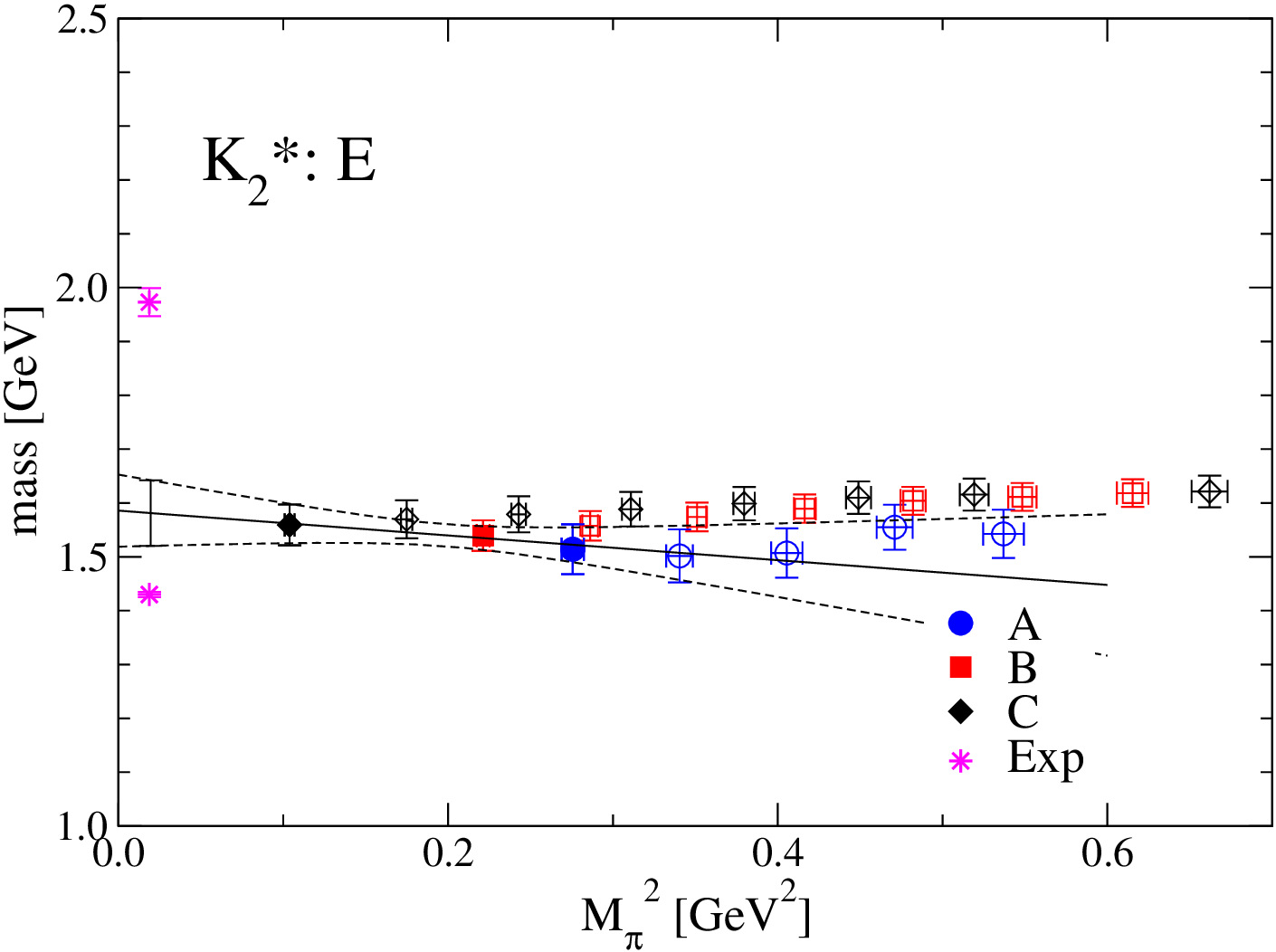}
\caption{Mass plot for the 2$^{+}$ channel ($K^*_2$) in irrep E, ground state.
Measured using interpolators (2,3,9) in A and C and (3,4,9) in B.}
\label{2+E}
\end{figure}

In the 2$^{+}$ channel the signal is somewhat better than in the  2$^{-}$
channel. Here we find the result of irrep T$_2$ closer to the experimental
$K^*_2$(1430) than the one of irrep E (see Figs. \ref{2+} and \ref{2+E}).
However, the negative slope of the chiral extrapolation in irrep E is clearly an
artifact of too small statistics (remember that only the three dynamical points
enter the linear fit). A more complete basis and additional ensembles are
expected to improve the calculation in this channel.

\subsubsection{The 1$^{--}$ channel: $\phi$}

The decay channels of $\phi$ suggest that it is an almost pure $s\bar{s}$-state.
Therefore, in the partially quenched analysis we can read off our results for
$\phi$ from the partially quenched data in the 1$^{--}$ ($\rho$) channel without
any chiral extrapolation (see Fig.~\ref{phi}). As already discussed, the ground
state reproduces $\phi$(1020) nicely, which can be seen as an affirmative
cross-check for our method to set the strange quark mass parameter. Reading off
the excited state we find some deviation from the experimental value,
explicitly shown in Fig.~\ref{collection_masses} in the summary. The origin of
this discrepancy may be due to neglected disconnected diagrams or just lie in
the weakness of the corresponding effective mass plateau. Difficulties with
excitations are found also in related channels. E.g., in the 1$^{-}$ channel we
miss the first excitation of $K{^*}$ and in the light 1$^{--}$ channel the
excitation of $\rho$ is quite noisy.

\subsection{Baryons with strange quarks}

\subsubsection{$\Sigma$ positive parity}

\begin{figure}[tbp]
\includegraphics[width=85mm,clip]{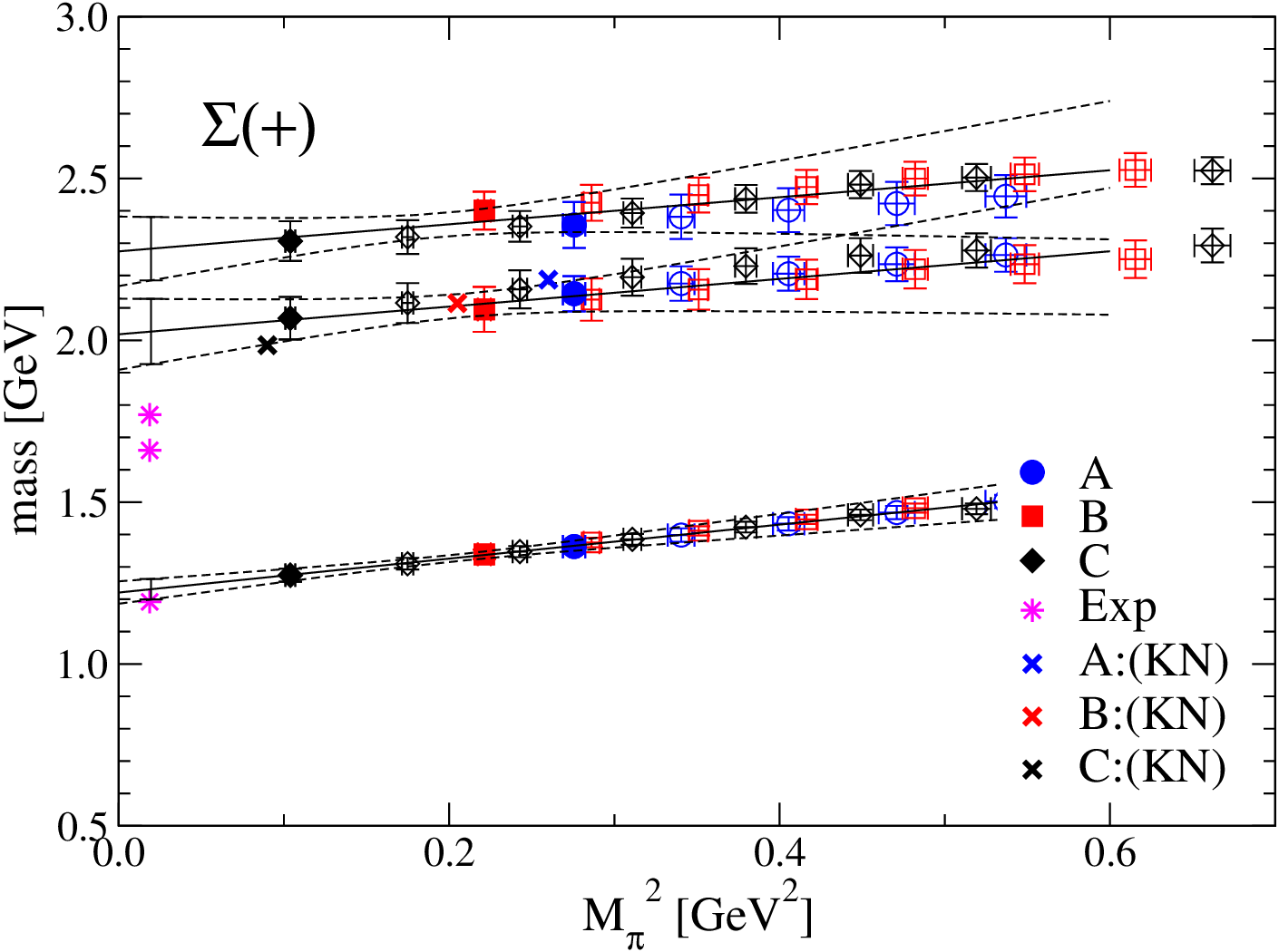}
\caption{Mass plot for the $\Sigma$ positive parity channel, ground state and two
excitations. Ground state measured using interpolators (1,6,17,20), first
excitation measured using (1,6,20), second excitation measured using
(3,4,8,9,13). The energy level of the $P$ wave scattering state $N  K$ is very
close to the first excitation in all three ensembles. For better identification,
we display the scattering states slightly shifted to the left.}
\label{sigma_pos}
\end{figure}

The interpolators of $\Sigma$ and $\Xi$ have the same Dirac structure as the
nucleon interpolator, they just differ by the flavor content. Hence we use
similar sets of interpolators in the variational method. We obtain a ground
state and two excited states in the $\Sigma$ positive parity channel (see
Fig.~\ref{sigma_pos}). The ground state result is in satisfactory agreement with
experimental data. This is another affirmative cross-check of the setting of the
strange quark mass parameter. The excitations are too high, which may be due to
finite volume effects, as we have discussed already. Analogously to previous
channels, we do not believe to see contributions from the $P$ wave
scattering state $K N$.

\subsubsection{$\Sigma$ negative parity}

\begin{figure}[tbp]
\includegraphics[width=85mm,clip]{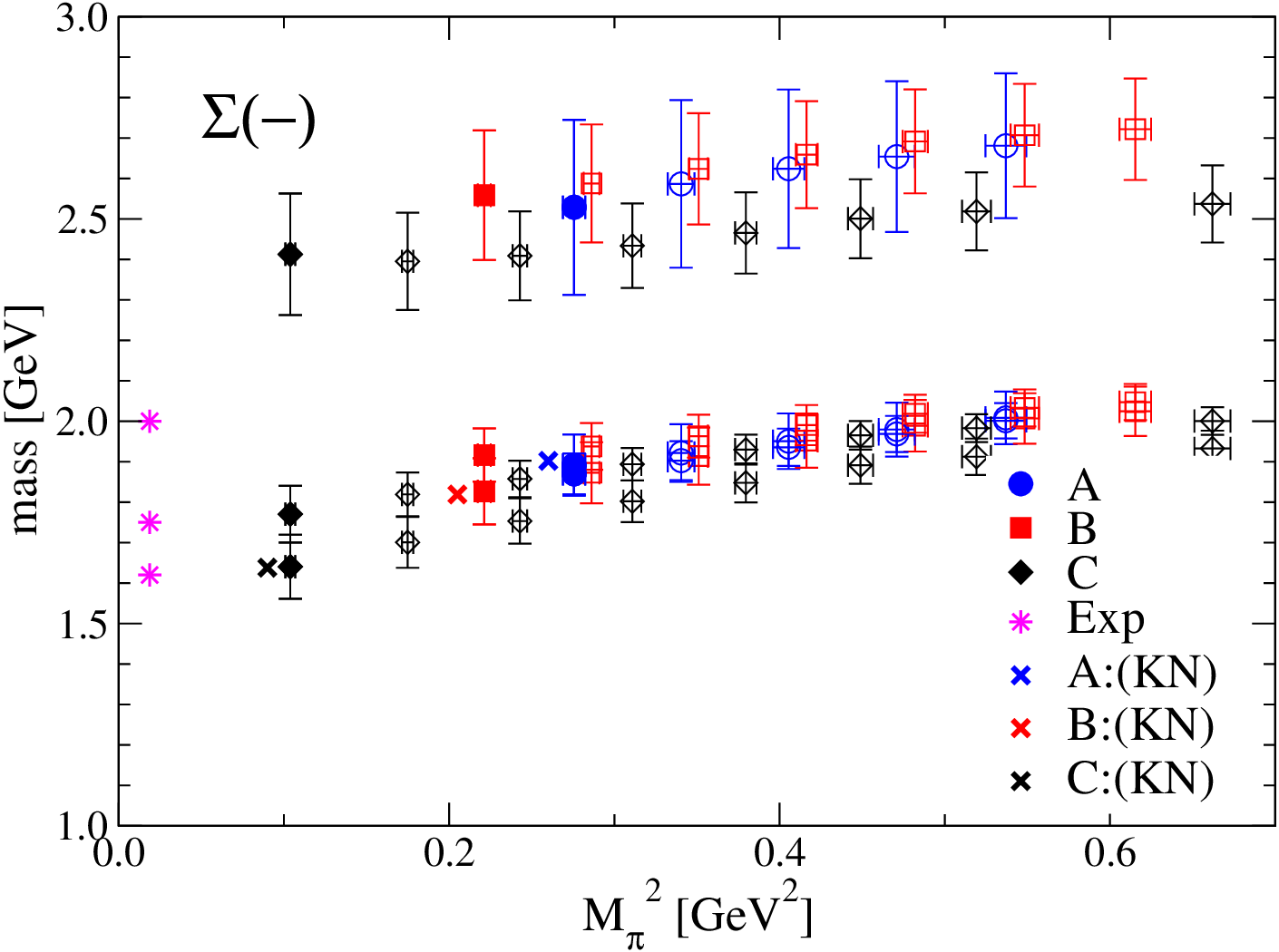}
\caption{Mass plot for the $\Sigma$ negative parity channel.
Measured using interpolators (1,9,10,12). The energy level of the $S$ wave
scattering state $K N$ is very close to the ground state in all three ensembles
(also to the first excitation in case of ensemble A). For better identification,
we display the scattering states slightly shifted to the left. The results would
be compatible with an interpretation in terms of a level crossing of the 1- and
2-particle states. However, analogously to the nucleon negative parity channel, the
eigenvectors contradict this picture. For clarity, the chiral
extrapolation is omitted in the figure.}
\label{sigma_neg}
\end{figure}

We find a ground state and two excitations in the $\Sigma$ negative parity
channel (see Fig.~\ref{sigma_neg}). Similar to the nucleon negative parity
channel, the first excitation is very close to the ground state, which is in
qualitative agreement with the experimental $\Sigma$(1620) and $\Sigma$(1750).
The ground state mass agrees with experiment within error bars, though being a
bit low, as in all other our negative parity baryon channels.  The second
excitation could be compatible with the observed $\Sigma$(2000), after
correction of finite volume effects. $\Sigma$(1620), $\Sigma$(1750) and 
$\Sigma$(2000) are classified with 2, 3 and 1 stars, respectively, by the
Particle Data Group \cite{PDG08}. We can confirm the two lower ones
qualitatively and the existence of an excitation in the vicinity of 2000 MeV.

In nature, the $S$ wave scattering state $K N$, which is lighter than the 1-particle
ground state, appears in the $\Sigma$ negative parity channel. We analyze the
eigenvectors analogously to the nucleon negative parity channel. Again, we find
that in all three ensembles the ground state is dominated by the second Dirac
structure ($\chi_2$), while the first excitation is an almost pure $\chi_1$
state (see Table \ref{tab:baryoninterpolators1}). The partially quenched points
show the same behavior. One may conclude that no level crossing of the lowest
two states is observed for pion masses in the range of 320 to 520 MeV. As in the
nucleon negative parity channel, we find that the low results for masses at small
pion masses may be explained by the presence of a scattering state, but the
eigenvectors do not confirm this picture. Since the argumentation based on the
eigenvectors seems to be more reliable, we believe to see an almost pure
1-particle state and quote the corresponding chiral extrapolation in the summary.

\subsubsection{$\Xi$ positive parity}

\begin{figure}[tbp]
\includegraphics[width=85mm,clip]{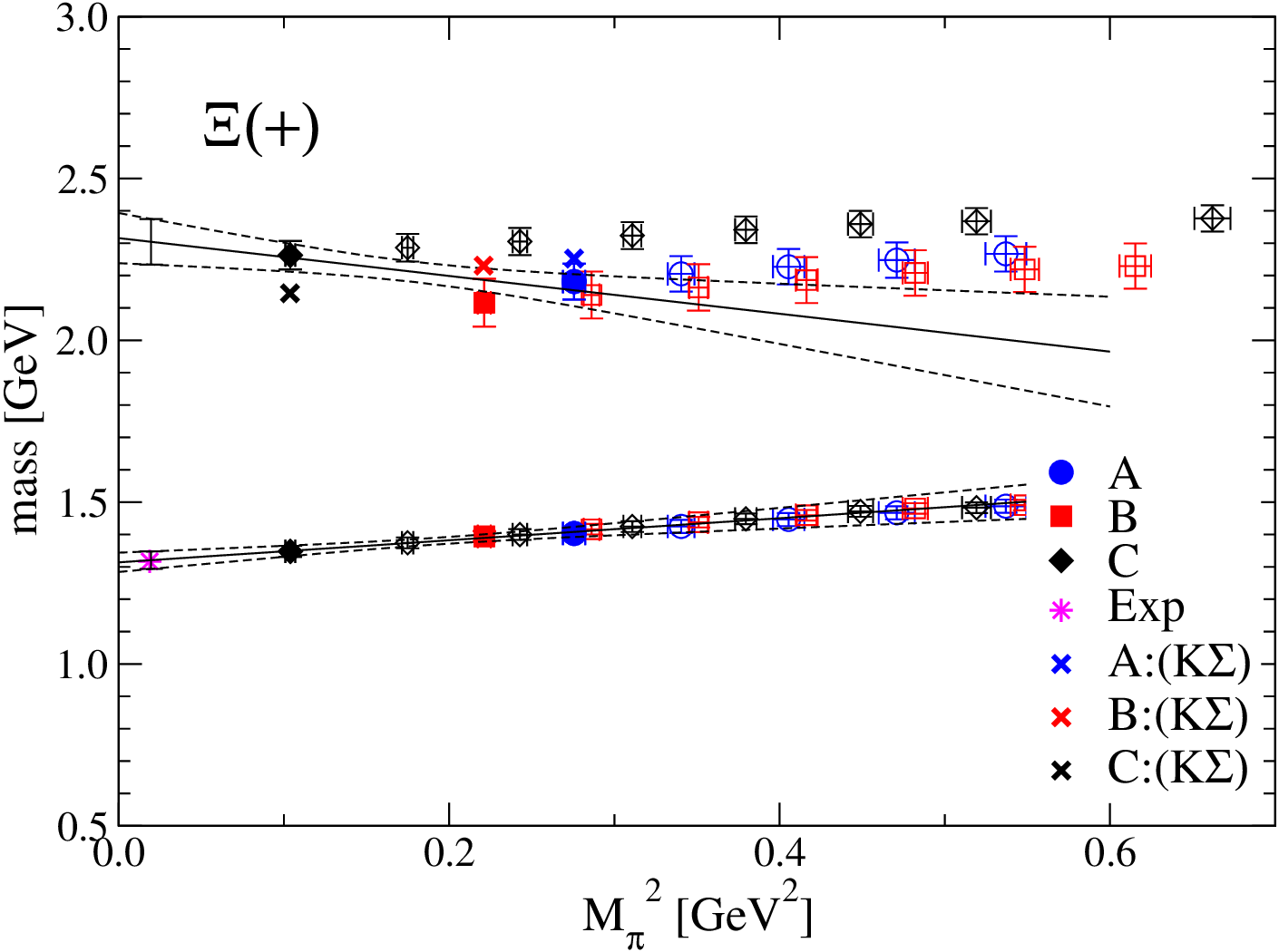}
\caption{Mass plot for the $\Xi$ positive parity channel.
Measured using interpolators (3,4,9,13,16).
The energy level of the $P$ wave scattering state $K \Sigma$ is very close to the first excitation in all three ensembles.}
\label{xi_pos}
\end{figure}

In the $\Xi$ positive parity channel we obtain a ground state with rather small
error which is in very good agreement with the experimental $\Xi$ ground state
(see Fig.~\ref{xi_pos}). We also get a prediction for a first excited state with
comparatively small error bar in the range of 2200 to 2400 MeV, which after
correction of finite volume effects  could be compatible with the $\Xi$(2120) (1
stars) or the $\Xi$(2250) (2 stars), where both states are listed by the
particle data group stating neither spin nor parity. Analogous to previous
cases, we do not believe to see measurable contributions from the $P$ wave
scattering state $K \Sigma$.

\subsubsection{$\Xi$ negative parity}

\begin{figure}[tbp]
\includegraphics[width=85mm,clip]{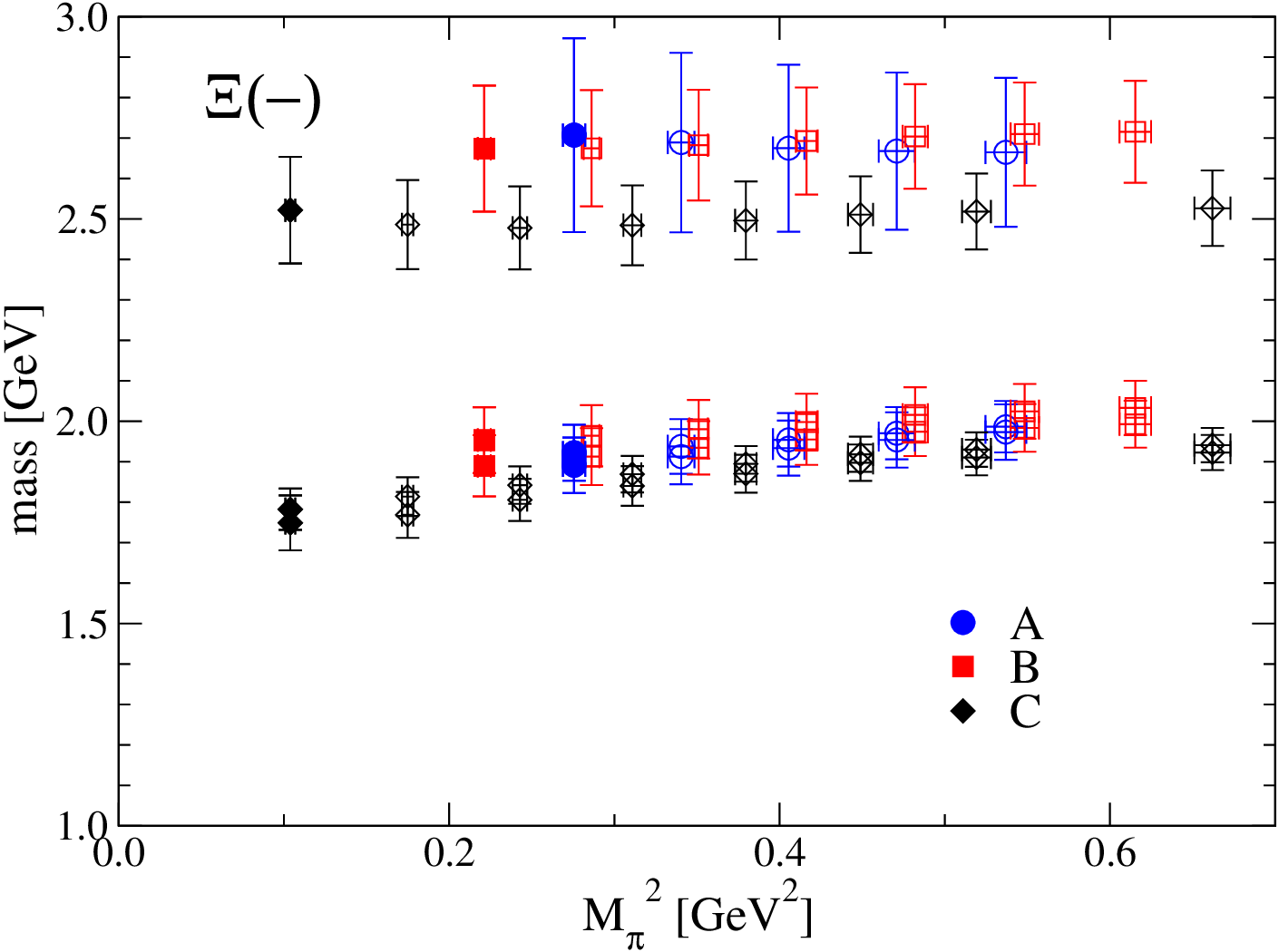}
\caption{Mass plot for the $\Xi$ negative parity channel. Measured using
interpolators (4,9,10,14,23) in A and B, (3,7,9,10,14) in C. For clarity, the
chiral extrapolation is omitted in the figure.}
\label{xi_neg}
\end{figure}

In the negative parity channel there is not even one established state listed by
the Particle Data Group. We observe a ground state and a close first excitation
in the range of 1550 to 1800 MeV, both with relatively small error (similar to
the quenched results \cite{Burch:2006cc}).  These states may match the
$\Xi$(1620) (1 stars) and the $\Xi$(1690) (3 stars). There is a signal of a
second excitation around 2400 MeV as well, albeit with rather large statistical 
uncertainty and suffering from finite volume effects.  We found it crucial to
include all three baryon Dirac structures in the variational method in order to
obtain both excitations.

Compared to quenched results of \cite{Burch:2006cc}, we extracted an 
additional excited state in the present work.

\subsubsection{$\Omega$ positive parity}

As already discussed, we use our result for $\Omega$(1672) to set the strange
quark mass parameter (see Fig.~\ref{omega}). Similar to the case of $\phi$, we
read off results for $\Omega$ from partially quenched data for  $\Delta$,
without any chiral extrapolation. In addition to $\Omega$(1672), we obtain a
first excitation in the positive parity channel (explicitely shown in
Fig.~\ref{collection_masses} in the summary). Since there is no extrapolation
involved, its statistical error is rather small, however, the true mass value
may be somewhat smaller due to finite volume effects. The Particle Data Group
lists $\Omega$(2250) (3 stars), $\Omega$(2380) (2 stars) and $\Omega$(2470) (2
stars), stating neither spin nor parity. Taking into account finite volume
effects, the result for the excitation may turn out to be compatible with
$\Omega$(2250).

\subsubsection{$\Omega$ negative parity}

\begin{figure}[tbp]
\includegraphics[width=80mm,clip]{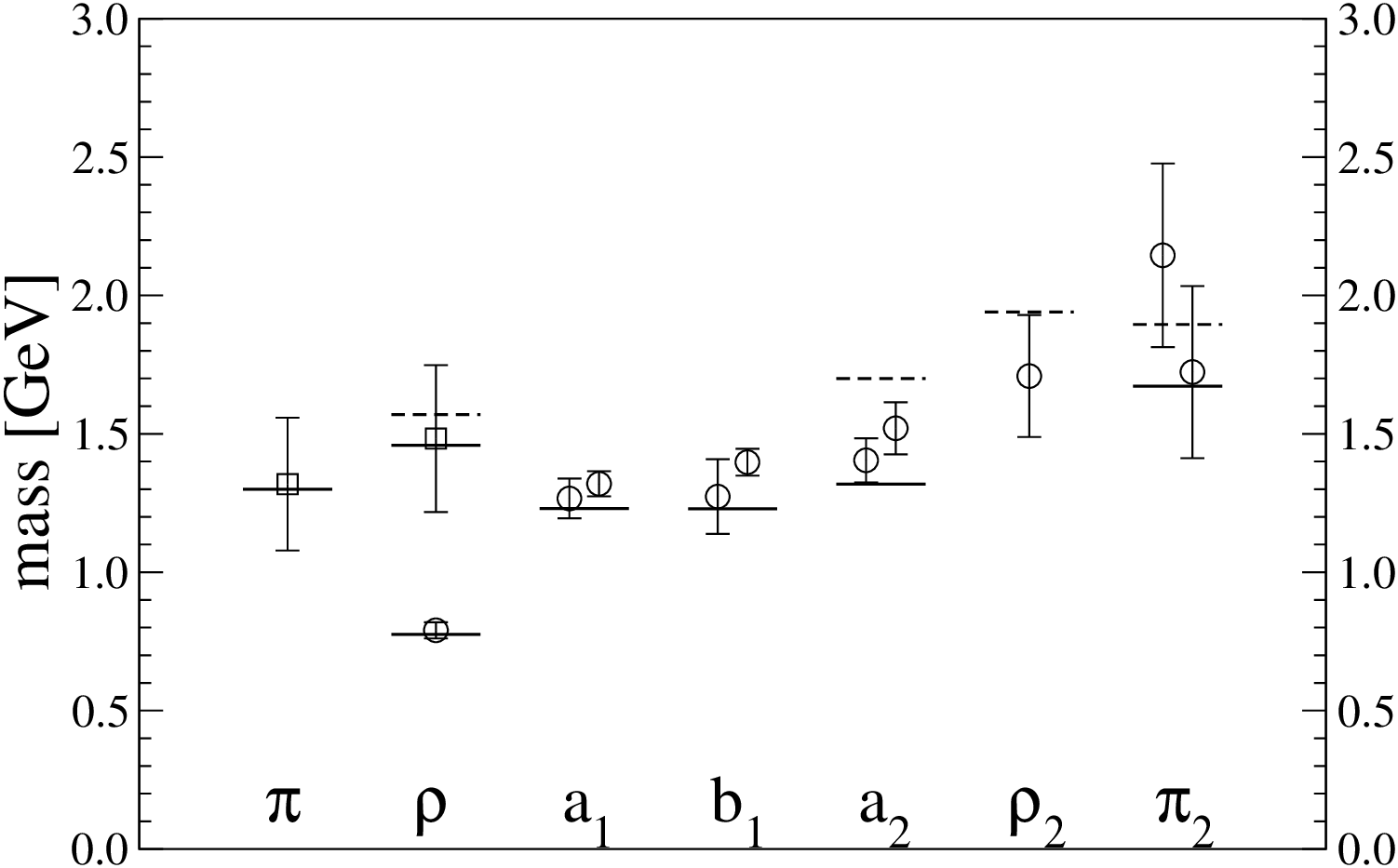}
\includegraphics[width=80mm,clip]{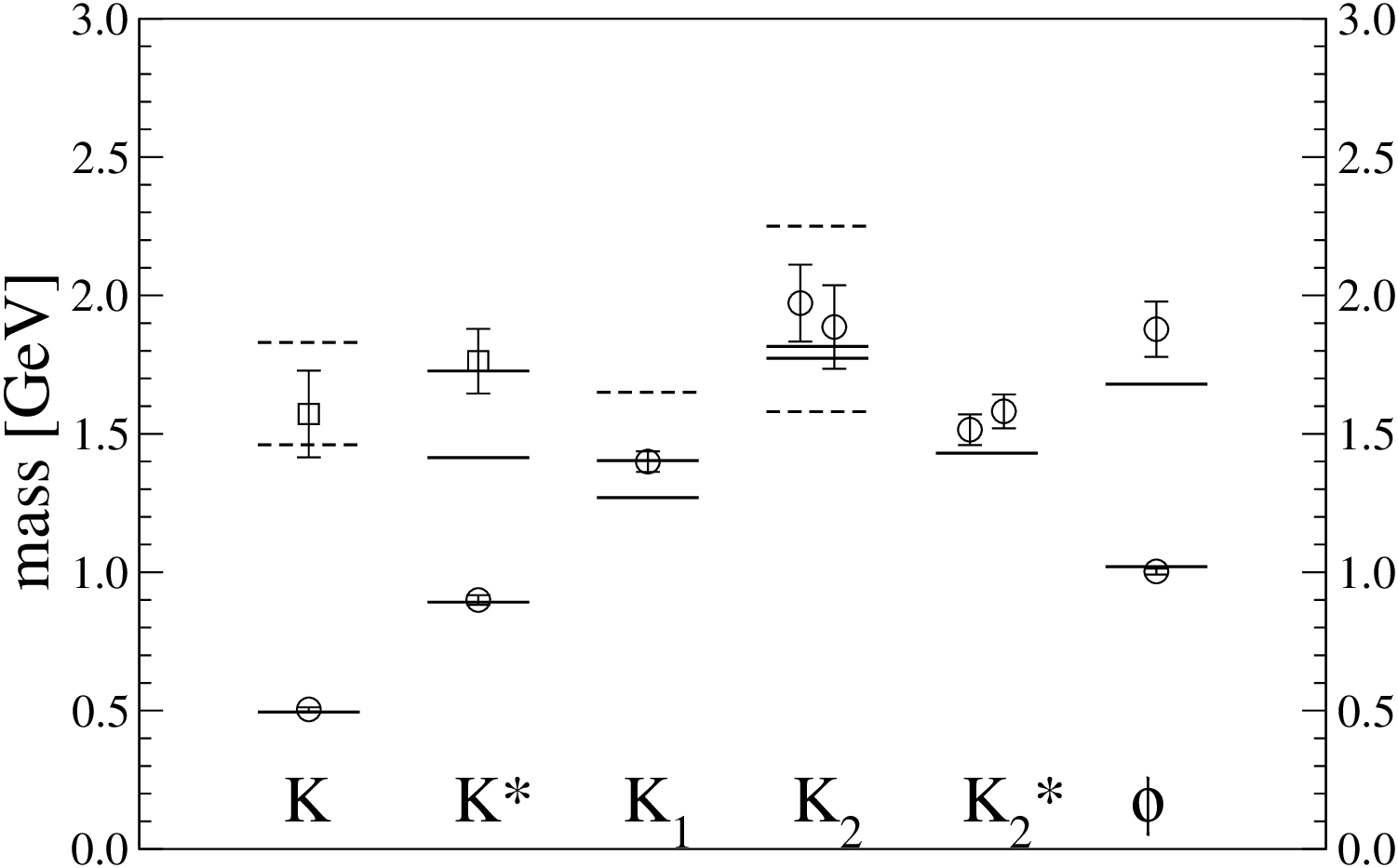}
\includegraphics[width=80mm,clip]{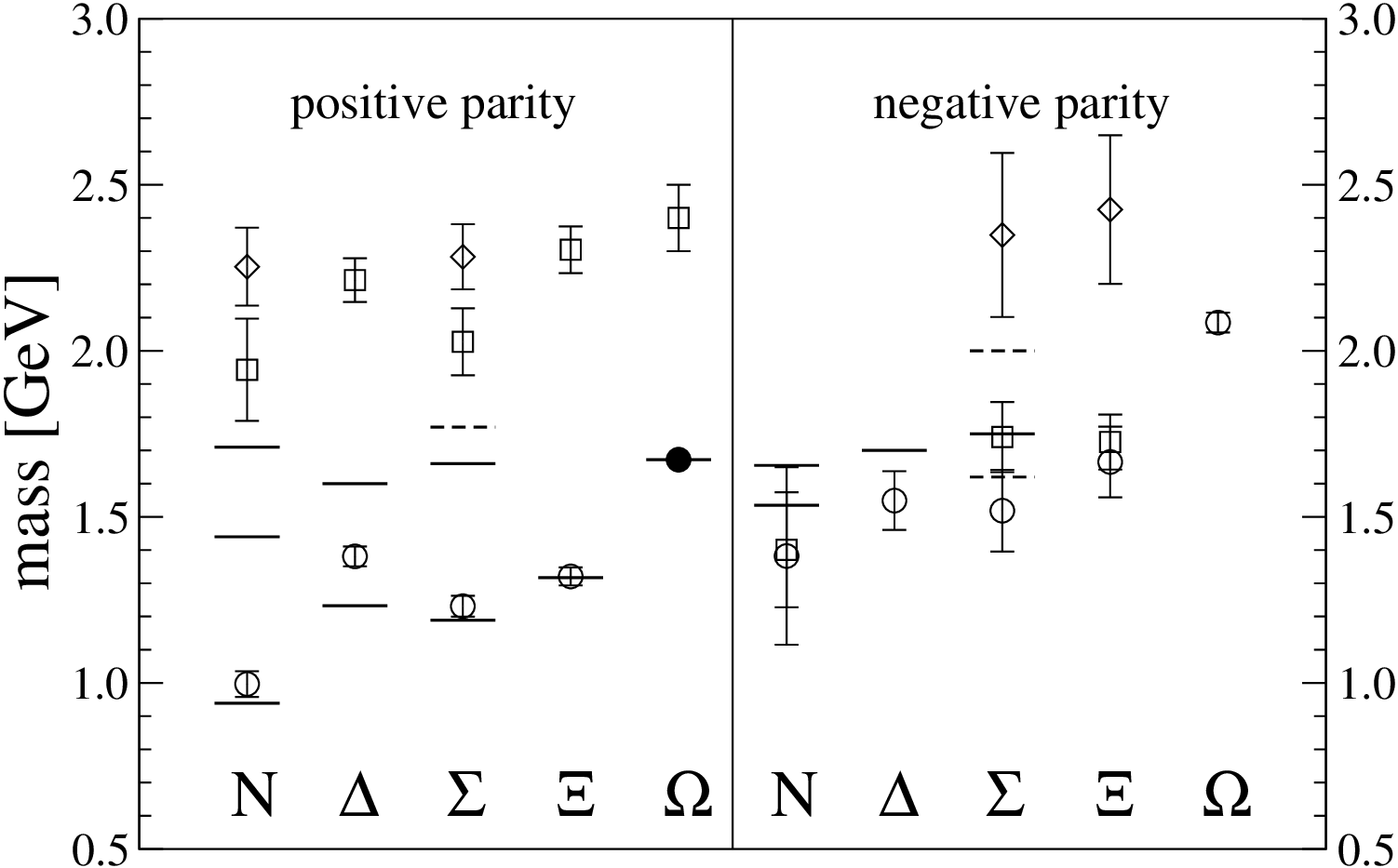}
\caption{Collection of the results for light meson masses (top), strange meson
masses (middle) and baryon masses (bottom).  Indicated errors are of purely
statistical nature.  Results are obtained by chiral extrapolation of dynamical
data linear in the pion mass squared (except for $\phi$ and $\Omega$ hadrons).
Experimental values listed by the Particle Data Group \cite{PDG08} are denoted
by horizontal lines, the ones needing confirmation by dashed lines.  Circles indicate
ground states, squares first excitations and diamonds second excitations.
Results shown aside each other are obtained using different sets of
interpolators aiming for the same state. In the spin 2 channels, results from
T$_2$ interpolators are shown on the left, results from E on the right. In the $\rho_2$ channel
we have results only from E interpolators. Strange quarks are implemented by
partial quenching, the strange quark mass parameter is set by identification of
a partially quenched $\Delta$ with $\Omega$(1670).  Excited baryons which come
out too high seem to systematically suffer from finite volume effects. Higher
statistics will be needed to establish if the results for the ground states in
the negative parity baryon channels are compatible with experiment or whether
the results are systematically too low. Results for scalar mesons are omitted in
this figure, since the particle content of the corresponding observed states is
too uncertain.}
\label{collection_masses}
\end{figure}

The Particle Data Group does not list any established state in the $\Omega$
negative parity channel.  We observe a ground state in the range of 2050 to 2100
MeV (compare Fig.~\ref{delta_neg}, explicitely shown in
Fig.~\ref{collection_masses} in the summary).  Such a state has also been
observed in the quenched study \cite{Burch:2006cc}. It does not fit to any state
listed by the Particle Data Group.


\section{Conclusion}
\label{Conclusion}

We have presented results of hadron spectroscopy using the Chirally Improved
Dirac operator on lattices of size $16^3 \times 32$ with two mass-degenerate
light sea quarks. Three ensembles with pion masses of $322(5)$, $470(4)$ and
$525(7)$ MeV and lattice spacings all close to $0.15\,$ fm have been
investigated.  This allows for a naive chiral extrapolation in the
mass-dependent scheme, but neither for a continuum nor a thermodynamic limit.
We have discussed possible systematic effects in Sect. \ref{SysEffects}. 
Systematic uncertainty due to discretization effects is not explored
explicitely  and may be non-negligible for some observables. Finite volume
effects are discussed only qualitatively, where we find some  indications in
case of excited baryons.

We have shown results for ground states and excited states for several meson
and baryon channels, including spin 2 mesons constructed by the use of
derivative sources.  The spectrum of strange hadrons was accessed by using
partially quenched strange quarks.  Possible effects from partial quenching
have been discussed briefly and they seem to play no important role  for most
of our final results.  However, including strange sea quarks in the simulation
would be desirable  in order to reduce possible sources of systematic errors.
The value of the strange quark mass parameter was set by identification of the
partially quenched positive parity $\Delta$ with the $\Omega$(1672).

The results are summarized in Fig.~\ref{collection_masses}.  Several of the
experimentally known ground states are reproduced with fairly high precision
($\rho$, $K^0$, $K^*$, $N$ and $\Sigma$ positive parity). In addition, various
radial excitations are found ($\pi$, $\rho$, $K^*$, $\Delta$ positive parity, 
$N$, $\Sigma$ and $\Xi$ both parities). 
We discussed the possible appearance of scattering states in various channels.
The coupling of our interpolators to many-particle  states seems to be weak and
such states are barely, if at all, identifiable. Only in the light scalar
channel the partially quenched data suggest a large contribution from an $S$
channel 2-particle state of pseudoscalars. However, at the dynamical point no
clear statement is possible. In the negative parity nucleon and $\Sigma$
channels, the eigenvectors do not confirm the picture of the $S$ wave 2-particle
states, either, although such an admixture cannot be completely
excluded.  A clear interpretation of the particle content of the observed
states is still missing.

Where possible, we discussed the influence of sea quarks by comparison to
quenched results obtained from the same action.  In particular the spectroscopy
in the light scalar channel seems to benefit from dynamical quarks, nevertheless
it remains a difficult channel. Also the results in the light vector channel are
in slightly better agreement with experiment when sea quarks are included.
However, in most channels we did not observe a significant difference between
quenched and dynamical simulations. We stress that in most cases comparison is
difficult since some other details of the simulations differ as well. Also we
have to emphasize that in many observables the overall effect of the sea quarks
may approximately cancel (e.g., by including their effect on the scale
setting).  In other words, different dynamics may partly show similar
phenomenology. A similar mechanism possibly works in some of the strange
hadrons, where our results are in good agreement with experiment, despite the
partial quenching approximation of the strange quark.

Several channels are expected to benefit from enlarged statistics, especially
radial and spin excitations, the pseudovector mesons and negative parity baryon
channels. In the nucleon positive parity channel we observe excitations which
are definitely higher than the expected Roper resonance. This problem may be due
to finite size effects or the lack of explicit 2-particle interpolators in the
analysis. Our data confirm the existence of some states of unclear status and
predict some states which are not listed by the Particle Data Group so far, 
albeit with large statistical errors and also non-negligible systematic errors
in some cases.

\acknowledgments
We would like to thank Christof Gattringer, Leonid Y.~Glozman and Sasa Prelovsek
for valuable discussions. The calculations have been performed on the SGI Altix
4700 of the Leibniz-Rechenzentrum Munich and on local clusters at ZID at the
University of Graz. We thank these institutions for providing	 support. M.L.~
and D.M.~have been supported by ``Fonds zur F\"orderung der Wissenschaftlichen
Forschung in \"Osterreich'' (DK W1203-N08).  D.M.~acknowledges support by
COSY-FFE Projekt 41821486 (COSY-105) and by Natural Sciences and Engineering
Research Council of Canada (NSERC) and G.P.E.~and A.S.~acknowledge support by
the DFG project SFB/TR-55.

\newpage
\begin{appendix}
\section{Tables of interpolators}
\label{Interpolators}
\subsection{Meson interpolators}
We list the used interpolators for each meson channel in Tables
\ref{interpolators0-} - \ref{interpolators2+}. In case of the light mesons,
symmetrization is performed for definite $C$-parity. Our strange meson
correlation calculation lacks cross correlation matrix elements according to
interpolators with different $C$-parity quantum numbers in the limit of
degenerate quark masses. Therefore, when analyzing strange mesons, we have to
restrict ourselves to subsets of interpolators sharing the same $J^{PC}$ quantum
numbers in the limit of degenerate quark masses.

The letters $a$ and $b$ denote light or strange quarks; $n$ ($w$) denotes a
Gaussian shaped narrow (wide) source and $\partial_i$ denotes a derivative
source in spatial direction $i$ (see Sec.~\ref{Methods}). $\gamma_t$ is the
gamma matrix in time direction, $\epsilon_{ijk}$ is the Levi-Civita symbol,
$Q_{ijk}$ are Clebsch-Gordon coefficients, where all elements are zero except
$Q_{111}=\frac{1}{\sqrt{2}}$, 
$Q_{122}=-\frac{1}{\sqrt{2}}$, 
$Q_{211}=-\frac{1}{\sqrt{6}}$, 
$Q_{222}=-\frac{1}{\sqrt{6}}$ and  
$Q_{233}=\frac{2}{\sqrt{6}}$.

\begin{table}[bp]
\begin{center}
\begin{tabular}{c|c|cl}
\hline
\hline
Interpolator 						& 0$^{-}$ ($K_0$) & 0$^{-+}$ & ($\pi$) Sym.    	  \\
\hline
 $\ov{a}_n \gamma_5 b_n			      		$&$   1_{0^-}  $&$1_{0^{-+}} $&$ =1_{0^-}	   $    		  \\
 $\ov{a}_n \gamma_5 b_w			      		$&$   2_{0^-}  $&$2_{0^{-+}} $&$ =2_{0^-}+3_{0^-}  $      	  \\
 $\ov{a}_w \gamma_5 b_n			      		$&$   3_{0^-}  $&$  	   $&$			   $    	  \\
 $\ov{a}_w \gamma_5 b_w			      		$&$   4_{0^-}  $&$3_{0^{-+}} $&$ =4_{0^-}			   $    		  \\
 $\ov{a}_n \gamma_t\gamma_5 b_n		      		$&$   5_{0^-}  $&$4_{0^{-+}} $&$ =5_{0^-}			   $    		  \\
 $\ov{a}_n \gamma_t\gamma_5 b_w		      		$&$   6_{0^-}  $&$5_{0^{-+}} $&$ =6_{0^-}+7_{0^-}  $   	  \\
 $\ov{a}_w \gamma_t\gamma_5 b_n		      		$&$   7_{0^-}  $&$  	   $&$			   $   	  \\
 $\ov{a}_w \gamma_t\gamma_5 b_w		      		$&$   8_{0^-}  $&$6_{0^{-+}} $&$ =8_{0^-}			   $    		  \\
 $\ov{a}_{\partial_i} \gamma_i\gamma_5 b_n	      	$&$   9_{0^-}  $&$7_{0^{-+}} $&$ =9_{0^-}+11_{0^-} $        	      \\ 
 $\ov{a}_{\partial_i} \gamma_i\gamma_5 b_w	      	$&$  10_{0^-}  $&$8_{0^{-+}} $&$ =10_{0^-}+12_{0^-}$        	    \\
 $\ov{a}_n \gamma_i\gamma_5 b_{\partial_i}	      	$&$  11_{0^-}  $&$  	   $&$			   $    	  \\
 $\ov{a}_w \gamma_i\gamma_5 b_{\partial_i}	      	$&$  12_{0^-}  $&$  	   $&$			   $   	  \\
 $\ov{a}_{\partial_i} \gamma_i\gamma_t\gamma_5 b_n   	$&$  13_{0^-}  $&$9_{0^{-+}} $&$ =13_{0^-}-15_{0^-}$    	  \\
 $\ov{a}_{\partial_i} \gamma_i\gamma_t\gamma_5 b_w   	$&$  14_{0^-}  $&$10_{0^{-+}}$&$ =14_{0^-}-16_{0^-}$   	  \\
 $\ov{a}_n \gamma_i\gamma_t\gamma_5 b_{\partial_i}   	$&$  15_{0^-}  $&$  	   $&$			   $    	  \\
 $\ov{a}_w \gamma_i\gamma_t\gamma_5 b_{\partial_i}   	$&$  16_{0^-}  $&$  	   $&$			   $   	  \\
 $\ov a_{\partial_i} \gamma_5 b_{\partial_i}	      	$&$  17_{0^-}  $&$11_{0^{-+}}$&$ =17_{0^-}			   $    		  \\
 $\ov a_{\partial_i} \gamma_t\gamma_5 b_{\partial_i} 	$&$  18_{0^-}  $&$12_{0^{-+}}$&$ =18_{0^-}			   $    		  \\
\hline
\hline
\end{tabular}
\end{center}
\caption{Pseudoscalar interpolators from irrep $A_1$.
The unsymmetrized interpolator, the corresponding number in the strange meson channel 0$^{-}$, 
the number and the symmetrization in terms of the  0$^{-}$ interpolators are given for the light meson channel 0$^{-+}$.}
\label{interpolators0-}
\end{table}

\begin{table}[tbp]
\begin{center}
\begin{tabular}{c|c|cl}
\hline
\hline
Interpolator & 0$^{+}$ ($K_0^*$)& 0$^{++}$  & ($a_0$) Sym.   	  \\
\hline
$\ov{a}_n b_n				       $&$  1_{0^+} $&$  1_{0^{++}} $&$=1_{0^+}		  $\\
$\ov{a}_n b_w				       $&$  2_{0^+} $&$  2_{0^{++}} $&$=2_{0^+}+3_{0^+}   $\\
$\ov{a}_w b_n				       $&$  3_{0^+} $&$             $&$  	  $\\
$\ov{a}_w b_w				       $&$  4_{0^+} $&$  3_{0^{++}} $&$=4_{0^+} 		  $\\
$\ov{a}_{\partial_i} \gamma_i b_n	       $&$  5_{0^+} $&$  4_{0^{++}} $&$=5_{0^+}-7_{0^+}   $\\
$\ov{a}_{\partial_i} \gamma_i b_w	       $&$  6_{0^+} $&$  5_{0^{++}} $&$=6_{0^+}-8_{0^+}   $\\
$\ov{a}_n \gamma_i b_{\partial_i}	       $&$  7_{0^+} $&$             $&$  	  $\\
$\ov{a}_w \gamma_i b_{\partial_i}	       $&$  8_{0^+} $&$             $&$  	  $\\
$\ov{a}_{\partial_i} \gamma_i\gamma_t b_n      $&$  9_{0^+} $&$  6_{0^{++}} $&$=9_{0^+}-11_{0^+}  $\\
$\ov{a}_{\partial_i} \gamma_i\gamma_t b_w      $&$ 10_{0^+} $&$  7_{0^{++}} $&$=10_{0^+}-12_{0^+} $\\
$\ov{a}_n \gamma_i\gamma_t b_{\partial_i}      $&$ 11_{0^+} $&$             $&$  	  $\\
$\ov{a}_w \gamma_i\gamma_t b_{\partial_i}      $&$ 12_{0^+} $&$             $&$  	  $\\
$\ov{a}_{\partial_i} b_{\partial_i}   	       $&$ 13_{0^+} $&$  8_{0^{++}} $&$ =13_{0^+}		  $\\
\hline
\hline
\end{tabular}
\end{center}
\caption{Scalar interpolators from irrep $A_1$.
The unsymmetrized interpolator, the corresponding number in the strange meson channel 0$^{+}$, 
the number and the symmetrization in terms of the  0$^{-}$ interpolators are given for the light meson channel 0$^{++}$.}
\end{table}

\begin{table}[tbp]
\begin{center}
\begin{tabular}{c|c|cl}
\hline
\hline
Interpolator & 1$^{-}$ ($K^*$) & 1$^{--}$  & ($\rho$) Sym.  \\
\hline
$\ov{a}_n \gamma_k b_n  			         	  $&$ 1_{1^-}  $&$  1_{1^{--}} $&$ =1_{1^-}		$\\
$\ov{a}_n \gamma_k b_w  			         	  $&$ 2_{1^-}  $&$  2_{1^{--}} $&$ =2_{1^-}+3_{1^-}	$\\
$\ov{a}_w \gamma_k b_n  			         	  $&$ 3_{1^-}  $&$  	       $&$ 			$\\
$\ov{a}_w \gamma_k b_w  			         	  $&$ 4_{1^-}  $&$  3_{1^{--}} $&$ =4_{1^-}		$\\
$\ov{a}_n \gamma_k\gamma_t b_n  		         	  $&$ 5_{1^-}  $&$  4_{1^{--}} $&$ =5_{1^-}		$\\
$\ov{a}_n \gamma_k\gamma_t b_w  		         	  $&$ 6_{1^-}  $&$  5_{1^{--}} $&$ =6_{1^-}+7_{1^-}	$\\
$\ov{a}_w \gamma_k\gamma_t b_n  		         	  $&$ 7_{1^-}  $&$  	       $&$ 			$\\
$\ov{a}_w \gamma_k\gamma_t b_w  		         	  $&$ 8_{1^-}  $&$  6_{1^{--}} $&$ =8_{1^-}		$\\
$\ov{a}_{\partial_k} b_n			         	  $&$ 9_{1^-}  $&$  7_{1^{--}} $&$ =9_{1^-}-11_{1^-}	$\\
$\ov{a}_{\partial_k} b_w			         	  $&$10_{1^-}  $&$  8_{1^{--}} $&$ =10_{1^-}-12_{1^-} $\\
$\ov{a}_n b_{\partial_k}			         	  $&$11_{1^-}  $&$ 	       $&$ 			$\\
$\ov{a}_w b_{\partial_k}			         	  $&$12_{1^-}  $&$ 	       $&$ 			$\\
$\ov{a}_{\partial_k} \gamma_t b_n		         	  $&$13_{1^-}  $&$  9_{1^{--}} $&$ =13_{1^-}+15_{1^-} $\\
$\ov{a}_{\partial_k} \gamma_t b_w		         	  $&$14_{1^-}  $&$ 10_{1^{--}} $&$ =14_{1^-}+16_{1^-} $\\
$\ov{a}_n \gamma_t b_{\partial_k}		       		  $&$15_{1^-}  $&$ 	       $&$ 			$\\
$\ov{a}_w \gamma_t b_{\partial_k}		        	  $&$16_{1^-}  $&$ 	       $&$ 			$\\
$\ov{a}_{\partial_i} \gamma_k b_{\partial_i}	        	  $&$17_{1^-}  $&$ 11_{1^{--}} $&$ =17_{1^-}		$\\
$\ov{a}_{\partial_i} \gamma_k\gamma_t b_{\partial_i}    	  $&$18_{1^-}  $&$ 12_{1^{--}} $&$ =18_{1^-}		$\\
$\ov{a}_{\partial_k} \epsilon_{ijk} \gamma_j\gamma_5 b_n 	  $&$19_{1^-}  $&$ 13_{1^{--}} $&$ =19_{1^-}-21_{1^-} $\\
$\ov{a}_{\partial_k} \epsilon_{ijk} \gamma_j\gamma_5 b_w 	  $&$20_{1^-}  $&$ 14_{1^{--}} $&$ =20_{1^-}-22_{1^-} $\\
$\ov{a}_n \epsilon_{ijk} \gamma_j\gamma_5 b_{\partial_k} 	  $&$21_{1^-}  $&$ 	       $&$ 			$\\
$\ov{a}_w \epsilon_{ijk} \gamma_j\gamma_5 b_{\partial_k} 	  $&$22_{1^-}  $&$ 	       $&$ 			$\\
\hline
\hline
\end{tabular}
\end{center}
\caption{Vector interpolators from irrep $T_1$.
The unsymmetrized interpolator, the corresponding number in the strange meson channel 1$^{-}$, 
the number and the symmetrization in terms of the  1$^{-}$ interpolators are given for the light meson channel 1$^{--}$.}
\end{table}

\begin{table*}[tbp]
\begin{center}
\begin{tabular}{c|c|cl|cl}
\hline
\hline
Interpolator & 1$^{+}$ ($K_1$)& 1$^{++}$ & ($a_1$) Sym. & 1$^{+-}$ & ($b_1$) Sym.  	  \\
\hline
$\ov{a}_n \gamma_k \gamma_5 b_n 		      		$&$ 1_{1^+}  $&$  1_{1^{++}} $&$=1_{1^+}		$&$		$&$			$\\
$\ov{a}_n \gamma_k \gamma_5 b_w 	        		$&$ 2_{1^+}  $&$  2_{1^{++}} $&$=2_{1^+}+3_{1^+}	$&$		$&$			$\\
$\ov{a}_w \gamma_k \gamma_5 b_n 	        		$&$ 3_{1^+}  $&$  	     $&$			$&$		$&$			$\\
$\ov{a}_w \gamma_k \gamma_5 b_w 	        		$&$ 4_{1^+}  $&$  3_{1^{++}} $&$=4_{1^+}		$&$		$&$			$\\
$\ov{a}_{\partial_k} \gamma_5 b_n	        		$&$ 5_{1^+}  $&$  4_{1^{++}} $&$=5_{1^+}+7_{1^+}	$&$4_{1^{+-}}	$&$=5_{1^+}-7_{1^+}	$\\
$\ov{a}_{\partial_k} \gamma_5 b_w	        		$&$ 6_{1^+}  $&$  5_{1^{++}} $&$=6_{1^+}+8_{1^+}	$&$5_{1^{+-}}	$&$=6_{1^+}-8_{1^+}	$\\
$\ov{a}_n \gamma_5 b_{\partial_k}	        		$&$ 7_{1^+}  $&$  	     $&$			$&$		$&$			$\\
$\ov{a}_w \gamma_5 b_{\partial_k}	        		$&$ 8_{1^+}  $&$  	     $&$			$&$		$&$			$\\
$\ov{a}_{\partial_k} \gamma_t \gamma_5 b_n      		$&$ 9_{1^+}  $&$  6_{1^{++}} $&$=9_{1^+}+11_{1^+}	$&$6_{1^{+-}}	$&$=9_{1^+}-11_{1^+}	$\\
$\ov{a}_{\partial_k} \gamma_t \gamma_5 b_w      		$&$10_{1^+}  $&$  7_{1^{++}} $&$=10_{1^+}+12_{1^+}	$&$7_{1^{+-}}	$&$=10_{1^+}-12_{1^+} $\\
$\ov{a}_n \gamma_t \gamma_5 b_{\partial_k}      		$&$11_{1^+}  $&$  	     $&$			$&$		$&$			$\\
$\ov{a}_w \gamma_t \gamma_5 b_{\partial_k}      		$&$12_{1^+}  $&$  	     $&$			$&$		$&$			$\\
$\ov{a}_{\partial_i} \gamma_k \gamma_5 b_{\partial_i} 		$&$13_{1^+}  $&$  8_{1^{++}} $&$=13_{1^+}		$&$		$&$			$\\
$\epsilon_{ijk}\ov{a}_{\partial_k} \gamma_j b_n         	$&$14_{1^+}  $&$  9_{1^{++}} $&$=14_{1^+}-16_{1^+}	$&$		$&$			$\\
$\epsilon_{ijk}\ov{a}_{\partial_k} \gamma_j b_w         	$&$15_{1^+}  $&$ 10_{1^{++}} $&$=15_{1^+}-17_{1^+}	$&$		$&$			$\\
$\epsilon_{ijk}\ov{a}_n \gamma_j b_{\partial_k}         	$&$16_{1^+}  $&$  	     $&$			$&$		$&$			$\\
$\epsilon_{ijk}\ov{a}_w \gamma_j b_{\partial_k}         	$&$17_{1^+}  $&$  	     $&$			$&$		$&$			$\\
$\epsilon_{ijk}\ov{a}_{\partial_k} \gamma_j\gamma_t b_n 	$&$18_{1^+}  $&$ 11_{1^{++}} $&$=18_{1^+}-20_{1^+}	$&$		$&$			$\\
$\epsilon_{ijk}\ov{a}_{\partial_k} \gamma_j\gamma_t b_w 	$&$19_{1^+}  $&$ 12_{1^{++}} $&$=19_{1^+}-21_{1^+}	$&$		$&$			$\\
$\epsilon_{ijk}\ov{a}_n \gamma_j\gamma_t b_{\partial_k} 	$&$20_{1^+}  $&$ 	     $&$			$&$		$&$			$\\
$\epsilon_{ijk}\ov{a}_w \gamma_j\gamma_t b_{\partial_k} 	$&$21_{1^+}  $&$	     $&$			$&$		$&$			$\\
$\ov{a}_n \gamma_k\gamma_t\gamma_5 b_n  		    	$&$22_{1^+}  $&$	     $&$			$&$1_{1^{+-}}	$&$=22_{1^+}		$\\
$\ov{a}_n \gamma_k\gamma_t\gamma_5 b_w  		    	$&$23_{1^+}  $&$             $&$			$&$2_{1^{+-}}	$&$=23_{1^+}+24_{1^+} $\\
$\ov{a}_w \gamma_k\gamma_t\gamma_5 b_n  		    	$&$24_{1^+}  $&$	     $&$			$&$		$&$			$\\
$\ov{a}_w \gamma_k\gamma_t\gamma_5 b_w  		    	$&$25_{1^+}  $&$	     $&$			$&$3_{1^{+-}}	$&$=25_{1^+}		$\\
$\ov{a}_{\partial_i} \gamma_k\gamma_t\gamma_5 b_{\partial_i}	$&$26_{1^+}  $&$	     $&$			$&$8_{1^{+-}}	$&$=26_{1^+}		$\\
\hline
\hline
\end{tabular}
\end{center}
\caption{Pseudovector interpolators from irrep $T_1$.
The unsymmetrized interpolator, the corresponding number in the strange meson channel 1$^{+}$, 
the number and the symmetrization in terms of the 1$^{+}$ interpolators are given for the light meson channels 1$^{++}$ and 1$^{+-}$.}
\end{table*}

\begin{table*}[tbp]
\begin{center}
\begin{tabular}{c|c|cl|cl}
\hline
\hline
Interpolator & 2$^{-}_{T_2}$ ($K_2$)& 2$^{--}_{T_2}$ & ($\rho_2$) Sym. & 2$^{-+}_{T_2}$ & ($\pi_2$) Sym.  	  \\
\hline
$|\epsilon_{ijk}| \bar{a}_{\partial_k} \gamma_j\gamma_5 b_n         $&$ 1_{2^-_{T_2}}  $&$    1_{2^{--}_{T_2}} $&$=1_{2^-_{T_2}}-3_{2^-_{T_2}} $&$		     $&$			$\\
$|\epsilon_{ijk}| \bar{a}_{\partial_k} \gamma_j\gamma_5 b_w	    $&$ 2_{2^-_{T_2}}  $&$    2_{2^{--}_{T_2}} $&$=2_{2^-_{T_2}}-4_{2^-_{T_2}} $&$		     $&$			$\\
$|\epsilon_{ijk}| \bar{a}_n \gamma_j\gamma_5 b_{\partial_k}	    $&$ 3_{2^-_{T_2}}  $&$  	       $&$			 $&$		     $&$			$\\
$|\epsilon_{ijk}| \bar{a}_w \gamma_j\gamma_5 b_{\partial_k}	    $&$ 4_{2^-_{T_2}}  $&$  	       $&$			 $&$		     $&$			$\\
$|\epsilon_{ijk}| \bar{a}_{\partial_k} \gamma_j\gamma_t\gamma_5 b_n $&$ 5_{2^-_{T_2}}  $&$  	       $&$			 $&$	1_{2^{-+}_{T_2}} $&$=5_{2^-_{T_2}}-7_{2^-_{T_2}}  $\\
$|\epsilon_{ijk}| \bar{a}_{\partial_k} \gamma_j\gamma_t\gamma_5 b_w $&$ 6_{2^-_{T_2}}  $&$  	       $&$			 $&$	2_{2^{-+}_{T_2}} $&$=6_{2^-_{T_2}}-8_{2^-_{T_2}}  $\\
$|\epsilon_{ijk}| \bar{a}_n \gamma_j\gamma_t\gamma_5 b_{\partial_k} $&$ 7_{2^-_{T_2}}  $&$  	       $&$			 $&$		     $&$			$\\
$|\epsilon_{ijk}| \bar{a}_w \gamma_j\gamma_t\gamma_5 b_{\partial_k} $&$ 8_{2^-_{T_2}}  $&$  	       $&$			 $&$		     $&$			$\\
\hline
\hline
\end{tabular}
\end{center}
\caption{Pseudotensor interpolators from irrep ${T_2}$.
The unsymmetrized interpolator, the corresponding number in the strange meson channel 2$^{-}_{T_2}$, 
the number and the symmetrization in terms of the 2$^{-}_{T_2}$ interpolators are given for the light meson channels 2$^{--}_{T_2}$ and 2$^{-+}_{T_2}$.}
\end{table*}

\begin{table*}[tbp]
\begin{center}
\begin{tabular}{c|c|cl|cl}
\hline
\hline
Interpolator & 2$^{-}_E$ ($K_2$) & 2$^{--}_E$ & ($\rho_2$) Sym. & 2$^{-+}_E$ & ($\pi_2$) Sym. 	  \\
\hline
$Q_{ijk} \ov{a}_{\partial_k} \gamma_j\gamma_5 b_n	    $&$ 1_{2^-_E}  $&$    1_{2^{--}_E} $&$=1_{2^-_E}-3_{2^-_E} $&$		     $&$			$\\
$Q_{ijk} \ov{a}_{\partial_k} \gamma_j\gamma_t\gamma_5 b_n   $&$ 2_{2^-_E}  $&$    2_{2^{--}_E} $&$=2_{2^-_E}-4_{2^-_E} $&$		     $&$			$\\
$Q_{ijk} \ov{a}_n \gamma_j\gamma_5 b_{\partial_k}	    $&$ 3_{2^-_E}  $&$  	       $&$			 $&$		     $&$			$\\
$Q_{ijk} \ov{a}_w \gamma_j\gamma_5 b_{\partial_k}	    $&$ 4_{2^-_E}  $&$  	       $&$			 $&$		     $&$			$\\
$Q_{ijk} \ov{a}_{\partial_k} \gamma_j\gamma_t\gamma_5 b_n   $&$ 5_{2^-_E}  $&$  	       $&$			 $&$	1_{2^{-+}_E} $&$=5_{2^-_E}-7_{2^-_E}  $\\
$Q_{ijk} \ov{a}_{\partial_k} \gamma_j\gamma_t\gamma_5 b_w   $&$ 6_{2^-_E}  $&$  	       $&$			 $&$	2_{2^{-+}_E} $&$=6_{2^-_E}-8_{2^-_E}  $\\
$Q_{ijk} \ov{a}_n \gamma_j\gamma_t\gamma_5 b_{\partial_k}   $&$ 7_{2^-_E}  $&$  	       $&$			 $&$		     $&$			$\\
$Q_{ijk} \ov{a}_w \gamma_j\gamma_t\gamma_5 b_{\partial_k}   $&$ 8_{2^-_E}  $&$  	       $&$			 $&$		     $&$			$\\
$Q_{ijk} \ov{a}_{\partial_j} \gamma_5 b_{\partial_k}	    $&$ 9_{2^-_E}  $&$  	       $&$			 $&$	3_{2^{-+}_E} $&$=9_{2^-_E}		$\\
$Q_{ijk} \ov{a}_{\partial_j} \gamma_t\gamma_5 b_{\partial_k}$&$10_{2^-_E}  $&$  	       $&$			 $&$	4_{2^{-+}_E} $&$=10_{2^-_E}		$\\
\hline
\hline
\end{tabular}
\end{center}
\caption{Pseudotensor interpolators from irrep $E$.
The unsymmetrized interpolator, the corresponding number in the strange meson channel 2$^{-}_E$, 
the number and the symmetrization in terms of the 2$^{-}_E$ interpolators are given for the light meson channels 2$^{--}_E$ and 2$^{-+}_E$.}
\end{table*}

\begin{table*}[tbp]
\begin{center}
\begin{tabular}{c|c|cl}
\hline
\hline
Interpolator & 2$^{+}_{T_2}$ ($K_2^*$) & 2$^{++}_{T_2}$ & ($a_2$) Sym. \\
\hline
$|\epsilon_{ijk}| \bar{a}_{\partial_k} \gamma_j b_n         $&$ 1_{2^+_{T_2}}  $&$    1_{2^{++}_{T_2}} $&$=1_{2^+_{T_2}}-3_{2^+_{T_2}} $\\
$|\epsilon_{ijk}| \bar{a}_{\partial_k} \gamma_j b_w	    $&$ 2_{2^+_{T_2}}  $&$    2_{2^{++}_{T_2}} $&$=2_{2^+_{T_2}}-4_{2^+_{T_2}} $\\	  
$|\epsilon_{ijk}| \bar{a}_n \gamma_j b_{\partial_k}	    $&$ 3_{2^+_{T_2}}  $&$	       $		        		  \\
$|\epsilon_{ijk}| \bar{a}_w \gamma_j b_{\partial_k}	    $&$ 4_{2^+_{T_2}}  $&$	       $		        		  \\
$|\epsilon_{ijk}| \bar{a}_{\partial_k} \gamma_j\gamma_t b_n $&$ 5_{2^+_{T_2}}  $&$    3_{2^{++}_{T_2}} $&$=5_{2^+_{T_2}}-7_{2^+_{T_2}} $\\
$|\epsilon_{ijk}| \bar{a}_{\partial_k} \gamma_j\gamma_t b_w $&$ 6_{2^+_{T_2}}  $&$    4_{2^{++}_{T_2}} $&$=6_{2^+_{T_2}}-8_{2^+_{T_2}} $\\	  
$|\epsilon_{ijk}| \bar{a}_n \gamma_j\gamma_t b_{\partial_k} $&$ 7_{2^+_{T_2}}  $&$	       $					  \\
$|\epsilon_{ijk}| \bar{a}_w \gamma_j\gamma_t b_{\partial_k} $&$ 8_{2^+_{T_2}}  $&$	       $			 		  \\
\hline
\hline
\end{tabular}
\end{center}
\caption{Tensor interpolators from irrep ${T_2}$.
The unsymmetrized interpolator, the corresponding number in the strange meson channel 2$^{+}_{T_2}$, 
the number and the symmetrization in terms of the 2$^{+}_{T_2}$ interpolators are given for the light meson channel 2$^{++}_{T_2}$.}
\end{table*}

\begin{table}[tbp]
\begin{center}
\begin{tabular}{c|c|cl}
\hline
\hline
Interpolator & 2$^{+}_E$ ($K_2^*$) & 2$^{++}_E$ & ($a_2$) Sym \\
\hline
$Q_{ijk} \bar{a}_{\partial_k} \gamma_j b_n	      $&$ 1_{2^+_E}  $&$    1_{2^{++}_E} $&$=1_{2^+_E}-3_{2^+_E} $\\
$Q_{ijk} \bar{a}_{\partial_k} \gamma_j b_w	      $&$ 2_{2^+_E}  $&$    2_{2^{++}_E} $&$=2_{2^+_E}-4_{2^+_E} $\\  
$Q_{ijk} \bar{a}_n \gamma_j b_{\partial_k}	      $&$ 3_{2^+_E}  $&$		 $&$			   $\\
$Q_{ijk} \bar{a}_w \gamma_j b_{\partial_k}	      $&$ 4_{2^+_E}  $&$		 $&$			   $\\
$Q_{ijk} \bar{a}_{\partial_k} \gamma_j\gamma_t b_n    $&$ 5_{2^+_E}  $&$    3_{2^{++}_E} $&$=5_{2^+_E}-7_{2^+_E} $\\
$Q_{ijk} \bar{a}_{\partial_k} \gamma_j\gamma_t b_w    $&$ 6_{2^+_E}  $&$    4_{2^{++}_E} $&$=6_{2^+_E}-8_{2^+_E} $\\
$Q_{ijk} \bar{a}_n \gamma_j\gamma_t b_{\partial_k}    $&$ 7_{2^+_E}  $&$		 $&$			   $\\
$Q_{ijk} \bar{a}_w \gamma_j\gamma_t b_{\partial_k}    $&$ 8_{2^+_E}  $&$		 $&$			   $\\
$Q_{ijk} \bar{a}_{\partial_j} b_{\partial_k}	      $&$ 9_{2^+_E}  $&$    5_{2^{++}_E} $&$=9_{2^+_E}		   $\\
\hline
\hline
\end{tabular}
\end{center}
\caption{Tensor interpolators from irrep $E$.
The unsymmetrized interpolator, the corresponding number in the strange meson channel 2$^{+}_E$, 
the number and the symmetrization in terms of the 2$^{+}_E$ interpolators are given for the light meson channels 2$^{++}_E$.}
\label{interpolators2+}
\end{table}

\subsection{Baryon interpolators}
The baryon interpolators are slightly more complicated, the construction in each channel is given by:
\begin{eqnarray}
N^{(i)} & = & \epsilon_{abc} \Gamma^{(i)}_1 u_a ( u_b^T  \Gamma^{(i)}_2 d_c - d_b^T  \Gamma^{(i)}_2 u_c )  \\
\Sigma^{(i)} & = & \epsilon_{abc} \Gamma^{(i)}_1 u_a ( u_b^T  \Gamma^{(i)}_2 s_c - s_b^T  \Gamma^{(i)}_2 u_c ) \\
\Xi^{(i)} & = & \epsilon_{abc} \Gamma^{(i)}_1 s_a ( s_b^T  \Gamma^{(i)}_2 u_c - u_b^T  \Gamma^{(i)}_2 s_c ) \\
\Delta_\mu & = & \epsilon_{abc}  u_a ( u_b^T C\gamma_\mu  u_c )\\
\Omega_\mu & = & \epsilon_{abc}  s_a ( s_b^T C\gamma_\mu  s_c )
\end{eqnarray}
Subsequent numbering of the interpolators with respect to gamma and smearing structure is performed in each channel, leading to the interpolator numbers given in Tables \ref{tab:baryoninterpolators1} and \ref{tab:baryoninterpolators2}.
Projection to definite parity is performed by the projection operator $P_{\pm}=(\mathds{1}\pm\gamma_t)/2$.
$C$ is the charge conjugation operator, in the chiral representation it can be written as $C=\imath\,\gamma_2\gamma_4$.
The $\Delta$ and $\Omega$ interpolators are projected to spin $\frac{3}{2}$ and averaged over the three spatial vector components \cite{Burch:2006cc}.

\begin{table}[tbp]
\begin{center}
\begin{tabular}{ccc|c|c|c}
\hline
\hline
$\chi^{(i)}$ 	& $\Gamma^{(i)}_1$ 	& $\Gamma^{(i)}_2$	& Smearing 	  &  N	  & $\Sigma$, $\Xi$ 	\\
\hline
$\chi^{(1)}$	& $\mathds{1}$  	& $C\,\gamma_5$           & $n(nn)$         & 1     & 1   \\
$\chi^{(1)}$	& $\mathds{1}$  	& $C\,\gamma_5$           & $n(nw)$         & 2     & 2   \\
$\chi^{(1)}$	& $\mathds{1}$  	& $C\,\gamma_5$           & $n(wn)$         &       & 3   \\
$\chi^{(1)}$	& $\mathds{1}$  	& $C\,\gamma_5$           & $w(nn)$         & 3     & 4   \\
$\chi^{(1)}$	& $\mathds{1}$  	& $C\,\gamma_5$           & $n(ww)$         & 4     & 5   \\
$\chi^{(1)}$	& $\mathds{1}$  	& $C\,\gamma_5$           & $w(nw)$         & 5     & 6   \\
$\chi^{(1)}$	& $\mathds{1}$  	& $C\,\gamma_5$           & $w(wn)$         &       & 7   \\
$\chi^{(1)}$	& $\mathds{1}$  	& $C\,\gamma_5$           & $w(ww)$         & 6     & 8   \\
\hline
$\chi^{(2)}$	& $\gamma_5$    	& $C$ 			& $n(nn)$	  & 7	  & 9	\\
$\chi^{(2)}$	& $\gamma_5$    	& $C$ 			& $n(nw)$	  & 8	  & 10  \\
$\chi^{(2)}$	& $\gamma_5$    	& $C$ 			& $n(wn)$	  &	  & 11  \\
$\chi^{(2)}$	& $\gamma_5$    	& $C$ 			& $w(nn)$	  & 9	  & 12  \\
$\chi^{(2)}$	& $\gamma_5$    	& $C$ 			& $n(ww)$	  & 10    & 13  \\
$\chi^{(2)}$	& $\gamma_5$    	& $C$ 			& $w(nw)$	  & 11    & 14  \\
$\chi^{(2)}$	& $\gamma_5$    	& $C$ 			& $w(wn)$	  &	  & 15  \\
$\chi^{(2)}$	& $\gamma_5$    	& $C$ 			& $w(ww)$	  & 12    & 16  \\
\hline
$\chi^{(3)}$	& $\imath\,\mathds{1}$ 	& $C\,\gamma_t\,\gamma_5$   & $n(nn)$         & 13    & 17  \\
$\chi^{(3)}$	& $\imath\,\mathds{1}$ 	& $C\,\gamma_t\,\gamma_5$   & $n(nw)$         & 14    & 18  \\
$\chi^{(3)}$	& $\imath\,\mathds{1}$ 	& $C\,\gamma_t\,\gamma_5$   & $n(wn)$         &       & 19  \\
$\chi^{(3)}$	& $\imath\,\mathds{1}$ 	& $C\,\gamma_t\,\gamma_5$   & $w(nn)$         & 15    & 20  \\
$\chi^{(3)}$	& $\imath\,\mathds{1}$ 	& $C\,\gamma_t\,\gamma_5$   & $n(ww)$         & 16    & 21  \\
$\chi^{(3)}$	& $\imath\,\mathds{1}$ 	& $C\,\gamma_t\,\gamma_5$   & $w(nw)$         & 17    & 22  \\
$\chi^{(3)}$	& $\imath\,\mathds{1}$ 	& $C\,\gamma_t\,\gamma_5$   & $w(wn)$         &       & 23  \\
$\chi^{(3)}$	& $\imath\,\mathds{1}$ 	& $C\,\gamma_t\,\gamma_5$   & $w(ww)$         & 18    & 24  \\
\hline
\hline
\end{tabular}
\end{center}
\caption{Baryon interpolators for nucleon, $\Sigma$ and $\Xi$ channels.
The Dirac structures, the quark smearings and the corresponding interpolator numbers are given.
$\chi^{(i)}$ labels the Dirac structure of the baryon interpolators. 
In the nucleon channel we prune interpolators which are very similar to others, 
obtaining six per Dirac structure, thus a total of 18 interpolators.}
\label{tab:baryoninterpolators1}
\end{table}

\begin{table}[tbp]
\begin{center}
\begin{tabular}{c|c}
\hline
\hline
Smearing 	&  $\Delta$, $\Omega$	\\
\hline
$n(nn)$ 	 & 1		 	\\
$n(nw)$ 	 & 2		 	\\
$w(nn)$ 	 & 3		 	\\
$n(ww)$ 	 & 4		 	\\
$w(nw)$ 	 & 5		 	\\
$w(ww)$ 	 & 6		 	\\
\hline
\hline
\end{tabular}
\end{center}
\caption{Baryon interpolators for $\Delta$ and $\Omega$ channels. The quark
smearings and the corresponding interpolator numbers are given. We prune
interpolators which are very similar to others, obtaining six interpolators.}
\label{tab:baryoninterpolators2}
\end{table}

\end{appendix}
\clearpage


\end{document}